\RequirePackage{rotating}
\documentclass[fleqn,usenatbib]{mnras}

\usepackage{newtxtext,newtxmath}
\usepackage[T1]{fontenc}
\usepackage{ae,aecompl}

\usepackage{graphicx}	% Including figure files
\usepackage{amssymb}	% Extra maths symbols
\usepackage{fixltx2e} % helps with sub and superscripting
\usepackage{textgreek} %gets me greek symbols without having to be in math mode.
\usepackage{url}		%gets the URLs to hyperlink
\usepackage{array}
\usepackage{pdflscape}
\usepackage{lastpage}
\usepackage[caption = false]{subfig}
\usepackage[flushleft]{threeparttable}
\usepackage{CJKutf8}

\def\teff{{T$_{\text{eff}}$\,}}

\title[GALAH-TESS Catalog]{The GALAH Survey: Using Galactic Archaeology to Refine our Knowledge of \textit{TESS} Target Stars}

\author[J.T. Clark et al.]{Jake T. Clark,$^{1}$\thanks{E-mail: jake.clark@usq.edu.au}
Mathieu Clert\'e,$^{1}$
Natalie R. Hinkel,$^{2}$
Cayman T. Unterborn,$^{3}$
\newauthor
Robert A. Wittenmyer,$^{1}$
Jonathan Horner,$^{1}$
Duncan J. Wright,$^{1}$
Brad Carter,$^{1}$
\newauthor
Timothy D. Morton,$^{4}$
Lorenzo Spina,$^{5}$
Martin Asplund,$^{6}$
Sven Buder,$^{6,7}$
\newauthor
Joss Bland-Hawthorn,$^{8,7}$
Andy Casey,$^{5}$
Gayandhi De Silva,$^{5}$
Valentina D'Orazi,$^{9}$
\newauthor
Ly Duong,$^{6}$
Michael Hayden,$^{8,7}$
Ken Freeman,$^{6}$
Janez Kos,$^{10,7}$
Geraint Lewis,$^{8}$
\newauthor
Jane Lin,$^{6}$
Karin Lind,$^{11}$
Sarah Martell,$^{12,7}$
Sanjib Sharma,$^{8}$
Jeffrey Simpson,$^{8,7}$
\newauthor
Dan Zucker,$^{13,14}$
Tomaz Zwitter,$^{10}$
Christopher G. Tinney,$^{15,16}$
\newauthor
Yuan-Sen Ting (\begin{CJK*}{UTF8}{gbsn}丁源森\end{CJK*}),$^{17,18,19,6}$ and
Thomas Nordlander$^{6,7}$
\\
$^{1}$Centre for Astrophysics, University of Southern Queensland, West Street, Toowoomba, QLD 4350, Australia\\
$^{2}$Space Science and Engineering Division, Southwest Research Institute, San Antonio, TX 78238, USA\\
$^{3}$School of Earth and Space Exploration, Arizona State University, Tempe, AZ 85287, USA\\
$^{4}$Department of Physics and Astronomy, University of Southern California, Los Angeles, CA 90089-0484, USA\\
$^{5}$Monash Centre for Astrophysics, School of Physics and Astronomy, Monash University, VIC 3800, Australia\\
$^{6}$Research School of Astronomy \& Astrophysics, Australian National University, ACT 2611, Australia\\
$^{7}$Center of Excellence for Astrophysics in Three Dimensions (ASTRO-3D), Australia\\
$^{8}$Sydney Institute for Astronomy, School of Physics, University of Sydney, NSW 2006, Australia\\
$^{9}$INAF Osservatorio Astronomico di Padova, vicolo dell'Osservatorio 5, 35122, Padova, Italy\\
$^{10}$Faculty of Mathematics and Physics, University of Ljubljana, Jadranska 19, 1000 Ljubljana, Slovenia\\
$^{11}$Max Planck Institute for Astronomy (MPIA), Koenigstuhl 17, 69117 Heidelberg, Germany\\
$^{12}$School of Physics, University of New South Wales, Sydney 2052, Australia\\
$^{13}$Department of Physics \& Astronomy, Macquarie University, Sydney, NSW 2109, Australia\\
$^{14}$Research Centre in Astronomy, Astrophysics \& Astrophotonics, Macquarie University, Sydney, NSW 2109, Australia\\
$^{15}$Exoplanetary Science at UNSW, University of New South Wales, Sydney, NSW 2052, Australia\\
$^{16}$Australian Centre for Astrobiology,University of New South Wales, Sydney, NSW 2052, Australia\\
$^{17}${Institute for Advanced Study, Princeton, NJ 08540, USA}\\
$^{18}${Department of Astrophysical Sciences, Princeton University, Princeton, NJ 08544, USA}\\
$^{19}${Observatories of the Carnegie Institution of Washington, 813 Santa Barbara Street, Pasadena, CA 91101, USA}\\
$^{20}${Theoretical Astrophysics, Department of Physics and Astronomy, Uppsala University, Box 516, SE-751 20 Uppsala, Sweden}\\
}

\date{Accepted XXX. Received YYY; in original form ZZZ}

\pubyear{2020}
\begin{document}
\label{firstpage}
\pagerange{\pageref{firstpage}--\pageref{lastpage}}
\maketitle

\begin{abstract}

\noindent
An unprecedented number of exoplanets are being discovered by the Transiting Exoplanet Survey Satellite (\textit{TESS}). Determining the orbital parameters of these exoplanets, and especially their mass and radius, will depend heavily upon the measured physical characteristics of their host stars.
We have cross-matched spectroscopic, photometric, and astrometric data from GALAH Data Release 2, the \textit{TESS} Input Catalog and \textit{Gaia} Data Release 2, to create a curated, self-consistent catalog of physical and chemical properties for 47,285 stars. Using these data we have derived isochrone masses and radii that are precise to within 5\%. We have revised the parameters of three confirmed, and twelve candidate, \textit{TESS} planetary systems. These results cast doubt on whether CTOI-20125677 is indeed a planetary system since the revised planetary radii are now comparable to stellar sizes.
Our GALAH-TESS catalog contains abundances for up to 23 elements. We have specifically analysed the molar ratios for C/O, Mg/Si, Fe/Si and Fe/Mg, to assist in determining the composition and structure of planets with $R_p < 4R_\oplus$. From these ratios, 36~\% fall within 2 sigma of the Sun/Earth values, suggesting that these stars may host rocky exoplanets with geological compositions similar to planets found within our own Solar system.

\end{abstract}


%%%%%%%%%%%%%%%%%%%%%%%%%%%%%%%%%%%%%%%%%%%%%%%%%%

%%%%%%%%%%%%%%%%% BODY OF PAPER %%%%%%%%%%%%%%%%%%

\section{Introduction}
% \newpage \noindent
Exoplanets (planets that exist beyond the Solar system) moved beyond science fiction and into the realm of hard science late in the 20th century \citep{Latham89,gammaCeph,Lich,Mayor1995}. Over the first 18 years of the exoplanet era, radial velocity detections dominated exoplanet discovery, leading to a wealth of massive planet discoveries around largely Sun-like stars \citep[e.g.][]{RV2,RV1,RV3,RV4,RV5}. 

Towards the end of the first decade of the 2000s, the transit technique became the numerically dominant method for making new exoplanet discoveries, and revealed an abundance of planets moving on very short period orbits \citep[e.g.][]{TR1,TR3,TR2,TR5,TR6}. The great advantage of transit observations over those using the radial velocity technique is that they permit surveys to target large numbers of stars simultaneously. The ultimate expression, to date, of the transit method as a tool for exoplanetary science came with the \textit{Kepler} space telescope, launched in (2009) \citep{Kepler}. %\textit{Kepler} was capable of monitoring the brightness of more than 150000 stars at once, maximising the likelihood that it would observe transit events.

At the time of writing, \textit{Kepler} has been by far the most successful exoplanet detection program, discovering 65.5\% of currently known exoplanets\footnote{\url{https://exoplanetarchive.ipac.caltech.edu/}; accessed 6 August 2020, counting discoveries from both \textit{Kepler}'s primary mission, and the K2 survey.}. These planetary discoveries have showcased the vast richness and diversity of exoplanets across our galaxy. The great diversity of exoplanets and exoplanetary systems is illustrated by the discovery of large numbers of multi-planet systems \citep[e.g.][]{Gillon17,Shallue18}, planets in extremely eccentric orbits \citep[e.g.][]{Naef01,Wittenmyer17,Kepler420}, and planets that some have argued might resemble the Earth \citep[e.g.][]{Torres15_KOIval,Kepler69lol}.

The \textit{Kepler} and K2 missions also revealed that planets larger than Earth, $> 1R_\oplus$, yet smaller than Neptune, $< 4R_\oplus$, are remarkably common -- despite there being no such planets in the Solar system. Indeed, of those planets that we can readily detect, these ``super-Earths'' and ``mini-Neptunes'' seem to be by far the most common \citep{Batalha13}. On 30 October 2018, the \textit{Kepler} spacecraft depleted all of its on-board fuel, immediately retiring the mission and leaving behind a legacy that is unmatched in exoplanetary science. Fortunately, NASA's Transiting Exoplanet Survey Satellite (\textit{TESS}) mission, launched in April 2018, has picked up where the K2 mission left off.

The \textit{TESS} mission \citep{TESS} is a space-based photometric survey that will cover the entire sky, except for the region within $\pm$6 degrees of the ecliptic plane. The mission is designed to find small planets ($R_p < 2.5 R_\oplus$) around nearby, bright, main sequence stars. As of 2020 August, there have been 66 confirmed planetary discoveries made as a result of \textit{TESS'} ongoing survey \citep[e.g.][]{Huang2018,Vanderspek2019,TOI120,TESS700d,TOI257,TOI677}. In addition to the 66 confirmed \textit{TESS} exoplanets, there are more than two thousand \textit{TESS} Targets of Interest (TOI) and Community Targets of Interest (CTOI)\footnote{\url{https://exofop.ipac.caltech.edu/tess/}; accessed 6 August 2020.} waiting for their exoplanetary status to be confirmed by ground-based teams \citep[e.g][]{minervaAus,TESSLCO,TOI120,SonghuTESS,TESSKeck,planethuntersTESS}.

Once potential planets have been identified by \textit{TESS}, the \textit{TESS} Input Catalog (TIC) \citep{TIC_CTL,revisedTIC} and Candidate Target List (CTL) are the key catalogues that enable follow-up teams to characterise -- for both stars and planets -- the members of \textit{TESS} candidate systems. In particular, radial velocity data are needed to measure the planetary mass, and spectroscopic observations are needed to refine mass and radius of the host star. These measurements, in combination with the transit radius measurement from \textit{TESS}, allow the bulk density of the planet in question, $\rho_p$\footnote{$\rho_p = M_{p}\left(\frac{4}{3} \pi R_p^3\right)^{-1}$; where $M_p$ and $R_p$ are the planet's mass and radius, respectively.}, to be determined, and thereby provide constraints on that planet's overall composition \citep{Unterborn2016,Seager2007,Valencia2006}. 

Whilst the bulk density of a planet does provide clues to its potential bulk composition, it does not provide enough information for us to determine the geological structure of a potentially rocky planet, or to precisely determine its true composition. This is clearly illustrated by the work of
\citet{legendoftherentwasway}, who demonstrate that a newly discovered ``Earth-like" planet (a planet observed to be both the same mass and the same size as the Earth; i.e. 1 $M_\oplus$, 1 $R_\oplus$) could have a wide variety of internal compositions. Distinguishing between the many possible compositions and structures of such a planet will be of great interest in the years to come, particularly in the context of the search for potentially habitable planets, and the selection of the most promising such planets for further study \citep[e.g.][]{WhichExoEarth}.

Recent studies, however, have demonstrated that planetary scientists could potentially unlock the viscera of distant rocky worlds by combining our knowledge of the planets themselves with detailed information on the chemical abundances of their host stars \citep[e.g.][]{UnterbornPanero2019,DornCaAlPaper,Hinkel2018,Unterborn2018,Unterborn17,Dorn2017,Unterborn2016,Dorn2015,Bond10,Bond2010b}. In particular, knowledge of the chemical abundances of refractory elements (such as Mg, Al, Si, Ca and Fe) and volatile elements (such as C and O) can help us to determine the likely structure and composition of exoplanets smaller than 4$R_\oplus$ \citep{DornCaAlPaper,exoplanetRockyComp}. 

The most crucial elements for such an analysis are C, O, Mg, Si, and Fe, as these elements will determine the core to mantle fraction (in particular Fe/Mg\footnote{Stellar abundance ratios are calculated by:\newline $\left(\frac{\text{X}}{\text{Y}}\right)_\star = 10^{\left(\left[\frac{\text{X}}{\text{H}}\right]_\star + \text{A}(\text{X})_{\odot}\right) -\left(\left[\frac{\text{Y}}{\text{H}}\right]_\star + \text{A}(\text{Y})_{\odot}\right)}$}) and the composition of a rocky exoplanet's mantle \citep[e.g. Mg/Si and C/O, as per ][]{Dorn2017,Dorn2015,Unterborn14,Madhusudhan12,Bond10}. Such models have recently proven vital in inferring the geological and chemical composition of the planets in the TRAPPIST-1 \citep{Unterborn2018}, 55 Cnc \citep{Dorn2017b}, HD\,219134 \citep{Ligi2019}, and other planetary systems. 

\begin{figure*}
\begin{center}
\includegraphics[width=\textwidth]{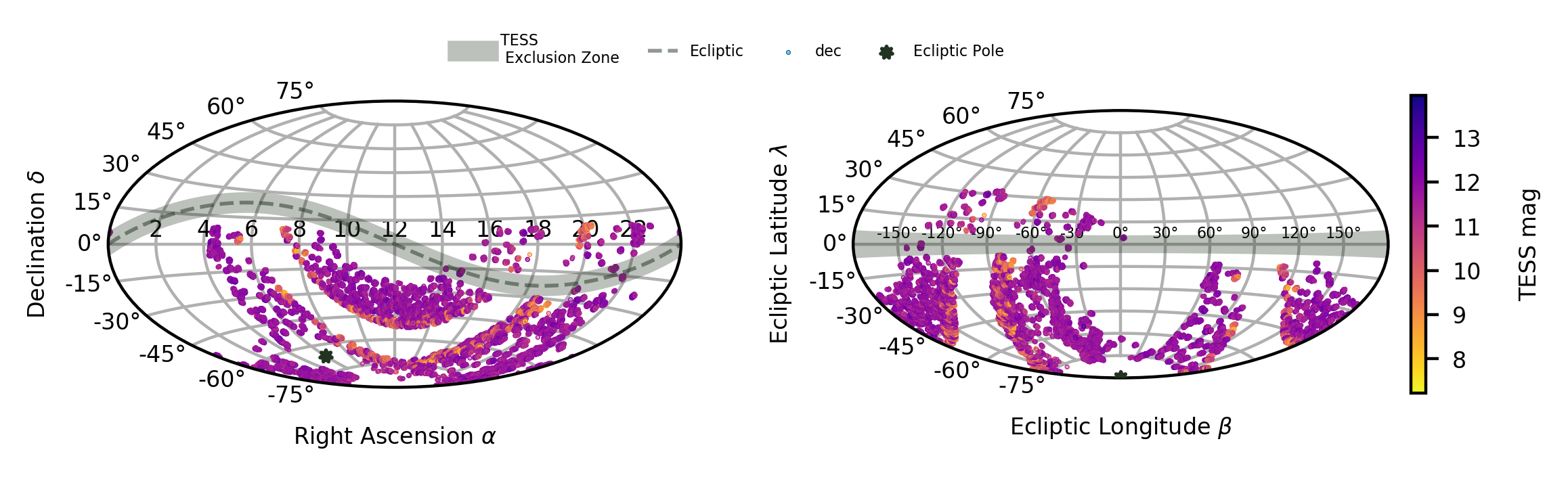}
\caption{Aitoff projection of GALAH-TESS stars in both equatorial (right ascension and declination) and ecliptic (latitude and longitude) co-ordinates. The ecliptic plane, southern ecliptic pole, and \textit{TESS} exclusion zone are shown in each (except for the ecliptic from the ecliptic co-ordinate plot as it corresponds to \textlambda = 0$^{\circ}$). Stars that are observed with HERMES within \textit{TESS}'s Continuous Viewing Zone are a part of the TESS-HERMES survey, and thus not observed with GALAH. Stars within the \textit{TESS} exclusion zone have been left within the GALAH-TESS catalog, as these stars may be observed in the future with \textit{TESS}.}\label{fig:galah_sky}
\end{center}
\end{figure*}

As the catalogue of known exoplanets has grown, it has becoming increasingly obvious that our understanding of the planets we find is often limited by the precision with which we can characterise their host stars. In particular, measurements of the elemental abundances of exoplanet host stars are becoming increasingly important in developing our understanding of the fundamental synergies between stars and the planets they host. As a result, there is an increasing amount of research within exoplanetary science that aims to understand the relationship between a star's chemical abundances and the types of planets and planetary systems that they can form \citep[e.g.][]{Fischer2005,Adibekyan12,Buchhave2014,Buchhave2015,Teske_metalrichhosts}. 

The relationship between planetary demographics and a star's measured photospheric iron abundance is a complex one. Over twenty years ago, studies showed that stars hosting hot-Jupiters (giant planets in very short period orbits) are %more likely to be 
typically iron-enriched compared to the Sun \citep{hotjupfeh97,hotjupfeh01,Fischer2005}. This trend has, however, weakened in more recent studies \citep{osbornhotjup,Teske_metalrichhosts}. Similarly, the relationship between a star's iron abundance and the number of planets it hosts remains the subject of significant debate \citep[e.g.][]{CKSIV,MgSiNeptunes}. Recent machine-learning work by \citet{Hinkel2019} has indicated that elemental abundances, including those of C, O, and Fe, can be used as a means to identify potential planet-hosting stars amongst the wider stellar population. 

In addition to potentially helping us to understand the interior structure and composition of newly discovered exoplanets, recent work has also suggested that measurements of the elemental abundances of stellar photospheres and planetary atmospheres could also aid our investigation of the formation and migration history of the exoplanets we study. For example, \citet{COhotjupBrew} describe how measurements of an enhanced C/O ratio and [O/H] abundance in the atmospheres of ten hot Jupiters, compared to the equivalent abundances in their stellar hosts, serve as evidence that those planets must have formed beyond the water ice line, and that they must have then migrated inwards to reach their current location. In a similar fashion,
%For example, measurements of an enhanced C/O ratio and [O/H] abundance in exoplanetary atmospheres (compared to their stellar counterparts) can confirm that some hot-Jupiters have formed beyond the water ice line and have migrated inwards to their current location \citep{COhotjupBrew}. Indeed, 
studies of the composition and isotopic abundances of the planets and small bodies have long been used to attempt to disentangle their formation locations and migration histories \citep[see e.g.][, and references therein]{SSreview}. In summary, this recent work reveals that, if we are to fully characterise the exoplanets we discover, it is vital that we consider the elemental abundances of their host stars.

The southern hemisphere's largest spectroscopic stellar abundance survey -- the Galactic Archaeology with HERMES (GALAH) survey -- is designed to investigate the stellar formation and chemical enrichment history of the Milky Way galaxy \citep{DeSilva15,GALAHDR1,GALAHDR2}. To do this, GALAH has collected high-resolution spectra for more than 600,000 stars, from which the abundances of up to 23 elements can be determined for each star. GALAH's latest public release, GALAH DR2 \citep{GALAHDR2}, contains the details of 342,682 stars for which both physical and chemical properties have been observed and derived. 

In this work, we make use of the data in GALAH DR2 to calculate revised values for the mass and radius of 47,285 stars that have been cross-matched between GALAH DR2 and the TIC. We then calculate the C/O, Fe/Mg and Mg/Si abundance ratios for those stars, providing a database of stellar abundances for potential planet hosting stars to facilitate future studies of the composition, structure, habitability, and migration history of exoplanets discovered by \textit{TESS}.
 
In Section~\ref{sec:method}, we describe how GALAH DR2 is cross-matched with the \textit{TESS} and \textit{Gaia} catalogs (Section \ref{sec:cross}), before describing how we derive the isochronic stellar mass, radius and age estimates for our stars (Section~\ref{sec:isocrone}). We then go on to discuss the derivation of elemental abundances and abundance ratios for GALAH-TESS stars using GALAH DR2 (Section~\ref{sec:methodabund}). The resulting physical and elemental parameters are then validated by comparison with other catalogs in Sections~\ref{sec:physparam} and \ref{sec:chem}. In our discussion section, we examine our refined stellar and planetary parameters for confirmed and candidate exoplanet host stars (Section~\ref{sec:refineplanetsystems}) and the abundance ratio trends in our stellar sample (Section~\ref{sec:abudancediscussion}). Finally, we summarise our findings and draw our conclusions in Section~\ref{sec:conclusion}.

\begin{figure*}
 \begin{center}
	\includegraphics[width=\textwidth]{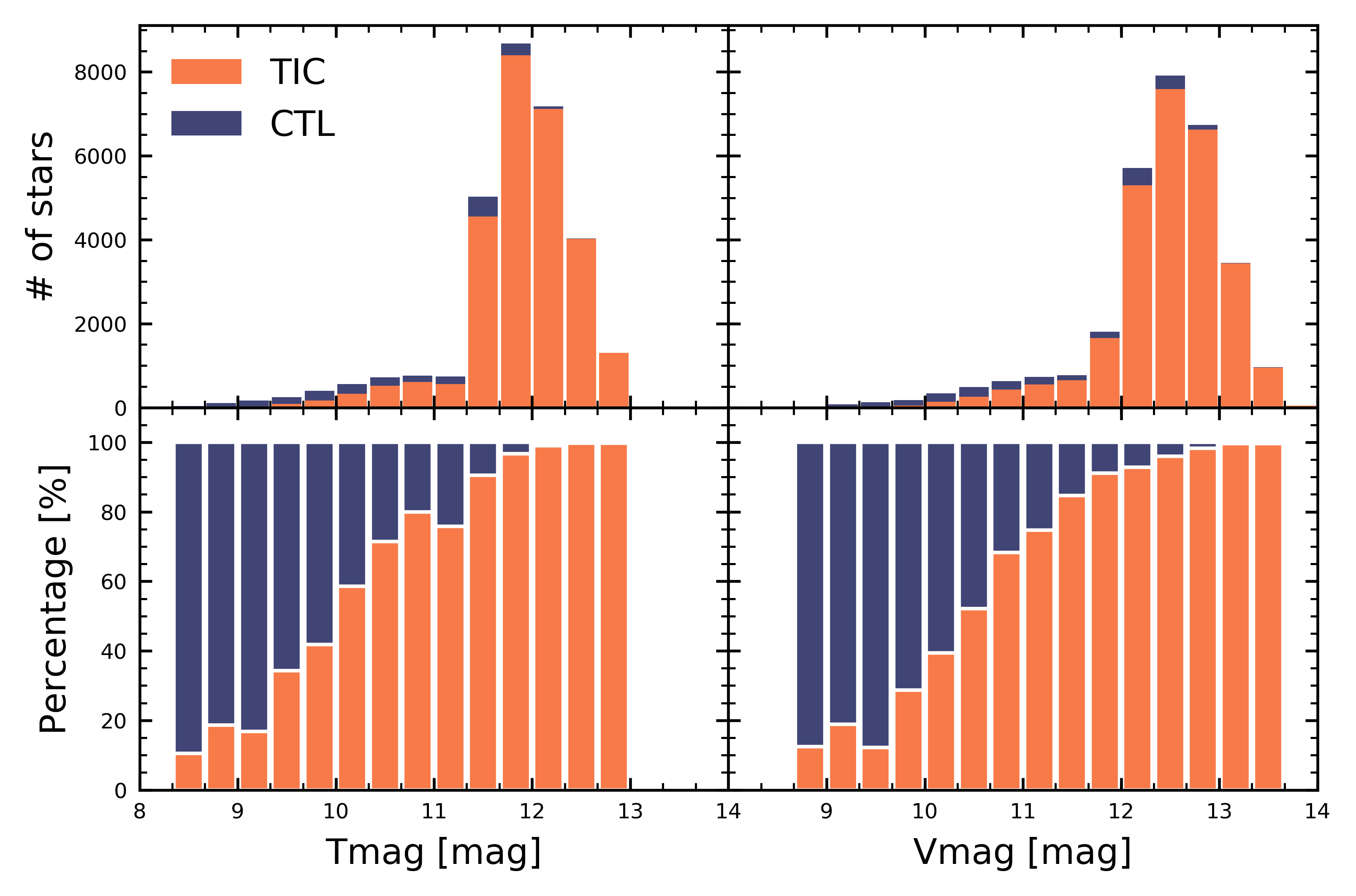}
 \caption{Of the 47,285 stars that are included in both the GALAH DR2 catalog and the \textit{TESS} input catalog (TIC), 2,260 are members of the Candidate Target List (CTL; shown here in purple), and are scheduled to be observed with a higher cadence relative to stars within the general TIC (orange). Left: Of the 2,260 CTL stars, 650 stars are brighter than a \textit{TESS} magnitude ($T_{\mathrm{mag}}$) of 10, 1527 lie between between $T_{\mathrm{mag}}$ 10--12, and 83 with a $T_{\mathrm{mag}}$ between 12--14. The median \textit{TESS} magnitudes for our CTL and TIC stars are 11.4 and 12.5 respectively. The top plot shows the number of TIC and CTL members in each bin whilst the bottom plot shows the percentage of stars in each magnitude bin that belong to the TIC and CTL, respectively. Right: Of the 2,260 CTL stars, 299 stars are brighter than a V magnitude of 10, 1099 lie between V magnitudes of 10--12, and 862 have a V magnitude between 12--14. The median V magnitudes for our CTL and TIC stars are 10.7 and 11.9, respectively. The slightly lower median values for CTL values compared to the TIC reflect \textit{TESS}'s primary mission objectives, prioritising brighter stars. The significant increase in the number of stars between Vmag 12.0--13.7 and $T_{\mathrm{mag}}$ 11.3--13.0 reflects GALAH's observing strategy.}\label{fig:VTmag}
 \end{center}
\end{figure*}

\section{Methodology and Data Analysis}\label{sec:method}

In this methodology section we describe how we cross-matched the GALAH-TESS catalog (Section \ref{sec:cross}), derived our physical stellar parameters including isochronic masses, radii and ages from GALAH DR2 (Section \ref{sec:isocrone}), and calculated our [X/H] and X/Y abundance ratios using GALAH DR2 data (Section \ref{sec:methodabund}).

\subsection{Cross-matching the CTL and GALAH Catalogs}\label{sec:cross}

The \textit{TESS} Input Catalog \citep[TIC;][]{TIC_CTL,revisedTIC} presents the physical characteristics of stars that are likely to be observed during the primary \textit{TESS} mission. Built before the launch of the spacecraft, the TIC uses photometric relationships to derive the physical properties of over 470 million point sources. Due to the large number of stars being observed by \textit{TESS}, there is a selection process that gives a higher priority to stars that better suit the \textit{TESS} mission goals, which are primarily to discover planets around bright, cool dwarfs \citep{TESS,TIC_CTL}. Stars within this subset of the TIC are a large component of the Candidate Target List (CTL), and are observed by \textit{TESS} at a two-minute cadence, whilst the remaining targets are recorded at a 30-minute cadence in the full-frame images (FFIs).

Several simulations of the exoplanetary outcomes of \textit{TESS} have been produced, including \cite{Sullivan15} and \cite{Barclay18}. \citet{Sullivan15} predicted that \textit{TESS} will discover 20,000 planets over the next two years (1,700 from CTLs and the rest from full-frame images). A more conservative yield prediction by \cite{Barclay18} estimates that 1,250 exoplanets will be discovered orbiting CTL stars, with an additional 3,100 being found orbiting stars within the full-frame images. Both sets of simulations suggest that a large number of planets will discovered by \textit{TESS}, from both the CTL and full TIC samples.

The most recent data release from GALAH \citep[DR2;][]{GALAHDR2} contains data derived from high resolution spectra for a total of 342,682 southern stars. Stars in the GALAH DR2 were first cross-matched with \textit{Gaia}'s second data release \citep[DR2;][]{gaiadeearetwo}, using the TOPCAT \citep{TOPCAT} tool to match GALAH and \textit{Gaia} sources with a position tolerance of $\pm$1\arcsec, providing \textit{Gaia}-band magnitudes and parallaxes for our isochronic models. The returned stars were then cross-matched against release 8.0 of the TIC\footnote{\url{https://filtergraph.com/tess_ctl}; accessed 6 August 2020.} using 2MASS \citep{TWOMASS} identifiers from the GALAH catalog, accessed through the Mikulski Archive for Space Telescopes \texttt{astroquery}'s API \citep{astroquery}. 

For our catalog, we selected GALAH DR2 stars with a high signal-to-noise (S/N) across all four of HERMES's CCDs, only accepting stars that had a S/N ratio value of 50 or higher in each wavelength band. We also omit stars with a \texttt{flag{\textunderscore}cannon} greater than zero, which indicates some problem in the data analysis, from our data set. The flagging scheme utilised in GALAH DR2 is described in greater detail in \citet{GALAHDR2}.

For completeness, we compared the \textit{Gaia} G band magnitude from the TOPCAT cross-match to the same value found in the CTL catalog. These values should be identical to one another, and hence serve as confirmation that we have the correct stars cross-matched within our catalog. We considered a match to be confirmed if the difference in a star's celestial coordinates was less than 0.0001 degrees and the difference in 2MASS J-H colour magnitudes (J-H) was also below 0.0001 mag for the exoplanet hosts.

We also wanted to include in our GALAH-TESS sample any stars that may have slightly lower S/N spectra in GALAH, but which are known to host either a confirmed exoplanet, a \textit{TESS} TOI, or a CTOI. We accessed the \textit{TESS} Follow-up Observing Program and NASA's EXOFOP-TESS databases, and cross matched them with GALAH DR2 and \textit{Gaia} DR2. Our cross-match approach was simpler for these targets, as we merely needed to match them by their TIC IDs. 

Taking all of the above into consideration, our newly formed GALAH-TESS catalog boasts 47,285 stars across the southern night sky, as shown in Figure \ref{fig:galah_sky}. Of these 47,285 stars, 2,260 are prioritised sufficiently highly by the \textit{TESS} mission that they are included in the TIC's CTL catalog, being observed with a higher cadence relative to other stars in the general TIC. Figure \ref{fig:VTmag} shows the distributions of our GALAH-TESS stars as a function of their \textit{TESS} and V-band magnitudes. The median \textit{TESS} magnitudes for our CTL and TIC stars are 11.4 and 12.5, respectively, whilst the median V-band magnitudes for our CTL and TIC stars are 10.7 and 11.9 respectively. The slightly lower median values for stars on the CTL compared to those for the general TIC reflect \textit{TESS}'s primary mission objectives, prioritising brighter stars.

Due to flexible constraints by which we cross-matched the catalogs, there are GALAH-TESS stars that are located within the ecliptic, with a \textit{TESS} priority of zero, that will not be observed within the initial two year \textit{TESS} primary mission. We have left those stars in our GALAH-TESS catalog, as they might be explored during the \textit{TESS} extended mission, following the conclusion of the primary survey. There is a large, deliberate absence of stars surrounding the \textit{TESS} Continuous Viewing Zone, with no star within our catalog being found at ecliptic latitudes south of -78$^{\circ}$, in order to avoid any crossover of stars being observed and analysed by the TESS-HERMES Survey \citep{TESSHERMES}. There are also no stars in our catalog which overlap fields observed as part of the K2 survey, in order to avoid any potential crossover with the K2-HERMES survey \citep{K2HERMESours,K2HERMESsanjib}.

\begin{figure*}
 \centering
	\includegraphics[width=\columnwidth]{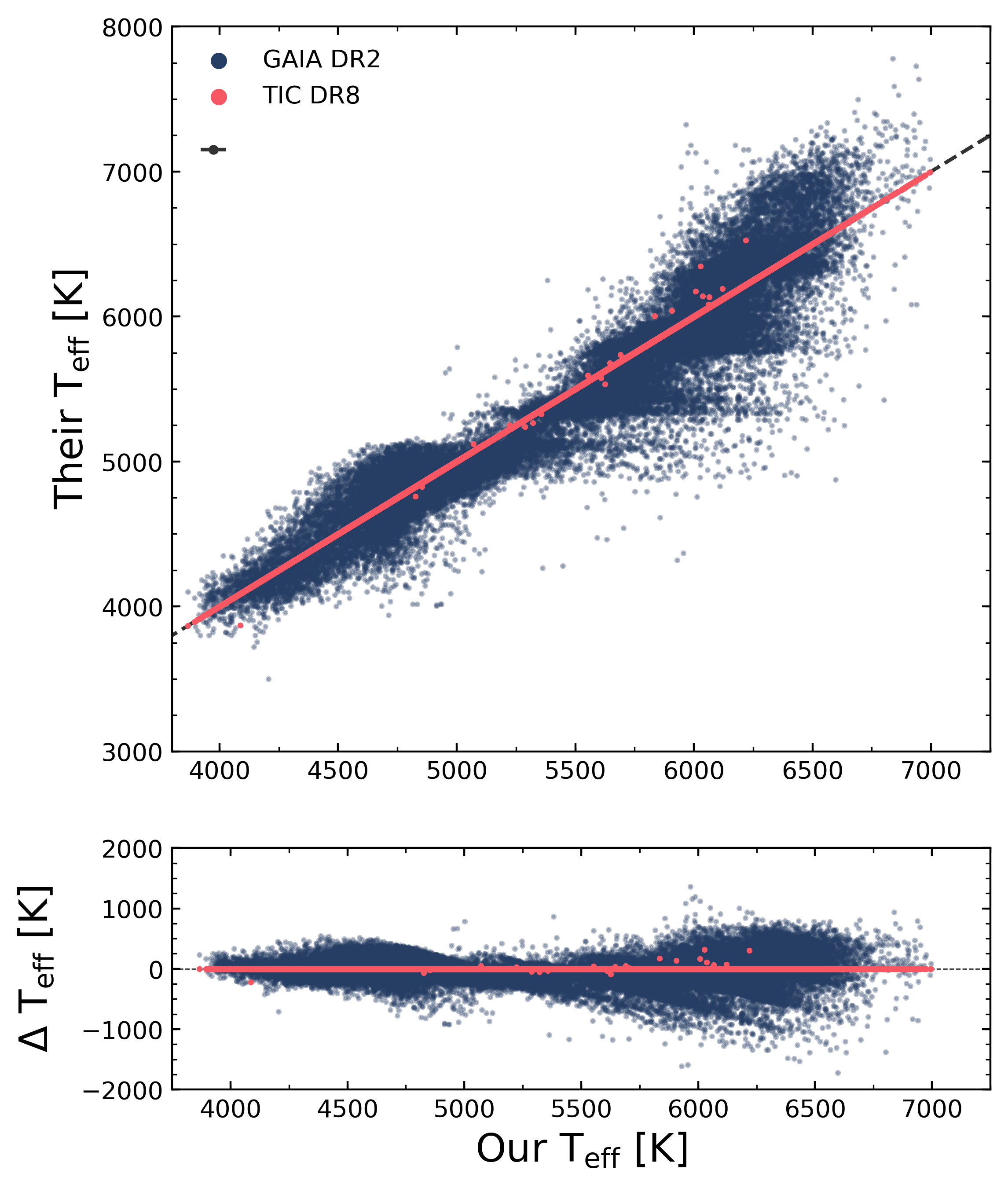}
	\includegraphics[width=\columnwidth]{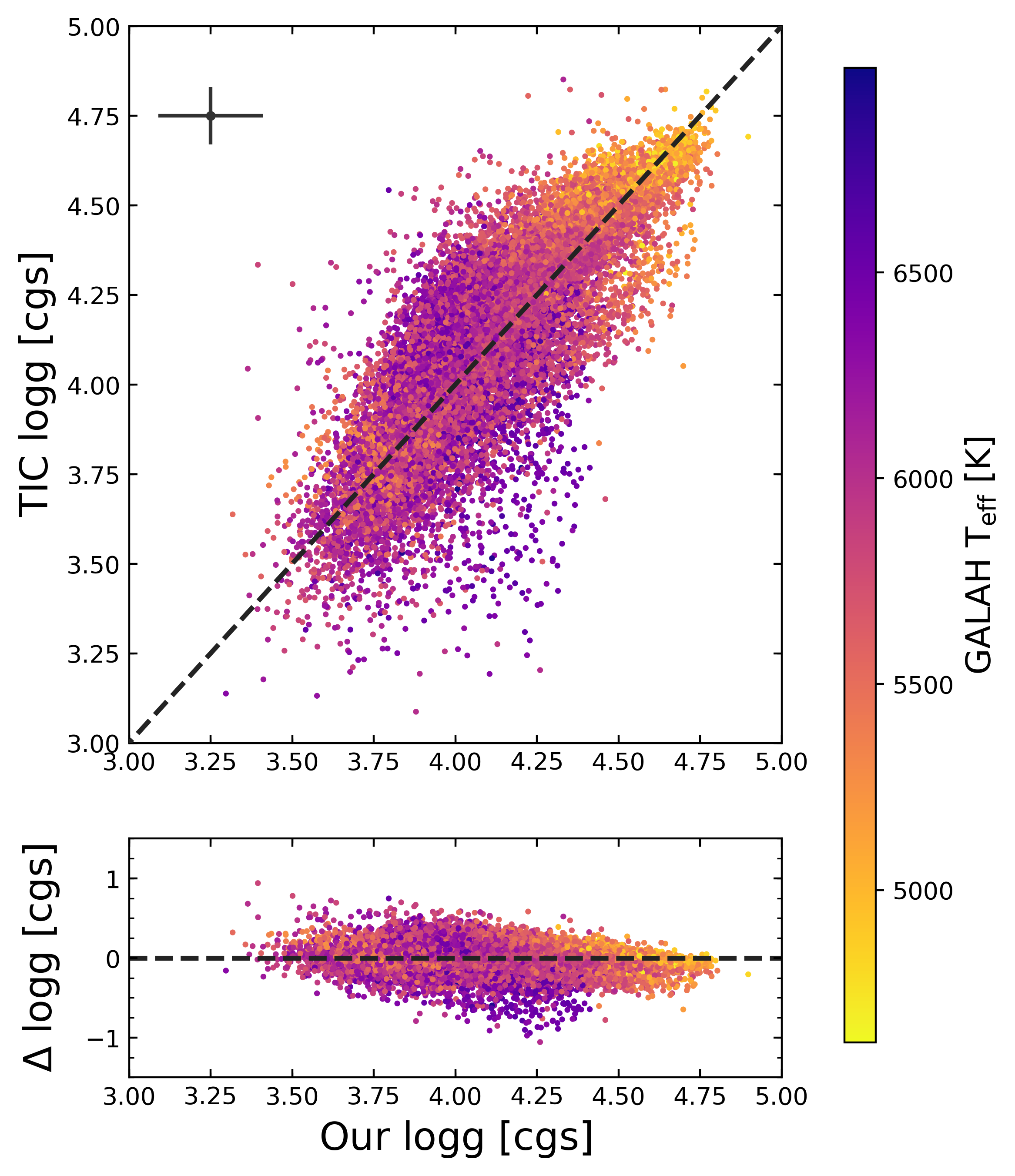}
 \caption{Left: Comparing the GALAH derived effective temperatures, with those published by \textit{Gaia} DR2 (blue) and TIC DR8 (pink) for our GALAH-TESS sample of 47.974 stars. The GALAH DR2 effective temperatures have been included within the TIC, indicated by scatter points lining up on top of the equality line (grey dashed line). There are horizontal structures between 5250--5750~K for \textit{Gaia} \teff values compared to those derived using GALAH data. Similar structures were also found within revised \teff values from \citet{KHU20}. These structures suggest that the \textit{Gaia} DR2 database has a tendency to obtain particular temperature values for these stars, which may be the result of \textit{Gaia}'s input training set. Right: Comparing the GALAH and TIC derived surface gravities with each star colour coded by its effective temperature. Only $\sim$60~\% of CTL stars within our GALAH-TESS sample have measured surface gravity measurements. Error bars for both figures have been suppressed due to clarity, however median error bars are given by single grey data points in the figures' top-left corners.}\label{fig:tefflogg}
\end{figure*}

\subsection{Deriving Stellar Radii, Masses and Ages from GALAH Stellar Parameters}\label{sec:isocrone}

Details on the observation strategy and data pipeline for GALAH DR2 can be found in \citet{GALAHpipeline,GALAHDR1} and \citet{GALAHDR2}. Briefly, all GALAH DR2 observations are acquired with the 3.9 metre Anglo-Australian Telescope situated at the Siding Spring Observatory, Australia. The two degree-field prime focus top-end \citep[2dF;][]{AAT2dF} with 392 science fibres is used to feed the High Efficiency and Resolution Multi Element Spectrograph (HERMES) \citep{HERMES15}, delivering high resolution (R $\approx$28,000) spectra in four wavelength arms covering 471.3-490.3\,nm, 564.8-587.3\,nm, 647.8-673.7\,nm and 758.5-788.7\,nm.

The spectra for each star are corrected for systematic and atmospheric effects and then continuum normalised. Detailed physical parameters, including effective temperature (\teff), surface gravity (log\,$g$), global metallicity ([M/H]), and individual abundances ([X/Fe]), have been determined for 10,605 selected stars using 1D stellar atmospheric models via the Spectroscopy Made Easy \citep{SMEold} package. Both the spectroscopic information and stellar parameters for these 10,605 stars then form a training set for the machine-learning algorithm \textit{The Cannon} \citep{thecannon}, which is used to train a data driven spectrum model algorithm on the entire GALAH DR2 survey. Flags are produced by \textit{The Cannon}'s processing for the ``quality" of the derived physical parameters in each star. For our analysis, we only include stars in the GALAH-TESS catalog if they have a ``0" \texttt{flag{\textunderscore}cannon} in the GALAH DR2 release. 

To derive the mass, radius, and ages of our GALAH-TESS stars, we used the Python package \texttt{isochrones} \citep{isochrones}. The \texttt{isochrones} code uses MESA Isochrones \& Stellar Tracks (MIST) \citep{MIST} stellar evolution grids to infer the physical characteristics of stars. For this analysis, we used as input observables: the star's effective temperature (\teff), surface gravity (log\,$g$), 2MASS ($J$, $H$, $K_s$) and \textit{Gaia} ($G$, $G_{RP}$, $G_{BP}$) photometric magnitudes, along with parallax values obtained by \textit{Gaia} DR2 \citep{gaiadeearetwo} where available.

Isochrone models rely on knowledge of a star's global metallicity, [M/H]. The assumption that the iron abundance [Fe/H] can be a proxy (or even equal) to [M/H] breaks down for metal-poor stars \citep[e.g.][]{RecioBlanco14,Adibekyan13,Adibekyan12,Reddy06,Fuhrmann98}. The radiative opacity of metal poor-stars can be heavily affected by Mg, Si, Ca and Ti (i.e. by $\alpha$--elements). Including these $\alpha$--elements in our calculations of global metallicity better predicts the physical parameters derived with \texttt{isochrones}. GALAH DR2 calculates an [$\alpha$/Fe] value for each star using Equation \ref{eq:alpha}:

\begin{equation}\label{eq:alpha}
 \mathrm{[\alpha/Fe] = \frac{\sum{\frac{[X/Fe]}{(e\_[X/Fe])^2}}}{\sum{(e\_[X/Fe])}^{-2}}}
\end{equation}

\noindent where X = Mg, Si, Ca, and Ti and e\_[X/Fe] is the abundance's associated uncertainty. [$\alpha$/Fe] will be calculated even if one or more of these elements are missing. From our iron abundance, [Fe/H], and [$\alpha$/Fe], we can then calculate [M/H] using \citet{FeHtoMH}:

\begin{equation}\label{eq:fehmh}
\mathrm{[M/H] = [Fe/H]} + \log_{10}{\big(0.638f_\alpha + 0.362\big)}\,\, ,
\end{equation}

\noindent where $f_\alpha$ is the $\alpha$-element enhancement factor given by $f_\alpha = 10^{\left[\frac{\alpha}{Fe}\right]}$. Our calculated [M/H] value is then used in the isochrone modelling of each star. \\

\begin{figure}
 \centering
	\includegraphics[width=\columnwidth]{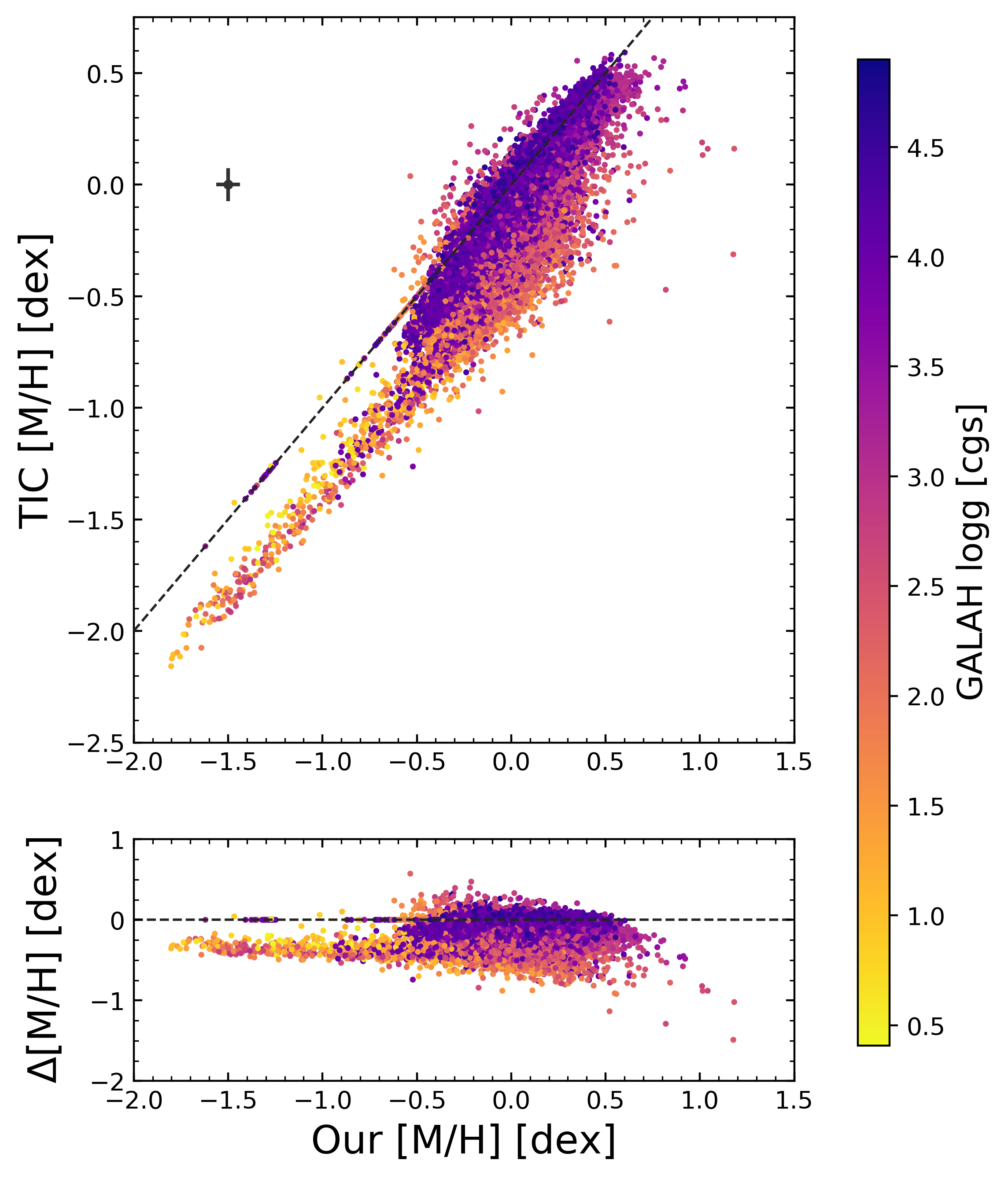}
 \caption{Comparing GALAH's global metallicity [M/H] to that of the TIC, with the colour of each star colour denoting its surface gravity, as derived by GALAH. The median error-bar is given by the grey point to the Figure's top-left corner, with an equality line given by the dark grey dashed line. The TIC directly uses GALAH DR2's [Fe/H] as [M/H], which this equality holds for thin-disc and alpha-poor stars. However, for thick-disc and alpha-rich stars, this equality does not hold true. From this figure, the median difference between [M/H] and [Fe/H] for alpha-rich stars is $\sim$0.3 dex. Since isochronic evolutionary tracks depend on [M/H], this assumption of [M/H] = [Fe/H] would have given less accurate mass, radius, and age results for our alpha-rich stars. For a small portion of our stars, there was no [$\alpha$/Fe] abundance, and hence we use [Fe/H] as [M/H] in our isochrone models for these specific stars.}\label{fig:fehmh}
\end{figure}

When using \texttt{isochrones}, if a star failed to converge, it was omitted from our catalog. Of our original 47,993 stars, 708 stars failed to converge, leaving the 47,285 stars that comprise the GALAH-TESS catalog. When the model reached convergence, the median output values of the stellar mass, radius, density, age, and equivalent evolution phase, as well as their corresponding 1-$\sigma$ uncertainties are calculated from the posterior distributions. We calculate stellar luminosity through the Stefan-Boltzmann relationship, and use those luminosities to derive the habitable zone boundaries for each star, as formulated by \citet{HZcalc}. GALAH DR2 rotational, radial and microturbulence velocities have been included in the GALAH-TESS catalog to assist ground-based radial velocity teams to better prioritise follow-up targets.

\begin{figure*}
 \centering
	\includegraphics[width=\columnwidth]{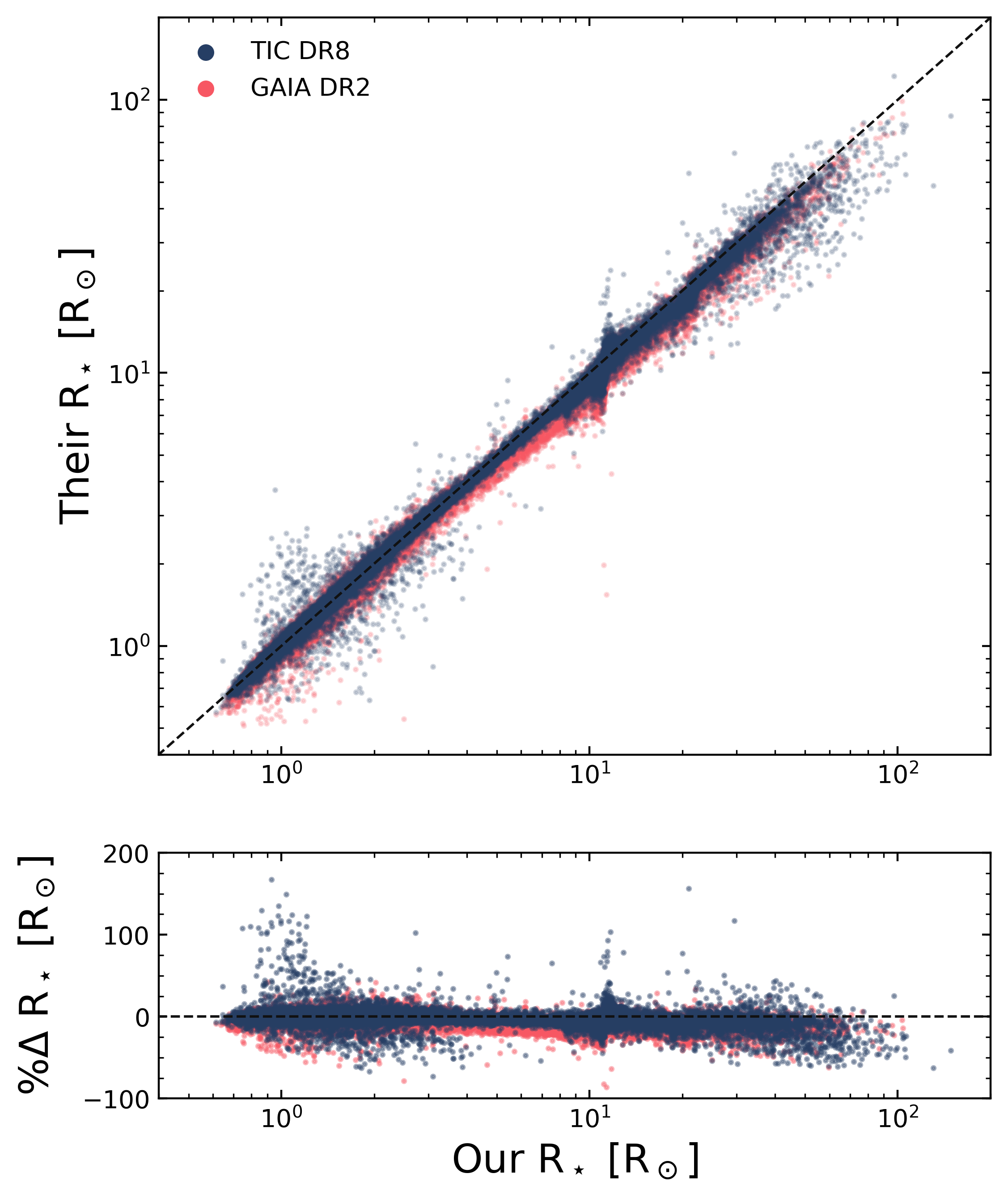}
	\includegraphics[width=\columnwidth]{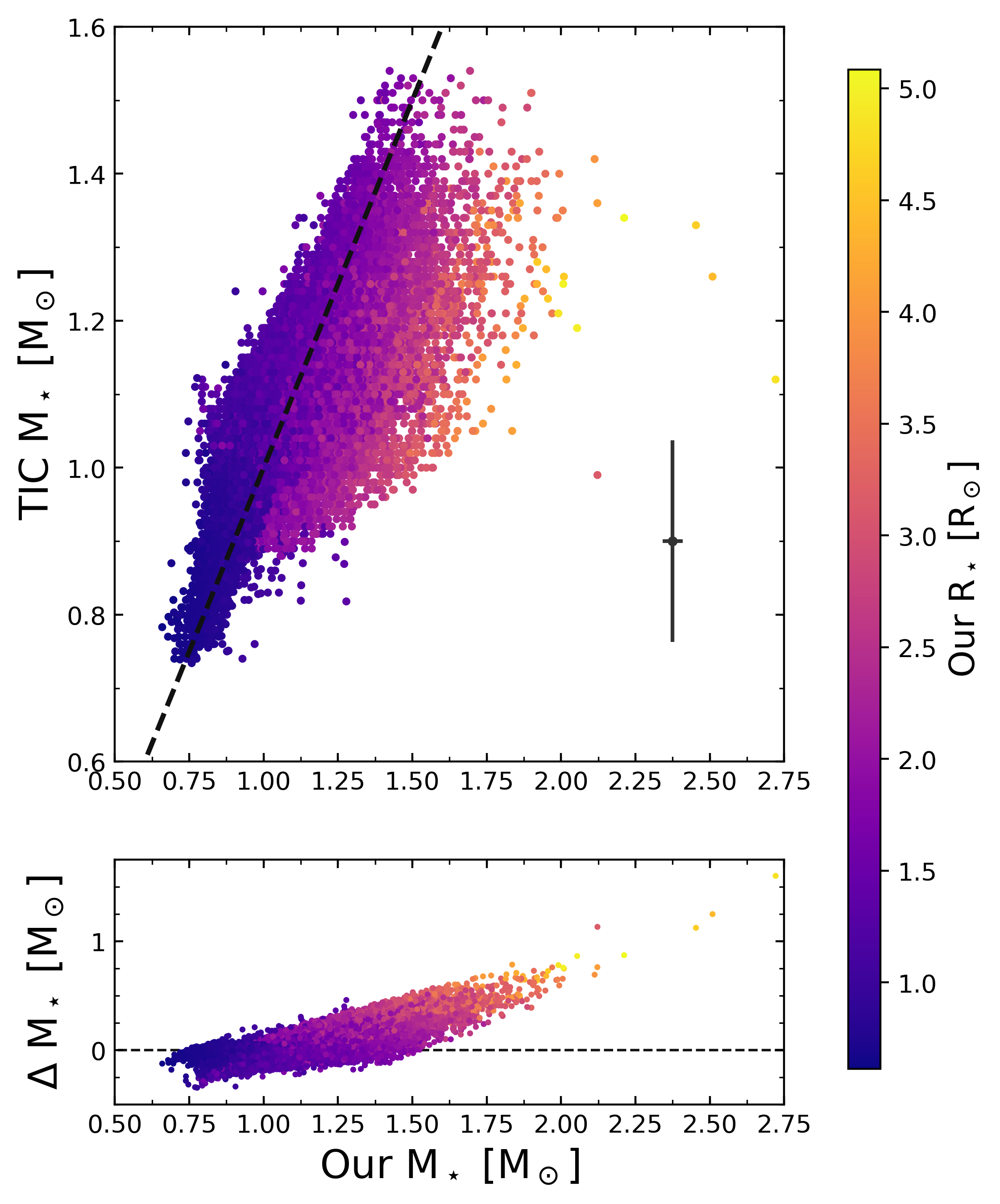}
 \caption{Left: Comparing the stellar radii of GALAH-TESS stars with the TIC (blue) and \textit{Gaia} (pink). There is good overall agreement between the derived radius values, with a relative RMS of 10\% and 14\% for \textit{Gaia} DR2 and TIC values respectively. An equality line is present in both plots, in the form of the dark grey dashed line. Right: Comparing our GALAH-TESS stellar masses with TIC-derived stellar masses. There is a good overall agreement between the derived \texttt{isochrone} masses and the TICs, with an RMS of 0.12 M$_\odot$. Only dwarf stars within the TIC have mass measurements, and thus these comparisons are only valid for this luminosity class. Each star is coloured by its stellar radius, with median error bars given in the bottom-right corner. }\label{fig:massrad}
\end{figure*}

\subsection{Deriving Stellar Abundances and Ratios for GALAH-TESS stars}\label{sec:methodabund}

In addition to providing the physical parameters for over 47,000 stars, our catalog also contains the chemical parameters that could prove vital in determining the composition of rocky planets potentially hosted by these stars. Stellar elemental abundances for 23 elements, as well as quality flags, are derived from \textit{The Cannon}, with the details of the derivation of these abundances their associated systematics discussed in detail in \citet{GALAHDR2}. To ensure that we deliver to the community a usable catalog, we have removed values with [X/Fe] flags not equal to zero. 

Whilst GALAH DR2 has its own internal Solar normalisation, we have converted our elemental abundances from a GALAH normalised scale to \citet{Lodders09}, and moved the abundances from being normalised by iron to hydrogen [X/H], since such values are more widely used within the current exoplanetary community.
The derived Mg/Si, Fe/Mg and C/O ratios were all calculated\footnote{Stellar abundance ratios, also known as stellar molar ratios, are calculated by:\newline $\left(\frac{\text{X}}{\text{Y}}\right)_\star = 10^{\left(\left[\frac{\text{X}}{\text{H}}\right]_\star + \text{A}(\text{X})_{\odot}\right) -\left(\left[\frac{\text{Y}}{\text{H}}\right]_\star + \text{A}(\text{Y})_{\odot}\right)}$} using our 
[X/H] stellar abundances and \citet{Lodders09} Solar normalisations, where available. 

\section{Results}

Our results section is split into two separate parts, which detail the in-depth results of both the physical (Section \ref{sec:physparam}) and chemical (Section \ref{sec:chem}) characteristics of stars within our GALAH-TESS catalog, and provide comparisons of those results to other surveys and catalogs.

\subsection{Physical Stellar Parameters}\label{sec:physparam}

The current TIC incorporates data from large, ground-based spectral surveys including LAMOST \citep{LAMOST}, RAVE \citep{RAVE}, TESS-HERMES \citep{TESSHERMES}, and GALAH. For the vast majority of stars in our sample, the TIC has incorporated GALAH DR2 effective temperatures, which can be see as a line of equality in Figure \ref{fig:tefflogg}. Our GALAH-TESS temperatures, which have a median error of 54 K, seem to be in reasonable agreement with \textit{Gaia}'s, with a larger scatter for hotter stars than for cooler stars.

There tends to be a slightly better agreement with \textit{Gaia}'s \teff for stars slightly cooler than the Sun( 4750 $\leqslant$ GALAH-TESS \teff $\leqslant$ 5500) with an RMS of 146 K and median bias of 50 K, compared to the hotter stars, (\teff $>$ 5500), and cooler stars, (\teff $<$ 4750), with RMS values of 168 K and 253 K and median bias values of 34 K and 25 K, respectively. The high scatter in results for the hotter stars is to be expected, with \citep{GALAHDR2} noting an underestimate of GALAH \teff values for hotter \textit{Gaia} benchmark stars, which might be due to GALAH's input training set preferentially favouring cooler temperatures. There are horizontal structures between 5250--5750~K for \textit{Gaia} \teff values compared to those obtained using GALAH data. Similar structures were found by \citet{KHU20} when comparing \textit{Gaia} \teff values with spectral values obtained with LAMOST. These structures suggest that the \textit{Gaia} temperature calculations in this range tend to certain preferred temperatures, which may be the result of \textit{Gaia}'s input training set.

\begin{figure}
 \centering
	\includegraphics[width=\columnwidth]{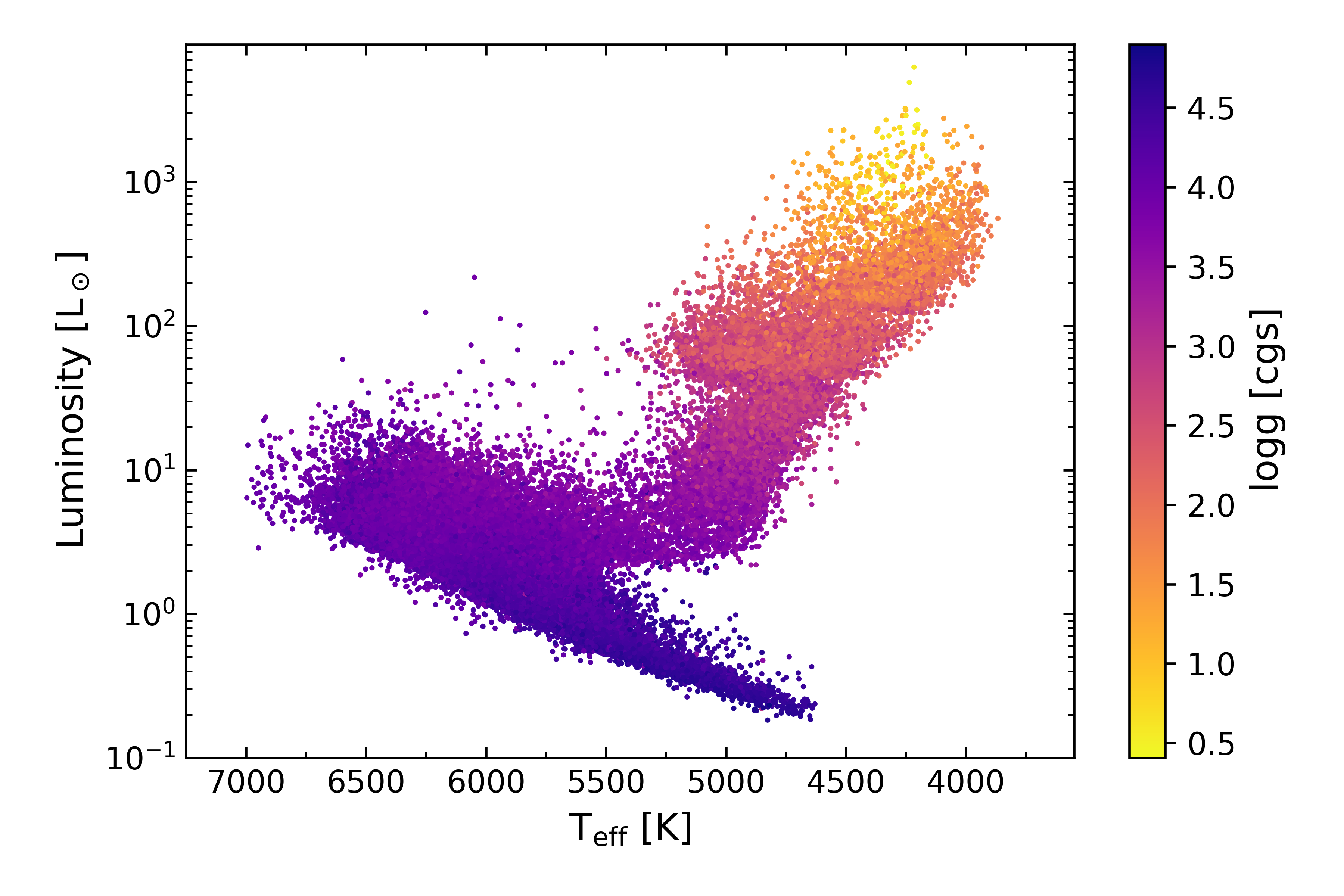}
 \caption{A Hertzsprung-Russell diagram of our GALAH-TESS cataloged stars using GALAH DR2's \teff and our isochrone derived luminosity values. Stars selected for our catalog include both those on the main sequence (lower right to mid-left; high log\,$g$) and evolved stars (mid-left to upper left; low log\,$g$). We have included giant stars within our catalog as these stars are also known to host exoplanets, and it seems likely that analysis of \textit{TESS}'s full frame images will yield a number of new discoveries of this type.}\label{fig:hrdiag}
\end{figure}

\begin{figure*}
 \centering
	\includegraphics[width=\textwidth]{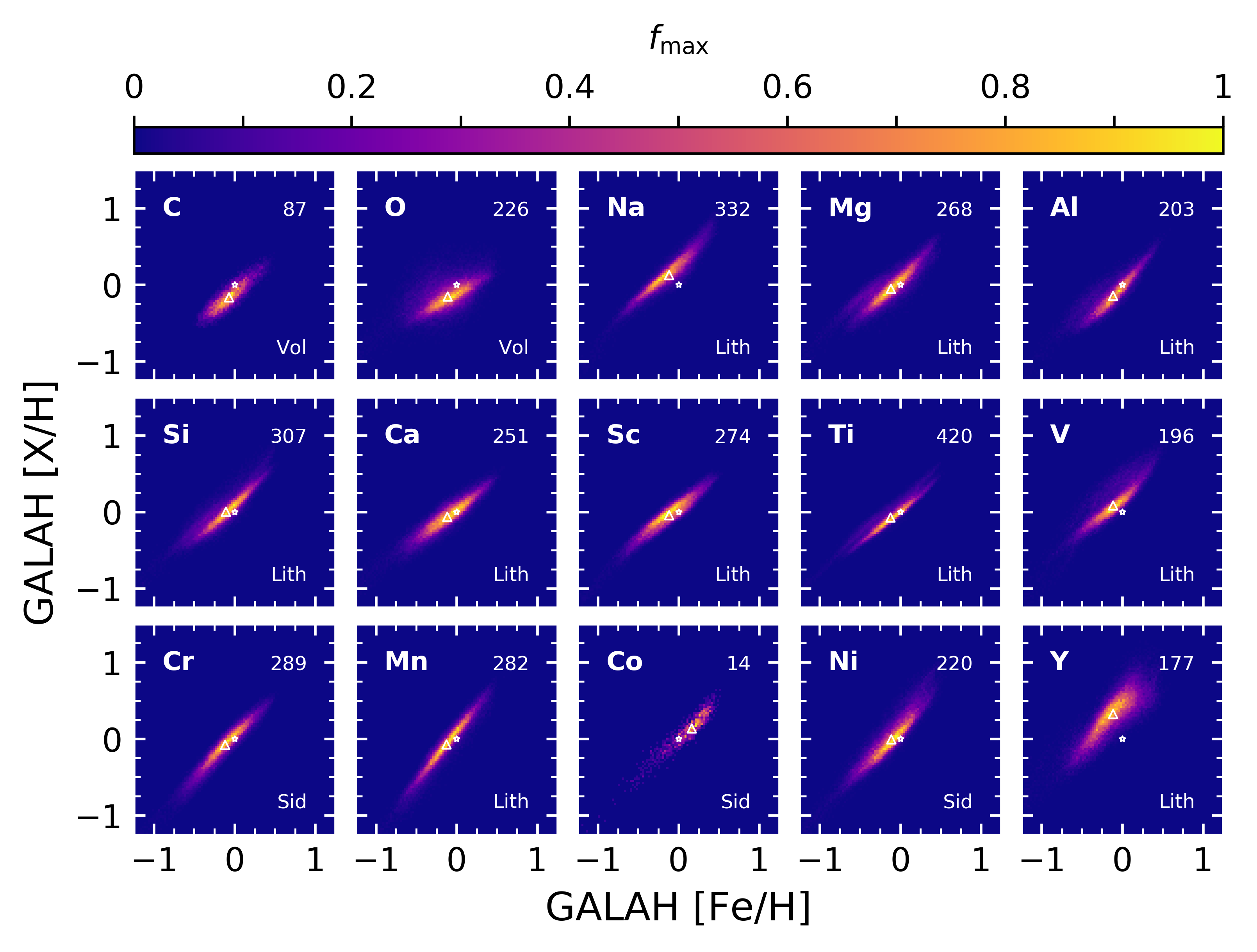}
 \caption{2-D histogram distributions of elemental abundances versus iron abundance for planet-building lithophile (Lith), siderophile (Sid) and volatile (Vol) elements. The Sun's values are represented on each plot by a white-bordered, hollow star, with the median values depicted by triangles. Since there are some elements that are easier to detect in a stellar photosphere than others, each bin is coloured by the fraction of the maximum bin value in each plot. The maximum bin value for each plot is given in the plot's top right-hand corner.}\label{fig:xhvsfeh}
\end{figure*}

Because the TIC prioritises stars being observed with a two-minute cadence (the CTL), surface gravities are only presented within the TIC for dwarf stars with a log\,$g$ $>$ 3. In addition, the TIC does not include include derived log\,$g$ values from other surveys, opting instead for a homogeneous dataset to ensure internal consistency with their mass and radius values. In our cross-matched sample, we include both dwarfs and giants, since giant stars are also known to be planet-hosts \citep{bigbois11,bigbois16,bigboisTOI197b,bigboisPPPS}. As a result, Figure \ref{fig:tefflogg} only shows the comparison for GALAH-TESS stars that have both measured log\,$g$ values in both catalogs. For our sample of dwarfs that have TIC log\,$g$ values, the agreement between their log\,$g$ values and ours appears reasonable, with an RMS and median bias of 0.14 and -0.03 dex respectively compared to the median GALAH log\,$g$ error of 0.16 dex. 

The TIC's global metallicity values, [M/H], have mostly been acquired from the large, ground-based surveys such as LAMOST, RAVE, etc. \citep{LAMOST,RAVE}. For those stars for which the TIC used GALAH DR2 parameters, they incorrectly assumed that the iron abundance, [Fe/H] is equal to the star's global metallicity, [M/H]. However, there is a large discrepancy between [M/H] and [Fe/H] for thick-disk and metal-poor stars that are enriched in $\alpha$-elements. These $\alpha$-elements affect the radiative opacity of iron-poor stellar surfaces, with the result that the overall metallicity and iron abundance equality breaks down. If the overall metallicity does not take into account the $\alpha$-abundance, [$\alpha$/Fe], for iron-poor stars, this could drastically alter the star's derived isochrone track. This in turn would alter the final stellar parameters that are produced with this model. 

If we wish to better characterise stars observed with \textit{TESS}, we therefore need to take [$\alpha$/Fe] into consideration, as we did in Section~\ref{sec:isocrone}. Figure~\ref{fig:fehmh} shows the comparison between the overall metallicities taken from the TIC, and those calculated using GALAH data. There are 317 stars that do not have a [$\alpha$/Fe] measurement, and for those stars, we simply equated their iron abundance to the overall stellar metallicity. The RMS and bias between the TIC and GALAH's overall metallicity is 0.18 and 0.08 dex respectively. As we expected, however, the RMS between the two datasets is significantly lower for alpha-poor stars ([$\alpha$/Fe] < 0.1), with an RMS and bias values being 0.08 and 0.05 dex respectively. There is a much larger difference in [M/H] for iron-poor/alpha-rich stars, which is to be expected, with a RMS and bias of 0.32 and 0.27 dex respectively. For comparison, the median error in the derived [M/H] values is 0.07 dex.

GALAH's \teff, log\,$g$ and [M/H] values together with the astrometric and photometric observables are fed into the \textsc{isochrones} code, producing the radius and mass values which are depicted in Figure \ref{fig:massrad}. Our radii show good overall agreement with both \textit{Gaia} DR2 and TIC. However, at large radii (giant stars), our calculated radii tend to be smaller than those taken from the TIC and \textit{Gaia}. The median relative error for our stellar radii is 2.7\%, with the relative RMS between our results and those of \textit{Gaia} DR2 and TIC found to be 10\% and 14\%, respectively. Our median stellar radius value is 1.89 R$_\odot$, which is comparable to the median values of the \textit{Gaia} and TIC data of 1.84 R$_\odot$ and 1.92 R$_\odot$, respectively.

The general agreement between our results and the radii derived by \textit{Gaia} and the TIC is not unexpected, since our \textsc{isochrones} models rely on \textit{Gaia} DR2's photometric magnitudes and parallax values. The TIC's methodology is similar in that it also relies on data from \textit{Gaia} to derive its stellar radii values. These stellar radii values will prove fundamental in calculating planetary radii for exoplanet host stars discovered by \textit{TESS} within our sample.

\begin{figure*}
 \centering
	\includegraphics[width=\columnwidth]{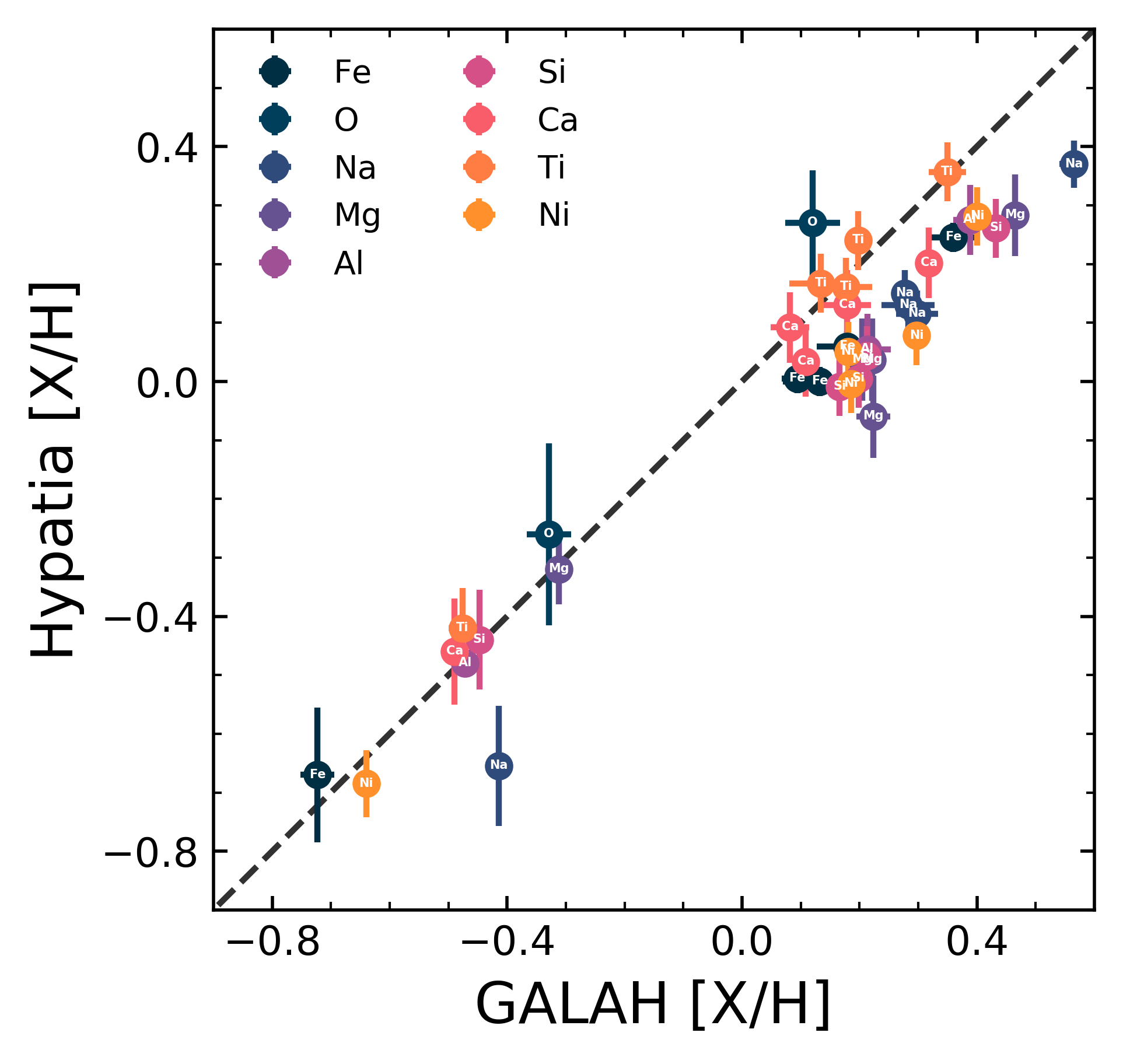}
	\includegraphics[width=\columnwidth]{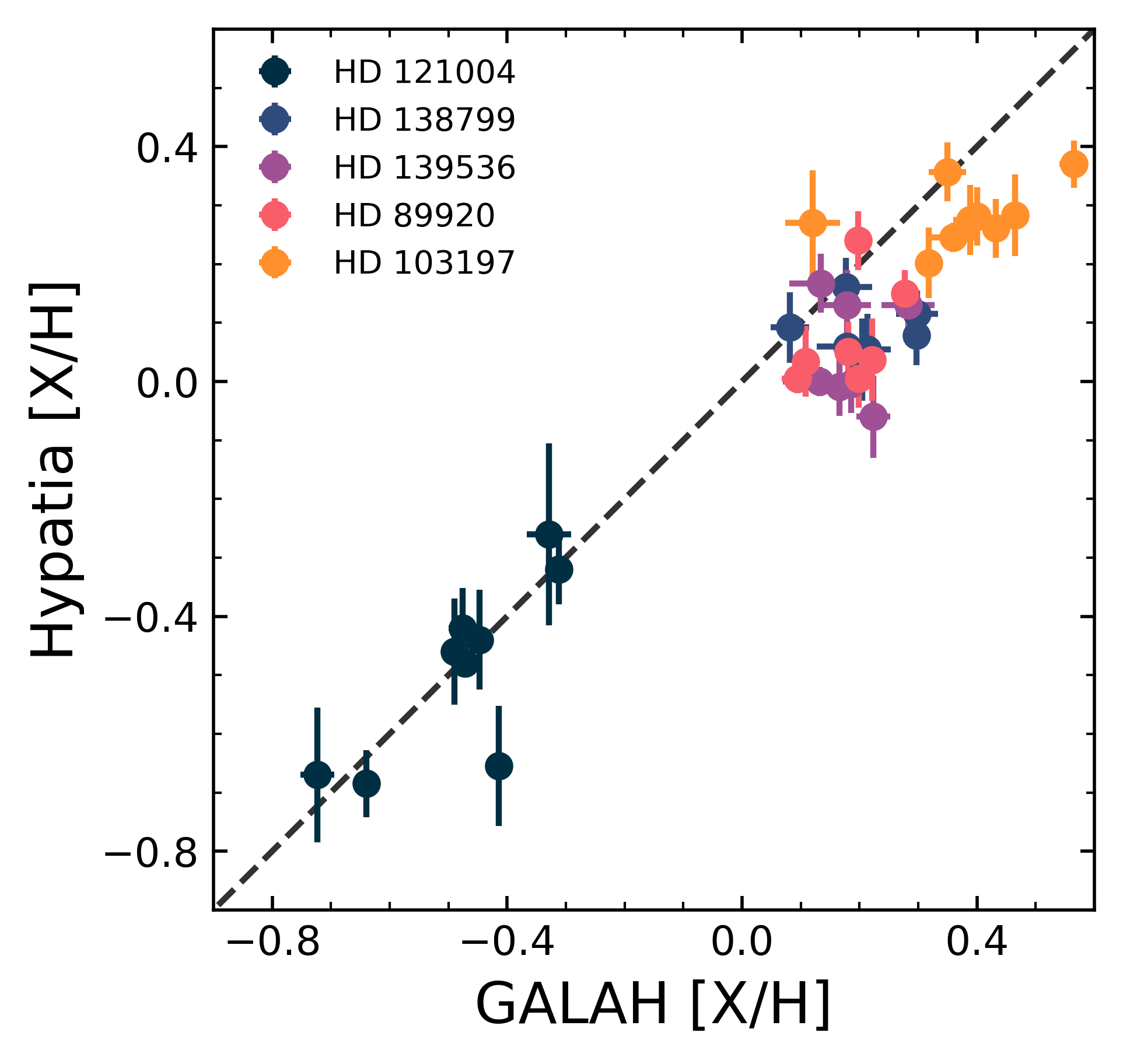}
 \caption{Both of these plots compare the elemental abundances for nine different elements across five stars cross-matched with the Hypatia Catalog \citep{HypatiaCat}. Left: Comparing by element with each element given a unique colour identifier Right: Same plot as the left, however abundances are now grouped by star, labeled by their Henry Draper catalog (HD) identifier.}\label{fig:hypcomparisons}
\end{figure*}

Ground-based follow up teams mostly rely upon the radial velocity method to confirm TOIs \citep[e.g][]{minervaAus,TESSLCO,TOI120,SonghuTESS,TESSKeck,planethuntersTESS}. From this methodology, it is possible to infer the planetary mass through the radial-velocity semi-amplitude. However, the planetary mass is inferred based on our knowledge of the mass of the host star. It is therefore important to not only determine and refine the stellar radii of GALAH-TESS stars, but to also refine their masses. Over 40\% of our sample do not have TIC stellar mass values as they are giant stars and prioritised less than their dwarf counterparts by \textit{TESS}.

Included within Figure \ref{fig:massrad} is the comparison between our derived isochronic masses and those contained within the TIC. In our total sample, the median stellar mass is 1.21 M$_\odot$, compared to a slightly smaller mass of 1.11 M$_\odot$ for the subset of stars with mass measurements in the TIC. This is to be expected, since the TIC only includes mass measurements for dwarf stars. Our masses are slightly larger than those within the TIC, with a median increase of 11\% between our mass measurements and those in the TIC. This increase is slightly larger than our median relative error in stellar mass, being roughly 4\%. However, our median uncertainty is significantly smaller than that found within the TIC, with their median relative uncertainty being 13\%.

A Hertzsprung-Russell diagram of our results is shown in Figure~\ref{fig:hrdiag}, based on GALAH DR2 \teff, log\,$g$, and \textsc{isochrones}-derived stellar luminosity. This sanity check confirms that none of our GALAH-TESS stars fall in unphysical regions of the H-R diagram parameter space. Using the definitions used in \citet{TESSHERMES}, hot dwarfs dominate the GALAH-TESS catalog, accounting for 62\% of the stars (with 38\% being giant stars). A very small fraction of our sample are cool dwarfs, with only 52 such stars. This number of cool dwarf stars is consistent with GALAH being a magnitude-limited survey and the \textit{TESS} goals of detecting exoplanets primarily around bright, nearby stars.

\subsection{Chemical Stellar Parameters}\label{sec:chem}

\begin{figure*}
 \centering
	\includegraphics[width=0.9\textwidth]{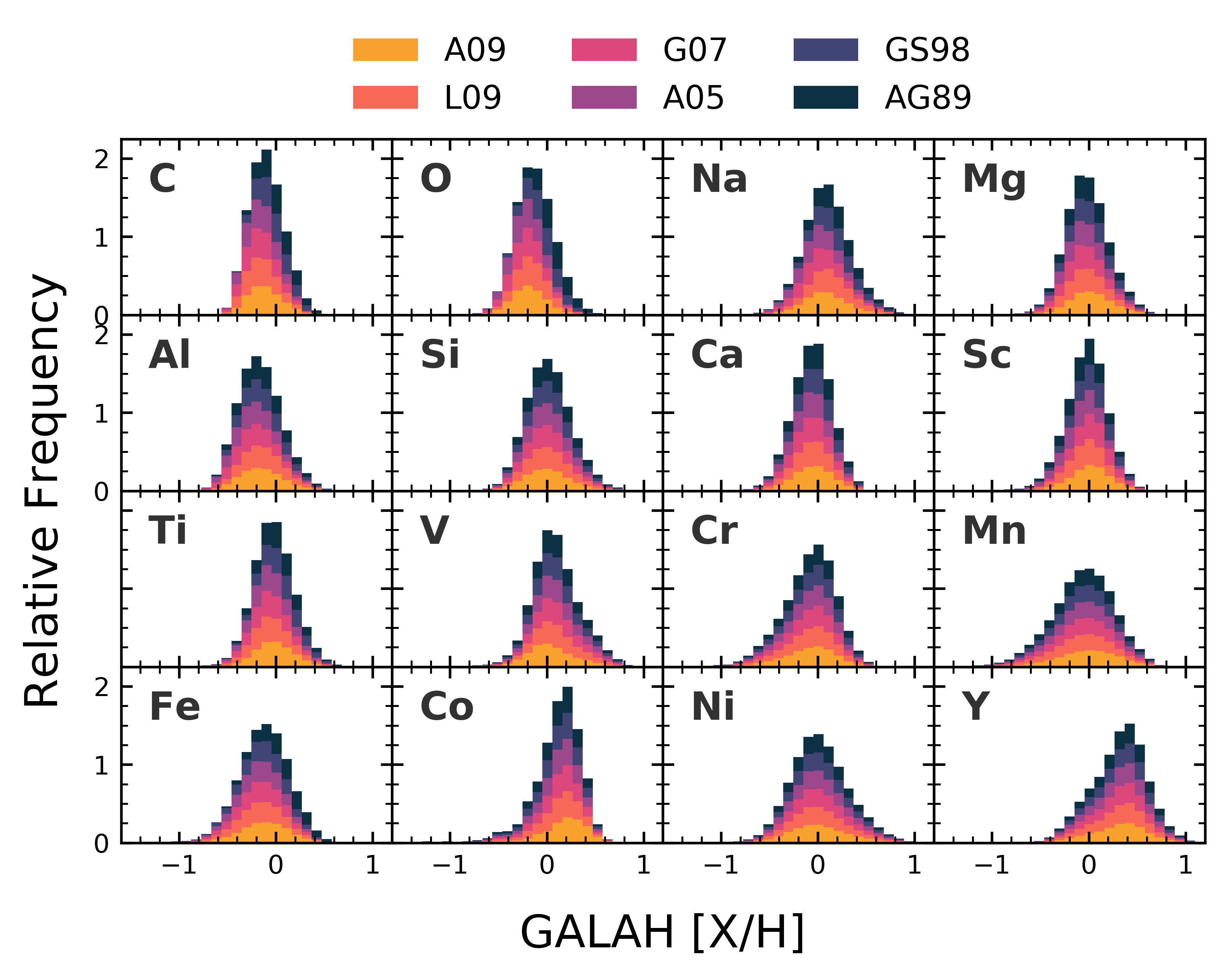}
 \caption{The same distribution of planet-building elements as found in Figure \ref{fig:xhvsfeh}. Here, however, we have normalised our [X/H] values to various Solar normalisations including \citet{Asplund09,Lodders09,Grevesse07,Asplund05,GrevesseSauval98,AndersGrevesse89}, displayed in yellow, orange, pink, purple, violet and navy respectively. Combining the stellar abundances in this manner with different Solar normalisations shows the general trends within a certain element, unbiased by using a specific Solar normalisation.}\label{fig:xhhistogram}
\end{figure*}

Our catalog of $\sim$47,000 stars provides elemental abundances for up to 23 unique species derived from GALAH DR2 abundances. It is not possible, however, to provide accurate elemental abundances for all 23 elements for all of our target stars -- and so we have only provided abundances for those species which can be reliably determined from each star's spectrum. As a result, 90\% of our sample have reliable O, Si, Mg, Si, Zn, and Y abundances, whilst just 2\% of the stars cataloged yield reliable Co abundances. In the most extreme case, only 23 stars in our catalog have reliable, measured Li abundances. Generally our catalog median values are near Solar, with C, O, Al, K, and Fe median values being significantly sub-Solar, and Li, Co, Y and La being significantly super-Solar (though Li suffers from small number statistics). Our distribution between selected elements and the measured Fe abundance is shown in Figure \ref{fig:xhvsfeh}. Given the paucity of Li measurements, we do not discuss the abundances of that element further in this work\footnote{We direct the interested reader to \citet{MartellLi}, and references therein, for a discussion of Li abundances from GALAH data, with a particular focus on the mechanisms by which different populations of stars can end up with dramatically different Li distributions.}.

\begin{table}
 \centering
 \begin{tabular}{lcclcc}
 \hline
 X & Num of & [X/H] & X & Num of & [X/H] \\
 & stars & [dex] && stars & [dex] \\
 \hline
 Li & 28 & 2.02 $\pm$ 0.06 & Cr & 38771 & -0.07 $\pm$ 0.04 \\
 C & 9716 & -0.16 $\pm$ 0.07 & Mn & 39214 & -0.07 $\pm$ 0.04 \\
O & 43297 & -0.15 $\pm$ 0.07 & Fe & 47289 & -0.12 $\pm$ 0.07 \\
Na & 44762 & 0.13 $\pm$ 0.05 & Co & 1057 & 0.14 $\pm$ 0.05 \\
Mg & 44972 & -0.05 $\pm$ 0.03 & Ni & 39450 & -0.00 $\pm$ 0.03 \\
Al & 24068 & -0.14 $\pm$ 0.06 & Cu & 22598 & 0.04 $\pm$ 0.04 \\
Si & 43164 & 0.00 $\pm$ 0.01 & Zn & 43976 & 0.09 $\pm$ 0.02 \\
K & 34258 & -0.29 $\pm$ 0.06 & Y & 43490 & 0.33 $\pm$ 0.04 \\
Ca & 41491 & -0.06 $\pm$ 0.05 & Ba & 28751 & 0.02 $\pm$ 0.06 \\
Sc & 41641 & -0.04 $\pm$ 0.05 & La & 8522 & 0.17 $\pm$ 0.05 \\
Ti & 39205 & -0.07 $\pm$ 0.06 & Eu & 5799 & -0.06 $\pm$ 0.05 \\
V & 27403 & 0.09 $\pm$ 0.04 &&&\\
\hline
 \end{tabular}
 \caption{ Here, we present the median and 1$\sigma$ error values for [X/H] abundances derived in our GALAH-TESS catalog normalised by \citet{Lodders09}. We also give, in the second column, the number of stars in our catalog for which a reliable value for the abundance in question was obtained. The 1$\sigma$ error values here quoted are the median 1$\sigma$ error values for each elemental abundance. The paucity of stars with a reliable Li abundance is particularly apparent.}
 \label{tab:xh}
\end{table}

To validate our stellar abundances, we made use of the online, interactive stellar abundance catalog, the Hypatia Catalog \citet{HypatiaCat}. The Hypatia Catalog is an amalgamation of stellar abundances, including physical and planetary parameters, for stars within 150 pc of the Sun \citep{hinkel14,Hinkel16,Hinkel17}. Comprised of mostly FGKM-type stars, the catalog is compiled from more than 190 literature sources that can be normalised by several Solar normalisations, particularly \citet{Lodders09}. By using the Hypatia Catalog alongside the abundances within our sample, we can directly compare our abundances that use the same Solar normalisation. We accessed the Hypatia Catalog on 6 August 2020 and cross-matched our GALAH-TESS stars with stars within Hypatia by directly comparing their 2MASS identifiers.

Our GALAH-TESS catalog contains data for 606 stars that are within 150 pc of the Sun, of which five matched with the Hypatia Catalog. Figure \ref{fig:hypcomparisons} shows the comparison of elemental abundances for the five cross-matched stars, namely HD\,121004, HD\,138799, HD\,139536, HD\,89920 and HD\,103197. HD\,121004 is the only metal-poor star within our sample that was cross-matched with Hypatia, with the other four stars boasting super-Solar abundances. HD\,121004, a G2V dwarf, has elemental abundances that show the best agreement with the abundances within Hypatia, with a median difference of 0.03 dex with those nine specific elements. The four iron-rich stars, which are all K dwarfs, show a minor discrepancy between their elemental abundances, with the GALAH abundances being enriched by 0.12 - 0.14 dex compared to Hypatia. 

In terms of the abundance difference per element between our data and those presented in the Hypatia catalog, the Ti abundances agree to within a median value of 0.03 dex, which is within the median 1-$\sigma$ error of GALAH-TESS and Hypatia Ti abundances for this sample, being 0.03 and 0.05 dex, respectively. The values for Ca, Al, and Na between the two catalogs differ by 0.08 dex, with the Fe, O, Si, and Ni abundances varying between the catalogs by between 0.12 and 0.16 dex. The largest discrepancy between abundances comes from Mg, which are overabundant in GALAH by 0.30 dex. The GALAH DR2 abundances include non Local Thermodynamic Equilibrium (non-LTE) effects for O \citep{ONLTE}, Na, Mg \citep{MgNLTEa,MgNLTEb}, Al, Si \citep{SiNLTE}, and Fe \citep{FeNLTE} \citep{GALAHDR2,GALAHNLTE}, whereas the Hypatia abundances \citep[from][]{Adibekyan12} do not take into account non-LTE affects, which may explain the discrepancy between the difference in elemental abundance values.

We calculated the Mg/Si, Fe/Mg, Fe/Mg and C/O abundance ratios using our GALAH-TESS [X/H] values and Solar values from \citep{Lodders09}. We only returned a ratio value if stars had both elements available to us, with 43,162 Fe/Si, 44,968 Fe/Mg, 41,741 Mg/Si and 9,521 C/O abundance measurements available. The limited C/O ratio measurements reflect the one atomic C line and two O lines available for reliable abundance measurements across HERMES' wavelength coverage and resulting detection limits. The median and 1$\sigma$ error values for our selected GALAH-TESS abundance ratios are presented in Table \ref{tab:xy}. For reference, the Solar values for Fe/Si, Fe/Mg, Mg/Si and C/O using \citet{Lodders09} are 0.85, 0.81, 1.05 and 0.46, respectively\footnote{Solar abundance ratios are calculated by $\log_{10}(X/Y)_\odot = A(X)_\odot - A(Y)_\odot$.}. Our abundance ratios all tend to have sub-Solar Fe/Si , Fe/Mg,  Mg/Si and C/O ratios. The distribution of our C/O and Mg/Si values are plotted against each other in Figure \ref{fig:covsmgsi}, and are discussed in more detail in Section \ref{sec:abudancediscussion}.

% Stellar elemental abundances can change slightly, depending upon the Solar normalisation used to derive such abundances. To illustrate this, in Figure~\ref{explainer}, we present the abundance of [C/H] for all stars in our data set, for each of the six canonical normalisations often seen in the literature. In the top panel, we show the six normalisations separately, which reveals the differences that occur as a result of the chosen method. In particular, it is clear that BLUE is skewed to higher metallicities. In the lower panel, we plot the cumulative number of stars in each bin summed across each normalisation. This obviously means that each star in the sample is counted six times. This allows the skewed distributions from the different methods to be assessed in a single figure, and forms the basis for Figure~\ref{fig:xhhistogram}, where we present the skews for all sixteen elements... %Jake to finish this paragraph 

Stellar elemental abundances can change slightly, depending upon the Solar normalisation used to derive such abundances. To illustrate this, we have then created Figure \ref{fig:xhhistogram} to show the distribution of our [X/H] abundances for planet-building elements scaled to the various Solar normalisations that are widely used within exoplanetary science. These rocky-planet building elements include the volatiles, which typically reside in the atmosphere (C, O), the lithophiles, which are present in the crust/mantle of rocky planets (Na, Mg, Al, Si Ca, Sc, Ti, V, Mn, Y), and the siderophiles, which easily alloy with Fe and primarily reside in the core (Cr, Fe, Co, Ni) \citep{Hinkel2019}. Having six different normalisations means that each star in the sample is counted six times. However, this allows the skewed distributions from the different methods to be assessed in a single figure, and forms the basis for Figure~\ref{fig:xhhistogram}, where we present the skews for all sixteen planet-building elements.

From Figure~\ref{fig:xhhistogram}, it is readily apparent that there is general overall agreement among our abundances normalised by \citep{Lodders09}, when compared to other distributions with the total median of the distributions falling within 1-$\sigma$ of our L09 values. Volatile elements such as C and O and lithophiles Na and Mg tend to negative [X/H] values in older normalisations compared to newer normalisations that instead peak towards super-Solar values. Larger changes can be seen in the spread of median C/O, Mg/Si and Fe/Mg abundance ratios for these different Solar normalisations. The spread of our median C/O values vary from 0.44 to 0.64, from 0.98 to 1.35 for Mg/Si, and from 0.58 to 1.39 for Fe/Mg depending upon what Solar normalisation is used. Changing the value of Mg/Si for a given planet would have the primary effect of altering the mantle mineralogy between olivine-rich and pyroxene-rich \citep{Hinkel2018,Unterborn17,Brewer16,Thiabaud2014,Thiabaud2015}. These differences in composition are known to change the degree of melting and crustal composition \citep{Brugman2020}, but the degree that that composition changes the interior behavior of a rocky exoplanet remains an area of active research. These results therefore highlight the importance of normalising abundances to the same Solar normalisations when comparing chemical abundances from different surveys and considering the implications those results might have on inferring the structure of rocky exoplanets.

\begin{table}
 \centering
 \begin{tabular}{rlll}
 
 \hline & Num. of stars & (X/Y) & (X/Y)$_\odot^{*}$ \\
 \hline
 Fe/Si & 43162 & 0.65 $\pm$ 0.22 & 0.85\\
 Fe/Mg & 44968 & 0.68 $\pm$ 0.23 & 0.81\\
 Mg/Si & 41741 & 0.98 $\pm$ 0.22 & 1.05\\
 C/O & 9521 & 0.44 $\pm$ 0.13 & 0.65 \\
 \hline
 \multicolumn{4}{l}{$^*$ Solar values from \citet{Lodders09}.}
 \end{tabular}
 \caption{Median and 1$\sigma$ error values for our GALAH-TESS abundance ratios. The majority of our stars have Mg/Si, Fe/Mg, and Fe/Mg values, however only 20\% have reliable C/O measurements.}\label{tab:xy}
\end{table}

\section{Discussion}

In this section we discuss the refinement of planetary systems with the newly derived GALAH-TESS stellar parameters (Section \ref{sec:refineplanetsystems}) and how the X/Y molar abundance ratios of stars within GALAH-TESS can inform us in forward predicting what possible planetary systems and makeups these stars may host (Section \ref{sec:abudancediscussion}).

\subsection{Refining Planetary System Parameters}\label{sec:refineplanetsystems}

\begin{table*}
\begin{center}
\caption{Our stellar physical parameters of matched confirmed and candidate exoplanet hosts. For CTOI hosts, since their CTOI ID is simply their TIC ID, we have omitted this column from the table. [M/H] in this table is the overall metallicity and not the host star's iron abundance, [Fe/H]. \label{tab:planethosts}}
\begin{tabular}{l l r r r r r r}
\hline
Catalog ID & TOI ID & TIC ID & \teff & [M/H] & log\,$g$ & M$_{\star}$ & R$_{\star}$\\
 & & & [K] & [dex] & [cgs] & [M$_\odot$] & [R$_\odot$]\\
\hline
WASP-61 & 439 & 13021029 & 6245 $\pm$ 58 & -0.06 $\pm$ 0.08 & 4.03 $\pm$ 0.17 & 1.20 $\pm$ 0.03 & 1.38 $\pm$ 0.02\\ 
UCAC4 238-060232 & 754 & 72985822 & 6096 $\pm$ 59 & 0.13 $\pm$ 0.08 & 4.16 $\pm$ 0.17 & 1.16 $\pm$ 0.04 & 1.21 $\pm$ 0.03\\
CD-43 6219 & 815 & 102840239 & 4954 $\pm$ 34 & 0.13 $\pm$ 0.05 & 4.46 $\pm$ 0.11 & 0.83 $\pm$ 0.01 & 0.76 $\pm$ 0.01\\
UNSW-V 320 & & 201256771 & 4979 $\pm$ 50 & 0.04 $\pm$ 0.07 & 3.42 $\pm$ 0.15 & 1.29 $\pm$ 0.09 & 3.22 $\pm$ 0.07\\
CD-57 956 & & 220402290 & 5817 $\pm$ 41 & 0.08 $\pm$ 0.06 & 4.33 $\pm$ 0.13 & 1.04 $\pm$ 0.03 & 1.10 $\pm$ 0.01\\
UCAC4 306-282520 & & 300903537 & 4841 $\pm$ 83 & 0.2 $\pm$ 0.09 & 4.41 $\pm$ 0.19 & 0.82 $\pm$ 0.02 & 0.80 $\pm$ 0.01\\
HD\,81655 & 1031 & 304021498 & 6415 $\pm$ 44 & -0.19 $\pm$ 0.06 & 3.88 $\pm$ 0.14 & 1.32 $\pm$ 0.04 & 1.89 $\pm$ 0.02\\
HD\,106100 & 777 & 334305570 & 6187 $\pm$ 35 & 0.12 $\pm$ 0.05 & 3.82 $\pm$ 0.11 & 1.28 $\pm$ 0.02 & 1.54 $\pm$ 0.02\\
WASP-182 & & 369455629 & 5615 $\pm$ 50 & 0.32 $\pm$ 0.07 & 4.15 $\pm$ 0.15 & 1.05 $\pm$ 0.03 & 1.25 $\pm$ 0.02\\
HD\,103197 & & 400806831 & 5223 $\pm$ 32 & 0.35 $\pm$ 0.04 & 4.43 $\pm$ 0.11 & 0.94 $\pm$ 0.02 & 0.90 $\pm$ 0.01\\
TYC 7914-01572-1 & 1126 & 405862830 & 5108 $\pm$ 55 & 0.09 $\pm$ 0.08 & 4.66 $\pm$ 0.17 & 0.82 $\pm$ 0.02 & 0.74 $\pm$ 0.01\\
\hline
\end{tabular}
\end{center}
\end{table*}

Within our GALAH-TESS sample, we cross-matched our GALAH-TESS sample with the catalog of known planetary systems on NASA's Exoplanet Archive and TOIs or CTOIs by accessing the Exoplanet Follow-up Observing Program for \textit{TESS} (ExOFOP-TESS)\footnote{\url{https://exofop.ipac.caltech.edu/}; accessed 6 August 2020.} website. At the time of writing, the GALAH-TESS catalog contains three confirmed single-planet systems: WASP-61 \citep{WASP61}, WASP-182 \citep{WASP182b} and HD\,103197 \citep{HD103197}. Our catalog also includes five single-planet candidate systems namely TOI-745, TOI-815, TOI-1031, TOI-777 and TOI-1126. We should note that WASP-61 b is also known as TOI 439.01. Lastly, there are also three CTOI planetary systems, two of which host two candidates, TIC 201256771 and TIC 220402290. The other CTOI system is a three-planet candidate system, TIC 300903537. A brief summary of the revised stellar parameters for these 11 confirmed and candidate exoplanet hosts are summarised in Table \ref{tab:planethosts}.

The calculated radius of an exoplanet planet is directly related to the radius of its host star - so any change in stellar radius will change the radius of the planet. All of our exoplanets and candidates have transit depth measurements from \textit{TESS}, which we obtain from ExOFOP-TESS, except for WASP-182 b and HD\,103197 b. For the short-period transiting exoplanet WASP-182 b, there is currently no transit data from \textit{TESS}. Instead we use the transit depth values from its discovery paper \citep{WASP182b} to refine its radius. Unfortunately, at the time of writing, the longer-period exoplanet HD\,103197 b has not been observed to transit its host, and no direct size determination is possible.

A brief summary of the revised planetary radii for the 14 confirmed and candidate exoplanets are summarised in Table \ref{tab:exoradius} along with the transit depth and literature planetary radii against which we are able to compare our results.

\begin{figure*}
 \centering
	\includegraphics[width=0.8\textwidth]{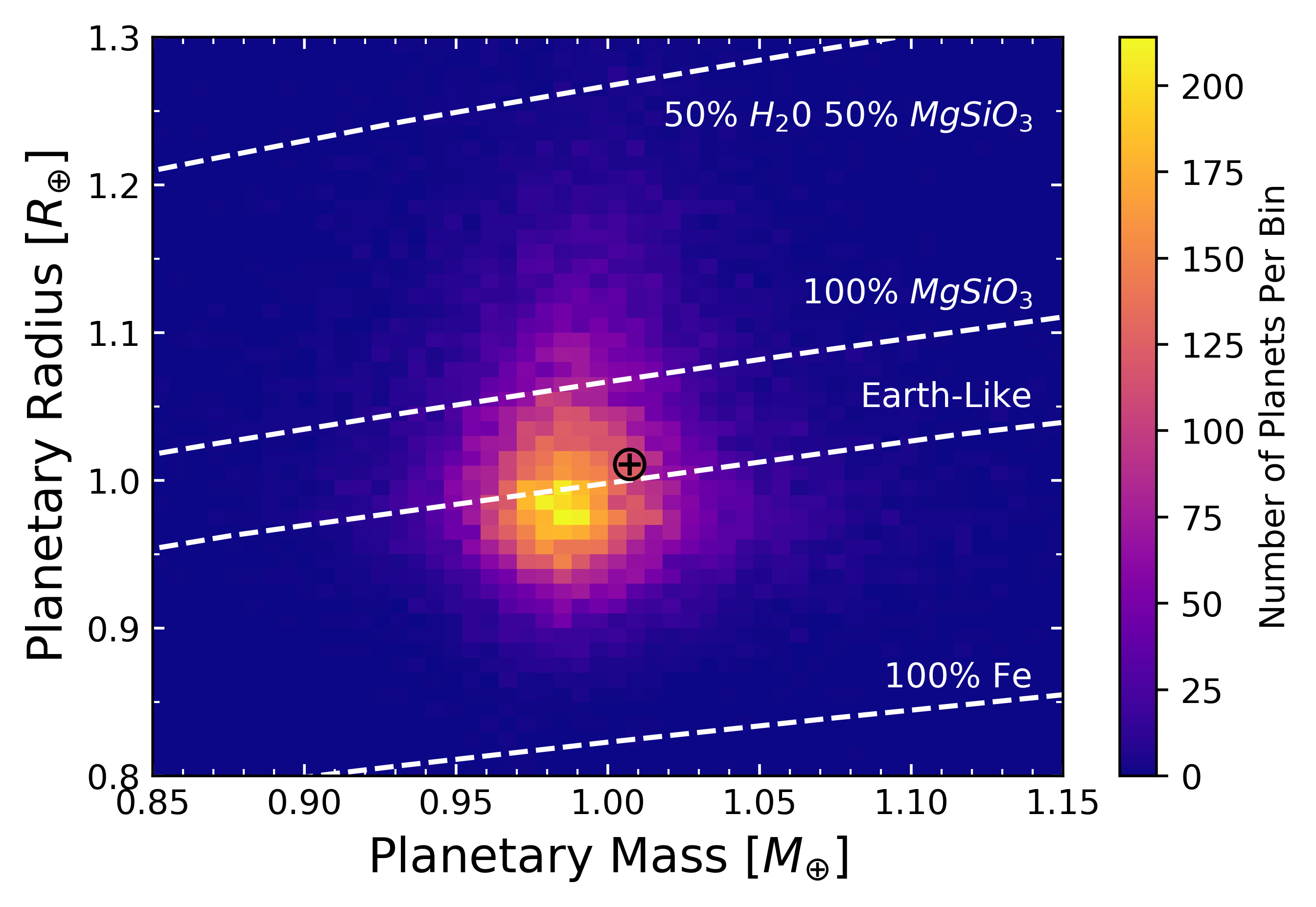}
 \caption{Simulation for the effects of parameter refinement on the mass and radius for a fictitious Earth-like planet discovered using the TIC catalog. The mass-radius relationships used for the dashed lines to show density curves for a 50~\% water-50~\%rocky, pure rocky (containing pure post-perovskite MgSiO\textsubscript{3}), "Earth-like" (33~\%Fe and 67~\% rock) and pure iron worlds are from \citep{zengPREM}. The black symbol ``$\oplus$" represents the Earth's mass and radius. In extreme cases, a putative ``Earth-like" planet varies between a scaled-up Enceladus-like world (i.e. dominated by layers of water and a silicate core), to a Fe-enriched, Mercury-like planet. This simulation shows the need for consistency and precision in exoplanetary mass and radius determination for meaningful comparative planetology.\label{fig:Earthlike}
 }
\end{figure*}

By far the most surprising result from our refinement of planetary radii, is the refinements of two planetary candidates orbiting the star TIC 201256771. Currently, TIC 201256771 hosts two CTOIs, 201256771.01 and 201256771.02, which are recorded on ExOFOP-TESS as having radii of 24.72 R$_\oplus$ and 26.51 R$_\oplus$, respectively. With our revised radii, these candidate events observed in TESS Sector 1 now have radii comparable with stellar radii \citep{ChenKipping} of 96.17$\pm$5.34 R$_\oplus$ and 103.15$\pm$5.69 R$_\oplus$, respectively. This casts serious doubts about the planetary nature of these candidate events, especially with their orbital periods being only separated by 17 minutes, with the orbital periods of CTOI-201256771.01 and CTOI-201256771.02's being stated as 3.754861 and 3.766667 days, respectively. Upon further investigation, this system is a known eclipsing binary that has an orbital period nearly equal to the candidates, being 3.76170 days \citep{UNSWV320}. From this data alone, we conclude that CTOI 201256771.01 and CTOI 201256771.01 are candidates of the same event, being the transit of the eclipsing companion to UNSW-V 320. Apart from this extreme example, the rest of our planetary radii fall nicely within the current literature values and their uncertainties, all of which can be found in Table \ref{tab:exoradius}. Upon the revision of this CTOI system, we re-checked the sensibility of the other CTOI systems within our planet-host sample. The orbital periods of CTOI 220402290.01 and CTOI 220402290.02 are 0.7833 and 0.7222 days respectively, or roughly 90 minutes. This would mean that their orbital separation would be comparable to their radii, which deems this system as extremely unstable. These transit events are likely caused by a single candidate, rather than two. Similarly, the orbital periods of CTOI 300903537.01 and CTOI 300903537.02 only differ by 36 minutes and are likely caused by the same candidate.

\begin{table*}
\begin{center}
\caption{Our refined planetary radii values for confirmed and TESS candidate exoplanets. All literature radius values and transit depth values come from ExOFOP-TESS except for WASP-182 b, where its literature planetary radius and transit depth values are from \citet{WASP182b}. We have flagged problematic planetary candidates in bold. From our revised planetary radii, CTOI 201256771.01 and CTOI 201256771.02 now have radii comparable to the Sun, and thus are not exoplanets. The orbital periods of CTOI 220402290.01, CTOI 220402290.02, CTOI 300903537.01 and CTOI 300903537.02 are problematic and are likely duplications of the same event. This is discussed further in Section \ref{sec:refineplanetsystems}. Since some of these planet candidates are comparable in scale to that of Jupiter, the conversion between Jupiter's radius to Earth's is R$_J$ = 11.209 R$_\oplus$.}\label{tab:exoradius}
\begin{tabular}{l r r r r r r}
\hline
TOI/CTOI ID & TIC ID & $\Delta$F & Our R\textsubscript{p} & Literature R\textsubscript{p} \\
& & [mmag] & [R$_{\oplus}$] & [R$_{\oplus}$] \\
\hline
\vspace{0.1cm}
439.01 & 13021029 & 9.04283 $\pm$ 0.00143 & 13.68 $\pm$ 0.20 & 13.27 $\pm$ 0.47 \\ \vspace{0.1cm}
754.01 & 72985822 & 8.93564 $\pm$ 0.50239 & 12.00 $\pm$ 0.48 & 13.90 $\pm$ 13.91 \\ \vspace{0.1cm}
815.01 & 102840239 & 1.25 $\pm$ 0.00155 & 2.81 $\pm$ 0.03 & 2.87 $\pm$ 0.13 \\ 
\textbf{201256771.01} & \textbf{201256771} & \textbf{84.34287} $\pm$ \textbf{8.98023} & \textbf{96.17} $\pm$ \textbf{5.34} & \textbf{24.72} \\ \vspace{0.1cm}
\textbf{201256771.02} & \textbf{201256771} & \textbf{97.59384} $\pm$ \textbf{10.39110} & \textbf{103.15} $\pm$ \textbf{5.69} & \textbf{26.51} \\ 
\textbf{220402290.01} & \textbf{220402290} & \textbf{21.84594} $\pm$ \textbf{2.32600} & \textbf{17.02} $\pm$ \textbf{0.93} & \textbf{17.15} \\ \vspace{0.1cm}
\textbf{220402290.02} & \textbf{220402290} & \textbf{44.09427} $\pm$ \textbf{4.69485} & \textbf{24.05} $\pm$ \textbf{1.30} & \textbf{24.25} \\ 
\textbf{300903537.01} & \textbf{300903537} & \textbf{94.16304} $\pm$ \textbf{10.02582} & \textbf{25.06} $\pm$ \textbf{1.33} & \textbf{25.10} \\
\textbf{300903537.02} & \textbf{300903537} & \textbf{11.02043} $\pm$ \textbf{1.17338} & \textbf{8.74} $\pm$ \textbf{0.48} & \textbf{8.75} \\ \vspace{0.1cm}
300903537.03 & 300903537 & 3.74782 $\pm$ 0.39904 & 5.10 $\pm$ 0.28 & 5.11 \\ \vspace{0.1cm}
1031.01 & 304021498 & 1.18 $\pm$ 0.00172 & 6.80 $\pm$ 0.08 & 6.91 $\pm$ 0.46 \\ \vspace{0.1cm}
777.01 & 334305570 & 2.80673 $\pm$ 0.08351 & 8.56 $\pm$ 0.16 & 7.32 $\pm$ 1.15 \\ \vspace{0.1cm}
WASP-182 b & 369455629 & 0.01067 $\pm$ 0.00000 & 8.90 $\pm$ 0.15 & 9.53 $\pm$ 0.34 \\ \vspace{0.1cm}
1126.01 & 405862830 & 1.06 $\pm$ 0.00144 & 2.53 $\pm$ 0.02 & 2.62 $\pm$ 0.11 \\ 
\hline
\end{tabular}
\end{center}
\end{table*}

Of our known confirmed and candidate exoplanets, only three have measured mass values. The most conventional way that an exoplanet's mass is determined is through the radial velocity technique. Specifically, an exoplanet's line-of-sight mass, $M_p \sin{i}$ is determined through measurement of the semi-amplitude of the host's radial velocities measurement, $K_{RV}$, orbital eccentricity, $e$, period $P$, and stellar mass $M_\odot$ \citep{RVmethod}. If the orbital inclination, $i$, of the system is known, traditionally found through fitting models to the photometric transit curve, we can then calculate the planet's true mass, $M_p$.

We use literature values for these planetary systems, namely WASP-182 b values from \citet{WASP182b} as well as WASP-61 b and HD\,103197 b values from \citet{WASP61bparams}. We combine these with the masses of their host stars in order to revise the planetary mass of the exoplanets. Our revised planetary mass values, along with the previous literature values, can be found in Table \ref{tab:exoMass}. As with the refined radii results, there is excellent overall agreement with our mass values compared to the literature. All three refined planetary mass values fall within 1-sigma error bars of the previous literature values. The biggest increase of planetary mass precision with our results comes from the Jovian type exoplanet HD\,103197 b. We have refined the mass of HD\,103197 b from a percentage error of 31\% down to 2\%, thanks largely due to the refinement in the stellar mass of HD\,103197. 

\begin{table*}
\begin{threeparttable}
\begin{center}
\caption{With our newly derived stellar mass values, we have refined the mass of three exoplanets, WASP-61 b, WASP-182 b and HD\,103197 b. In this table we have used our new stellar mass values, along with literature semi-amplitude (K) and orbital eccentricity (e), period (P), and, inclination (i) values to derive the new planetary mass values.\label{tab:exoMass}}
\begin{tabular}{l r r r r r r r}
\hline
Planet Name & TIC ID & K\textsubscript{RV} & P & e & i & Our M\textsubscript{p} & Literature M\textsubscript{p} \\
 & & [ms$^{-1}$] & [days] & & [deg] & [M$_\oplus$] & [M$_\oplus$] \\
\hline
WASP-61 b & 13021029 & 233 $\pm$ 0 & 3.8559 $\pm$ 3.00e-06 & 0 & 89.35 $\pm$ 0.56 & 646.01 $\pm$ 9.82 & 851.784 $\pm$ 266.977\\
WASP-182 b & 369455629 & 19 $\pm$ 1.2 & 3.376985 $\pm$ 2.00e-06 & 0 & 83.88 $\pm$ 0.33 & 46.41 $\pm$ 3.05 & 47.039 $\pm$ 3.496\\
HD\,103197 b & 400806831 & 5.9 $\pm$ 0.3 & 47.84 $\pm$ 0.03 & 0 & & 32.06 $\pm$ 1.67$^*$ & 28.605 $\pm$ 6.357$^*$\\
\hline
\end{tabular}
\begin{tablenotes}
\small
\item Literature values for WASP-182 b come from \citet{WASP182b} and WASP-61 b and HD\,103197 b's values are from \citet{WASP61bparams}. * denotes that HD\,103187 b's mass is actually M\textsubscript{p}$\sin{i}$ in this current form as there is yet to be any inclination data retrieved from this particular planetary system. Since some of these exoplanets are comparable in scale to that of Jupiter, the conversion between Jupiter's mass to Earth's is M$_J$ = 317.83 M$_\oplus$.
\end{tablenotes}
\end{center}
\end{threeparttable}
\end{table*}

Overall, our refined planetary mass and radius results are in good agreement with their literature values. This also validates the overall good agreement with our refined stellar mass and radius values. Even though the change in planetary mass or radius of 10-20\% might intuitively be insignificant in re-characterising Jovian worlds, it does however have larger implications for smaller planets like our own.

For example if an Earth-like planet in mass and radius (1.0 R$_\oplus$,1.0 M$_\oplus$), characterised by the TIC, was discovered orbiting around any of our GALAH-TESS stars, would this planet still be ``Earth-like" with our revised stellar parameters? Using a similar approach to that of \citet{GaiaRevisedRadii}, we can refine the planetary radius and mass of this fictitious Earth using both GALAH-TESS and TIC catalog values of stellar and planetary mass and radius values. 

Our refined radius and mass values for these fictitious Earth-like exoplanets are displayed in Figure \ref{fig:Earthlike}. Roughly 85\% of our planets fall within $\pm$10\% of Earth-like mass and radius values. Beyond this $\pm$10\%, there is a wide variety of mass and radius values throughout the plot, which would suggest that these exoplanets that were once thought to be Earth-like, are now anything but. From Figure \ref{fig:Earthlike}, there are varying degrees of bulk composition for these ``Earth-like'' worlds. In extreme cases, a putative ``Earth-like" planet's bulk density varies between a scaled-up Enceladus-like world (i.e. dominated by layers of water and a silicate core) \citep{Enceladus11,Enceladus07}, to a possible remnant Jovian-world core dominated by iron \citep{Mocquet2014,mercuryorigin} with the habitability of such worlds still up for debate \citep{waterworld19,waterworld18,waterworld17}. This shows that not only do we need better precision for stellar masses and radii, which better constrain the planetary mass and radius values, but there also needs to be a level of consistency across these fundamental parameters for future follow-up characterisation.

There are already a wide variety of planetary radius and mass values for known super-Earth and Earth sized worlds and thus there will be a wide variety of planetary compositions. A fundamental problem with inferring planetary compositions through mass-radius or ternary/quaternary diagrams \citep{Brugger2017,Rogers10} is that they cannot uniquely predict the interior composition of a given exoplanet. A variety of different interior compositions can lead to identical mass and radius values \citep{UnterbornPanero2019,legendoftherentwasway,Unterborn2016,Dorn2015}. This gives rise to an inherent density degeneracy problem. A wide variety of planetary compositions are allowed, especially if the models used have three or more layers. This is typical for most that assume a three (core, mantle, ocean) or four layered planet (core, mantle, ocean, atmosphere). Current Bayesian inference \cite{Dorn2015} and forward models \citet{Unterborn2018,exoplex} break down this degeneracy using stellar abundance ratios to infer an exoplanet's composition. These abundance ratios and their importance are described in Section \ref{sec:abudancediscussion}

\subsection{Importance of Stellar Abundances to Exoplanetary Science}\label{sec:abudancediscussion}

Within our own Solar system, observations show that the relative abundances of refractory elements such as Fe, Mg and Si, elements crucial in forming rocky material for planets like ours to build upon, are similar within the Sun, Earth, the Moon and Mars \citep{EarthWang,Lodders2003,McDonough95,Wanke1994}. The bulk planetary and stellar ratios of these elements during planetary formation are also similar, suggesting that stellar Fe/Mg and Mg/Si can assist with determining the building blocks of the planets they host \citep{Thiabaud2015,Thiabaud2014,Bond10}. These elemental abundances can help us understand what elements favour certain planetary architectures and can also provide constraints on the internal geological composition of exoplanets \citep{Unterborn2018,Brugger2017,Dorn2017,Dorn2015}.

In particular the elemental abundance ratios of Mg/Si, Fe/Mg and C/O are fundamental for probing the mineralogy and structure of rocky exoplanets. The formation, structure and composition of exoplanets is extremely complex, with these generalisations not taking into account planetary migration or secondary processes such as giant impacts. A more comprehensive analysis of GALAH DR2's abundances trends and implications for planet-building elements can be found in \citet{Bitsch20}.

\subsubsection{Estimating the size of a Rocky Planet's Core Through Stellar Fe/Si Ratios}

The amount of mass contained within a rocky exoplanet's core is determined by its Fe/Si ratio \citep{Dorn2015,Unterborn2018,Brugger2017}. An increasing Fe/Si ratio would result in a larger core mass fraction compared to a larger mantle core fraction for smaller values of Fe/Si. Within our Solar system, Earth \citep{McDonough2003,McDonough95} and Mars \citep{Wanke1994} have comparable bulk Fe/Si values to that of photospheric Solar values \citep{Lodders09,Lodders2003}. Mercury, however, is an anomaly with its bulk Fe/Si value estimates ranging from $\sim$5-10, corresponding to a core mass fraction of $\sim$45--75\% compared to a Fe/Si ratio near $\sim$1.00 and a core mass fraction of 32\% for Earth \citep{MercuryComposition,MercuryFesSi,EarthWang}.

It is possible for the majority of iron to be contained within silicate material including bridgmanite (MgSiO\textsubscript{3}/FeSiO\textsubscript{3}), magnesiow\"ustite (MgO/FeO), olivine (Mg\textsubscript{2}SiO\textsubscript{4}/Fe\textsubscript{2}SiO\textsubscript{4}) and pyroxenes (Mg\textsubscript{2}Si\textsubscript{2}O\textsubscript{6}/Fe\textsubscript{2}Si\textsubscript{2}O\textsubscript{6}) for bulk Fe/Si values less than 1.13 \citep{Alibert14}. For Fe/Si $>$ 1.13, models suggest that an iron core needs to be present within a rocky exoplanet to explain such a high ratio. This limit is calculated by simple stoichiometry and may not reflect the actual distribution of iron throughout a rocky exoplanet's core and mantle. The oxygen fugacity can also affect the distribution of a planet's iron distribution \citep{Bitsch20}, oxidising with mantle constituents instead of being differentiated into a core if the oxygen fugacity is too high \citep{ElkinsTanton08}. This would result in a lower core mass fraction compared to situations of lower fugacity. Current models show that iron can be taken up in the mantle \citep{Dorn2015,Unterborn2018} as well as silicon being taken up within an iron core \citep{Hirose2013}. Thus, Fe/Mg is a better proxy for core-to-mantle ratio and is produced within the GALAH-TESS catalog.

\begin{figure*}
\begin{center}
\includegraphics[width=0.8\textwidth]{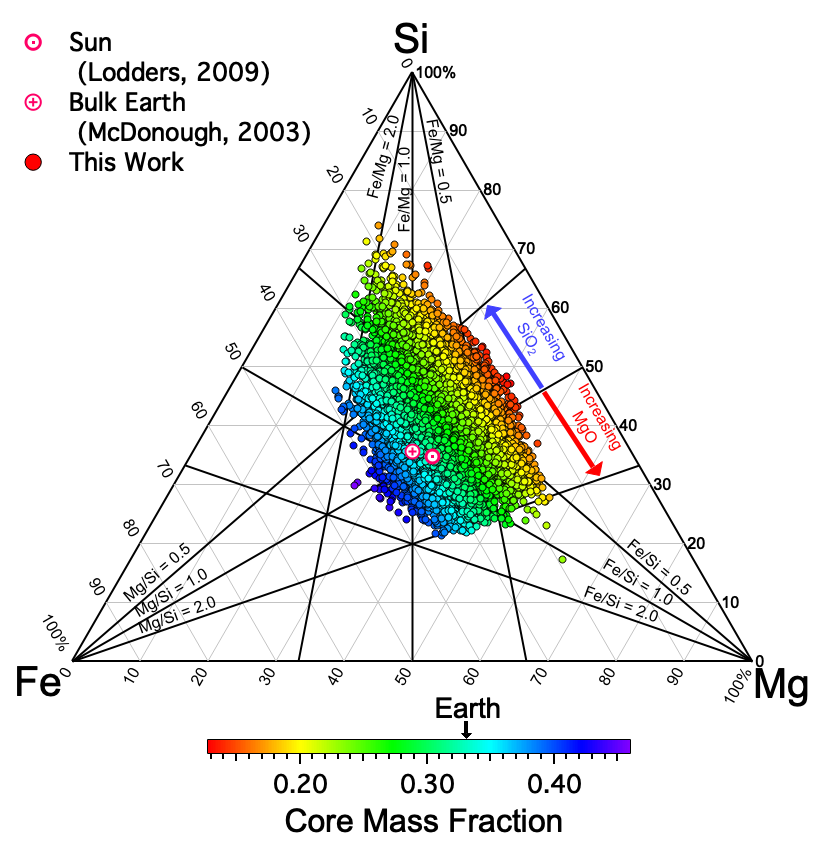}
\caption{Ternary diagram of the Fe, Mg and Si abundances from the for our GALAH-TESS stars assuming a Solar abundance model of \citet{Lodders09}. In general, the closer to an individual corner of the ternary a data point falls the greater the proportion of that element in the resulting planet assuming stellar composition roughly reflects planetary composition \citep{Bond2010b,Thiabaud14,Unterborn17}. Individual points are color-coded to show the maximum core mass fraction (CMF) of the planet assuming all Fe is present in the core and Mg and Si are in their oxide forms (MgO, SiO\textsubscript{2}). CMF is therefore calculated by $\#Fe*\mu_{Fe}/(\#Mg*(\mu_{Mg}+\mu_{O}) + \#Si*(\mu_{Si}+2*\mu_{O})+ \#Fe*\mu_{Fe})$ where $\#X$ represents the molar abundance of element $X$ and $\mu_X$ is the molar weight of that element. The Earth \citep{McDonough2003} and Solar \citep{Lodders09} abundances are shown for reference.}
\label{fig:tern}
\end{center}
\end{figure*}

\begin{figure}
\begin{center}
\includegraphics[width=\columnwidth]{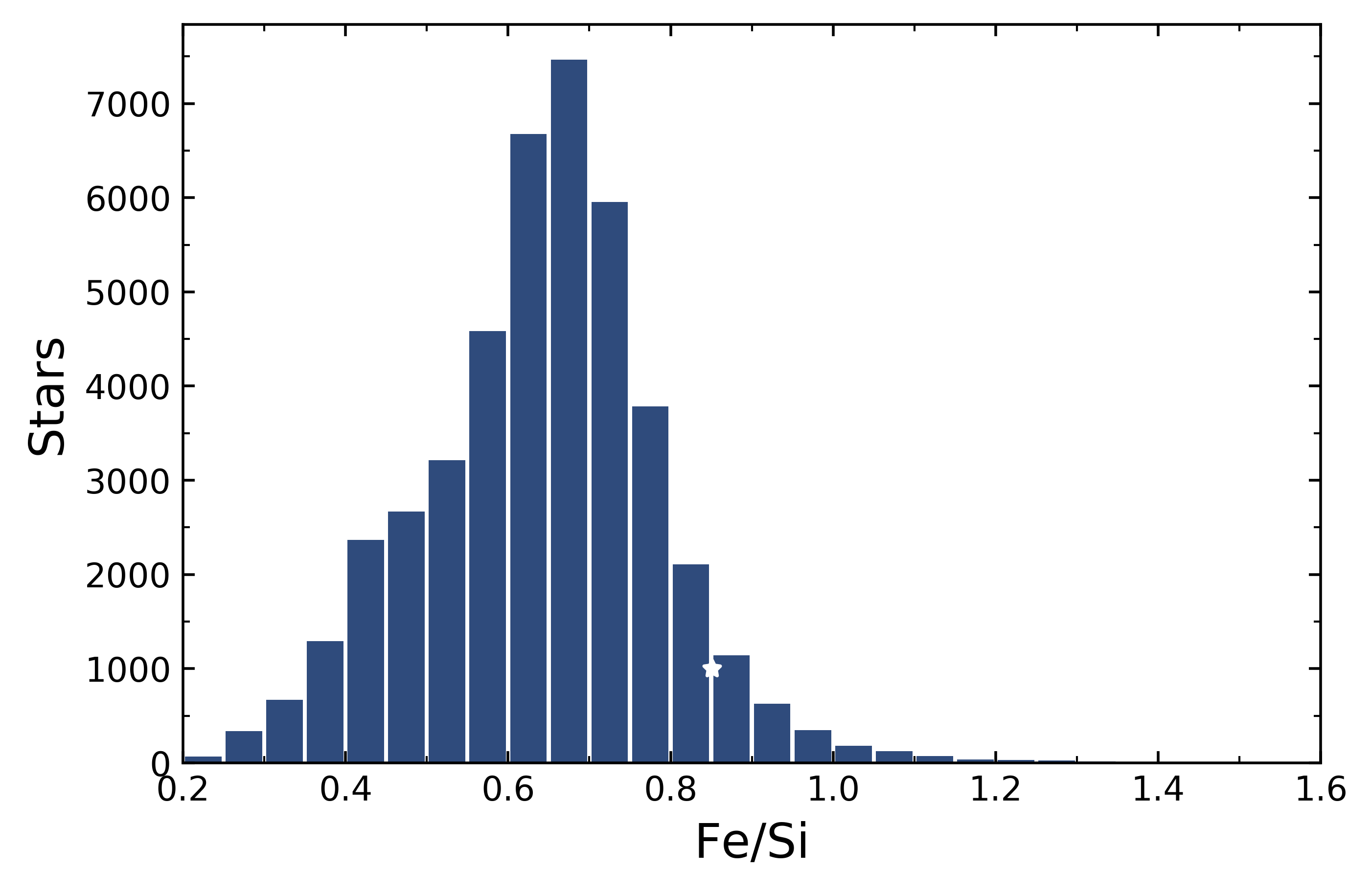}
\caption{Of the 47,285 stars within our sample, only 134 have Fe/Si values greater than 1.13 which would indicate the vast majority of our possible rocky worlds will have their iron content distributed between their iron and mantle layers. An iron-core must be present beyond Fe/Si values of 1.13 to explain such a high Fe/Si ratio. The white star within the histogram depicts the Sun's photospheric Fe/Si value of 0.85 \citep{Lodders09}. \label{fig:fesi}}
\end{center}
\end{figure}

Figure \ref{fig:tern} shows the distribution of Fe, Mg and Si for our sample of GALAH-TESS stars. We can calculate the core mass fraction of potential rocky planets hosted by GALAH-TESS stars using stiochiometry by the equation:
\begin{equation}
    \#Fe\times\mu_{Fe}/(\#Mg\times(\mu_{Mg}+\mu_{O}) + \#Si\times(\mu_{Si}+2\times\mu_{O})+ \#Fe\times\mu_{Fe})
    \label{eq:stoich}
\end{equation}
where $\#X$ represents the molar abundance of element $X$ and $\mu_X$ is the molar weight of that element. We are able to use this estimation as Fe, Mg and Si all have similar condensation temperature \citep{Lodders09} and thus thermal processes are unlikely to fractionate the elements relative to each other. That is while a planet may have significantly fewer atoms of Fe and Mg than the host star, the Fe/Mg ratio of the star and planet may only be different by $\sim$10\% \citep{Bond2010b, Thiabaud14,Unterborn17}. While mantle stripping by large impacts may increase the planet's Fe/Mg ratio \citep[e.g., ][]{bonomo19}. Equation \ref{eq:stoich} represents a reasonable upper-bound for CMF for most systems. As mentioned above, changes in oxygen fugacity will convert some core Fe into mantle FeO, which will lower the CMF for a given bulk composition. From this ternary we can see that stellar abundances outline a wide range of CMF compared to the Earth and Sun, with their abundances falling near the middle of the distribution (Figure \ref{fig:fesi}). Less than 0.3~\% of our stars have Fe/Si $>$ 1.13 (Figure \ref{fig:fesi}), therefore the rocky planets possibly orbiting GALAH-TESS stars may have their iron content distributed between both core and mantle layers with marginally lower CMF than predicted in Figure \ref{fig:tern}.

\subsubsection{Mantle Compositions of Rocky Exoplanets Through Stellar Host Mg/Si and C/O Ratios}

\begin{figure*}
\begin{center}
 \includegraphics[width=0.7\linewidth]{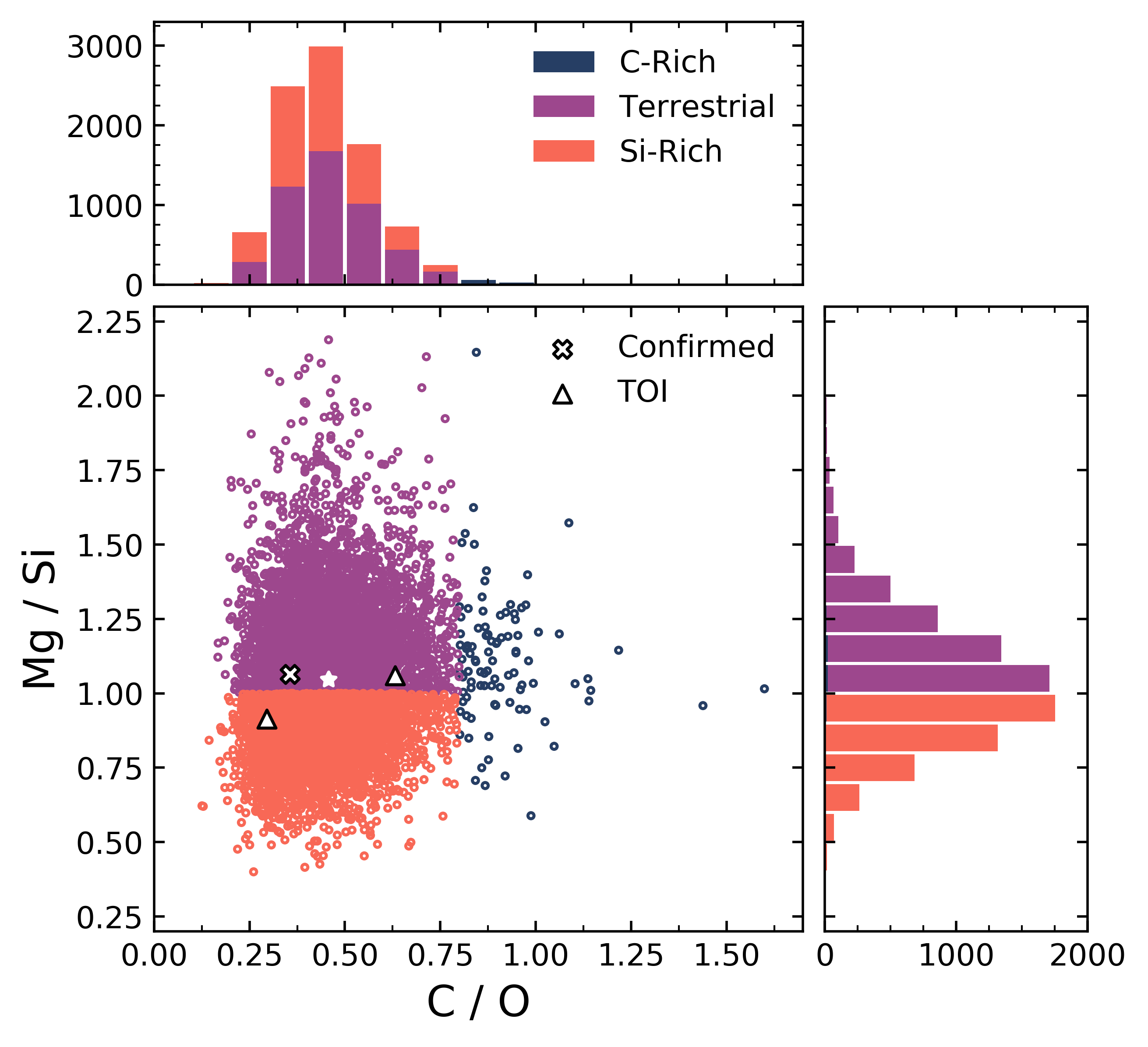}
 \caption{Distribution of all stars that have both measured C/O and Mg/Si ratios. These ratios can help inform astronomers on the likely composition of probable rocky worlds these stars may host. A total of 53.6~\% of these stars have C/O $<$ 0.8 and Mg/Si $>$ 1 values that would suggest that these stars may potentially host planets that would be similar in geological composition to Earth and Mars. These potential rocky worlds would host olivine and pyroxene within their upper mantle and bridgmanite and magnesiow\"{u}stite (or ferropericlase) in their lower mantle. That leaves 46.4~\% of stars that will host planets unlike any worlds within our Solar system. These include 45.4~\% of stars potentially hosting ``Silicon-Rich" rocky worlds with stellar abudnace ratios C/O $<$ 0.8 and Mg/Si $<$ 1, indicating that these worlds could contain pyroxene + SiO within both their upper and lower mantles. Only 1~\% of GALAH-TESS stars have C/O $>$ 0.8 indicating that they might host rocky worlds with carbon-rich mantles. Planet-hosting (cross) and candidate stars (TOI (upright triangle)) measured Mg/Si and C/O values are displayed on the figure with the Sun's Mg/Si and C/O values depicted with a white star \citep{Lodders09}.}\label{fig:covsmgsi}
\end{center}
\end{figure*}

The structure and composition of super-Earths and sub-Neptunes can be constrained through theoretical models using their host's Mg/Si and C/O elemental ratios. The stellar C/O abundance chemically controls the silicon distribution amongst oxides and carbides \citep{Bond10,Bond2012,Duffy15}. For those stars with C/O values less than 0.8, Mg/Si controls the mantle chemistry by varying the relative proportions of olivine, pyroxenes and oxides. However, within this realm of low C/O values, there are two distinct regimes in which the Mg and Si are distributed within the mantle: 

\begin{itemize}
 \item In a ``silicon-rich'' environment, whereby the Mg/Si $<$ 1, the upper mantle will be dominated by ortho- and clino-pyroxene, majoritic garnet (Mg\textsubscript{3}(MgSi)(SiO\textsubscript{4})\textsubscript{3}) as well as SiO\textsubscript{2} (either as quartz or coesite) with the lower mantle consisting of bridgmanite ((Mg,Fe)SiO\textsubscript{3}) and stishovite (SiO\textsubscript{2}). As Mg/Si decreases, the proportion of stishovite will increase at the cost of brigmanite in the lower mantle. 
 
 \item For larger values of Mg/Si, where Mg/Si $>$ 1, a rocky planet's upper mantle will mostly comprise of olivine (Mg\textsubscript{2}SiO\textsubscript{4}), pyroxenes and majoritic garnet, with bridgmanite and magnesiow\"ustite (or ferropericlase) ((Mg,Fe)O) in lower mantle. As the Mg/Si ratio increases, so does the amount of olivine and ferropericlase within the rocky planet's upper and lower mantle respectively. This regime of planetary composition is akin to rocky worlds (i.e Mars and Earth) within our Solar system and thus labelled as ``terrestrial-like'' mantle compositions within our paper. \citep{Unterborn17,Duffy15,Bond2012,Bond10}. As Mg/Si increases, the proportion of magnesio\"wustite will increase at the cost of brigmanite in the lower mantle. 
\end{itemize}

However these compositions only extend for C/O $<$ 0.8. For C/O $>$ 0.8, exotic mantle compositions of graphite and the carbides including SiC can start to dominate the geological composition of an exoplanet's core and mantle, when planets form within a protoplanetary disk's innermost region \citep{Bond2012,carbonworldEoS,Kuchner05,Wilson14,Nisr2017,Unterborn14}. These ``carbon-rich'' worlds can extend out through carbon-rich disks and can even form with C/O ratios as low as 0.67 \citep{Moriarty14}. However, the habitability of such worlds is still under debate, with some studies suggesting that habitability is unlikely. This is because theoretical models suggest that these worlds would likely be geodynamically inactive planets and would limit the amount of carbon-dioxide degassing into its atmosphere \citep{Unterborn14}. %However, stagnant-lid exoplanets with CO\textsubscript{2} budgets ranging from 0.001--1 times that of Earth could outgas CO\textsubscript{2} long enough to maintain habitability for $\sim$1--5 Gyr \citep{stagnetlid}.

\begin{figure*}
\begin{center}
\includegraphics[width=0.8\textwidth]{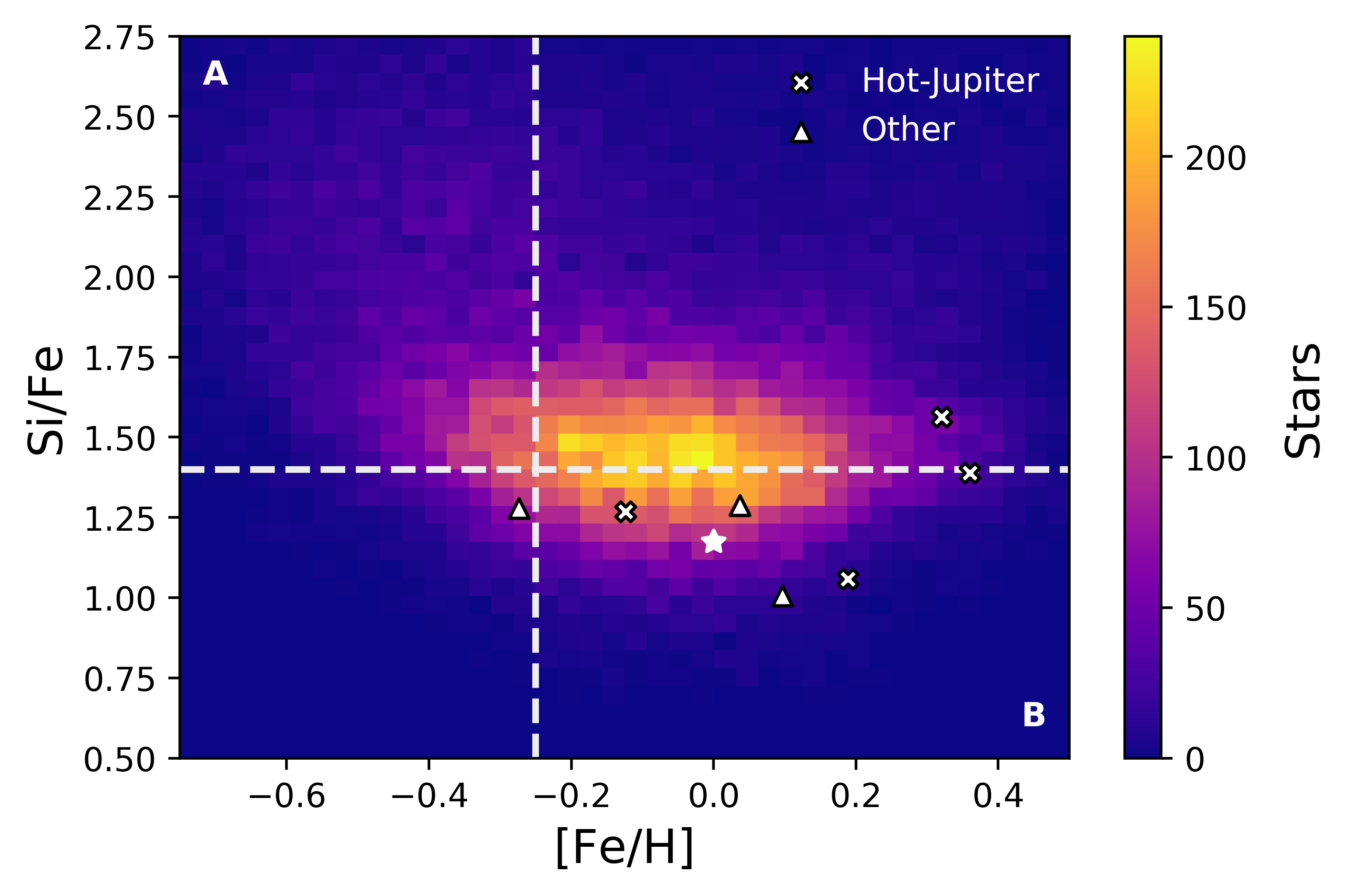}
\caption{In \citet{Brewer2018}, the authors discovered that compact multi-planet systems favoured iron-poor, silicon rich stars with a higher population of multiplanetary systems favouring the A quadrant in this Figure. A wider range of planetary systems, including single-systems consisting of hot-Jupiters were more common within quadrant B of this Figure. This figure shows our [Fe/H] and Si/Fe abundance ratios for 43162 GALAH-TESS stars. Based upon our results, a more diverse range of planetary systems will be uncovered around GALAH-TESS stars, with the majority of our stars lying in quadrant B. The white star represents the Sun's [Fe/H] and Si/Fe values with known confirmed or candidate hot-Jupiter (cross) and other (triangle) planetary systems shown for comparison.}\label{fig:fehsife}
\end{center}
\end{figure*}

Our C/O and Mg/Si distribution for the GALAH-TESS stars are found in Figure \ref{fig:covsmgsi}. Of our 47,000$+$ sample, only 8832 stars have C/O and Mg/Si ratios as most stars' C or O abundances were flagged by \textit{The Cannon}. This sample also includes exoplanet host WASP-61 and candidate hosts UCAC4 238-060232 (TOI-754) and  HD\,81655 (TOI-1031). A total of 53.6\% of these stars have C/O $<$ 0.8 and Mg/Si $>$ 1 values, suggesting that these stars may potentially host exoplanets that would have compositions akin to planets found within our own Solar system, including both known exoplanet-hosting stars WASP-61 and TOI-754. Both WASP-61 and TOI-754 however are only known to host Jupiter-sized worlds that would have significantly different core structures to that of smaller super-Earth and sub-Neptune exoplanets \citep{Mocquet2014,Fortney10,Buhler16}. However, future studies may discover smaller worlds around these stars. Within our GALAH-TESS sample, 46.4~\% of stars have Mg/Si and C/O ratios suggesting that these stars could possibly host rocky planets that are ``silicon-rich" compared to planets found within our Solar System. The candidate exoplanet host TOI-1031 is such a system that could boast Silicon-rich worlds with a Mg/Si value of 0.91 $\pm$ 0.20.

Distributions of Mg/Si similar to the ones we find within our sample have also been discovered with other surveys: $\sim$60\% of the \citet{Brewer16} sample of FGK dwarfs in the local neighbourhood also falls between 1 $<$ Mg/Si . Photospheric measurements of planet-hosting stars show a range of Mg/Si values ranging from 0.7 to 1.4 \citep{DelgadoMena10,Brewer16}, while our planet host and candidate stars Mg/Si values range from 0.9 to 1.1. Our median Mg/Si value is 0.98$\pm$0.22 which is lower than \citet{Brewer16}'s Mg/Si median value of 1.02. The larger spread of Mg/Si values in other surveys might be due to different Solar normalisations but seems more likely that this is due to a different stellar sample and methodologies to derive chemical abundances. \citep{hinkel14} showed that even for iron, the spread in for the same stars gathered from various groups was 0.16 dex. Thus, more work is needed to better understand the underlying systematics and variations of stellar abundances from various surveys and research groups.

Surprisingly, less than 1~\% of GALAH-TESS stars have a C/O ratio greater than 0.8, suggesting that these stars may host ``Carbon-Rich" worlds, that will have geological structures unlike any object within our Solar system. Our median C/O value is 0.44 $\pm$ 0.13 which is somewhat comparable to other stellar surveys \citep{Brewer16,Petigura11,DelgadoMena10} and population statistics \citep{Fortney12} -- but could be an overestimate from galactic chemical evolution models \citep{Fortney12}. The discrepancies between these surveys are likely due to different stellar populations, methodologies used to derive stellar abundances or Solar normalisations used as discussed in Section \ref{sec:chem}.

We should note that GALAH's [O/H] abundances do account for non-LTE effects but are only taken from at most four lines, with non-LTE effects for [O/H] abundance taken for the O\,\textsc{i} triplet near $\sim$777.5 nm \citep{GALAHDR2,GALAHNLTE,ONLTE}. This triplet is known to over-estimate abundances if non-LTE effects are not taken into account \citep{Teske13}. \citet{Brewer16}'s approach considers molecular OH lines and numerous more carbon lines, such that our results might be overestimated with respect to theirs. \citet{Teske14} found that there is currently no significant trend between planet-hosts, in particular the occurrence of hot-Jupiters, and their C/O values.

\subsubsection{How are Stellar Abundances Linked to Planetary Formation?}

There is theoretical evidence suggesting that the abundance ratios of refractory materials stay relatively constant throughout a protoplanetary disk, but it is misleading to suggest that volatile abundance ratios will be constant through the disk. Elemental abundance ratios can change through a protoplanetary disk depending upon the concentration of material and temperature profile of the disk \citep{Bond10,Bond2012,Unterborn17}. There are studies that suggest that estimates of the devolatilisation process within a protoplanetary disk could aid in determining the bulk elemental abundances of rocky worlds, assuming they have formed where they are currently situated within their own planetary system \citep{EarthWang}.

If we want to determine if a world has bulk composition as the earth, studies suggest that the errors with the elemental abundances themselves need to be further refined with uncertainties better than $\sim$ 0.04 dex needed for such a comparison \citep{Hinkel2018,EarthWang}. Even further, if we want to differentiate between unique planetary structures within a rocky exoplanet population, the uncertainties for Fe, Si, Al, Mg and Ca abundances need to be less than 0.02, 0.01, 0.002, 0.001 and 0.001 dex respectively \citep{Hinkel2018}. These uncertainties, especially for Al, Mg and Ca are unobtainable with current detection methods and Solar abundance normalisations. Hence, if we do want to accurately determine an exoplanet's interior and composition, which has vast implications for its habitability, then precision on spectroscopic abundances and Solar normalisations themselves also have to significantly increased.

The relationship between elemental abundances and planetary architectures is a complex one. There is an overall trend that hot-Jupiter systems favour iron-rich hosts \citep{Fischer2005,Mortier13} and early evidence that super-Earths are predominantly found around metal-poor and $\alpha$--rich stars \citep{Adibekyan12} and new work with machine-learning algorithms suggest elemental indicators for hot-Jupiter hosting stars apart from Fe are O, C, and Na \citep{Hinkel2019}. Work by \citet{Adibekyan2016,Sousa2019} shows super-Earths orbiting metal-rich stars have orbits that extended beyond their metal-poor hosted peers but contradicts \citet{CKSIV} who indicate that short-period super-Earths orbit metal-rich stars. \citet{Sousa2019} also suggest the mass of planets increases with the host star metallicity, but contradicts \citet{Teske_metalrichhosts}, which did not find such a correlation. \citet{Brewer2018} found that compact-multi systems are more common around metal-poor stars, showing a large [Fe/H] vs Si/Fe parameter space unfilled by single hot-Jupiters but filled with compact-multi systems for planet hosts with [Fe/H] values below 0.2 and Si/Fe values higher than 1.4. We have created a similar figure for our small exoplanetary sample, to somewhat forward predict the types of planetary architectures our GALAH-TESS stars might host. Figure \ref{fig:fehsife} shows that the majority of planet hosts and candidates fill quadrant B of this phase-space, where \citet{Brewer2018} found a diverse range of planetary architectures occupying this space. All of our confirmed and candidate systems favour iron-rich, silicon-poor stars, where a diverse range of exoplanetary architectures are likely to be found. This matches our current, though very small sample with single-planet systems hosting sub-Neptune to Jovian like worlds.

\section{Conclusion}\label{sec:conclusion}

The aim of this paper is to aid \textit{TESS} follow-up teams with a catalog of high precision physical and chemical stellar parameters for stars being observed with the space-based exoplanet survey satellite. We have cross-matched GALAH DR2 with the TIC to provide the physical and chemical characteristics for over $\sim$47,000 stars, eleven of which confirmed planet-hosts or planetary candidates discovered by the \textit{TESS} mission. The refinement of stellar radii and masses of those planet-hosting stars have improved the mass and radius measurements of the confirmed and candidate exoplanets they host, with a median relative uncertainty for our planetary mass and radius values being 5\% and 4\%, respectively. From these refinements, we have increased the planetary radii of CTOI-201256771.01 and CTOI-201256771.02 to near Solar values of 96.17 R$_\odot$ and 103.15 R$_\odot$, and with further investigation, have indicated that these transit events were likely caused by the eclipsing binary companion of UNSW-V 320 A. We also cast serious doubts over the candidate events CTOI-220402290.01, CTOI-220402290.02, CTOI-300903537.01 and CTOI 300903537.02 as their orbital periods alone suggest that these candidate systems are likely coming from one source and not two. Our updated mass and radius values changed on the order of 10--20\% from literature values, which have minor implications for the large exoplanets currently within the GALAH-TESS catalog, but would have profound impacts on the refinement of a fictitious ``Earth-like" world orbiting these stars, with a range of densities that would render some uninhabitable by current theories of habitability. 

Our catalog contains the elemental abundances for 23 elements which have been normalised by \citep{Lodders09} to not only drive consistency within the community, but to also make it easier for comparisons of elemental abundances from other abundance driven, stellar surveys to ours. The GALAH-TESS catalog includes the elemental abundance ratios for C/O, Mg/Si, Fe/Mg and Fe/Si which can help astronomers and planetary scientists make predictions about the composition and structure of potential rocky worlds orbiting our GALAH-TESS stars. Our stellar C/O and Mg/Si distributions suggest that the majority of GALAH-TESS stars will likely host worlds similar in composition to that of Earth and Mars, with over 54~\% of stars hosting Mg/Si $>$ 1, and C/O $<$ 0.8. However, 46~\% of stars have atmospheric abundance ratios of either Mg/Si $<$ 1 and C/O $<$ 0.8 or C/O $>$ 0.8, suggesting that these stars may host rocky worlds with geological compositions unlike any planet found within our Solar system. These values will change dependent upon the Solar normalisation used, hence the need for a standard Solar normalisation within the exoplanetary community. It is important in our language that a truly Earth-like planet has yet to be discovered \citep{Tasker2017}, but with our catalog and \textit{TESS}'s extended mission planned for the Southern Hemisphere in 2021, we can move closer to answering humankind's grandest question --- are we truly alone?

\section*{Software}
\textsc{AstroPy} \citep{astropy}, \textsc{Astroquery} \citep{astroquery}, \textsc{Isochrones} \citep{isochrones}, \textsc{Matplotlib} \citep{matplotlib}, \textsc{MultiNest} \citep{multinest19,multinest09,multinest08}, \textsc{Multiprocessing} \citep{multiprocessing}, \textsc{NumPy} \citep{numpy1,numpy2}, \textsc{OpenBLAS} \citep{openblas12,openblas13}, \textsc{Pandas} \citep{pandas}, \textsc{SciPy} \citep{scipy}

\section*{Acknowledgements}

Our research is based upon data acquired through the Australian Astronomical Observatory. We acknowledge the traditional owners of the land on which the AAT stands, the Gamilaraay people, and pay our respects to elders past, present and emerging.

This research has made use of the NASA Exoplanet Archive, which is operated by the California Institute of Technology, under contract with the National Aeronautics and Space Administration under the Exoplanet Exploration Program

This paper makes use of data from the first public release of the WASP data \citep{Butters2010} as provided by the WASP consortium and services at the NASA Exoplanet Archive, which is operated by the California Institute of Technology, under contract with the National Aeronautics and Space Administration under the Exoplanet Exploration Program.

This work has made use of the TIC and CT Stellar Properties Catalog, through the \textit{TESS} Science Office's target selection working group (architects K. Stassun, J. Pepper, N. De Lee, M. Paegert, R. Oelkers). The Filtergraph data portal system is trademarked by Vanderbilt University.

This research has made use of the Exoplanet Follow-up Observation Program website, which is operated by the California Institute of Technology, under contract with the National Aeronautics and Space Administration under the Exoplanet Exploration Program.

This work has made use of data from the European Space Agency (ESA) mission \textit{Gaia} (\url{http://www.cosmos.esa.int/gaia}), processed by the \textit{Gaia} Data Processing and Analysis Consortium (DPAC, \url{http://www.cosmos.esa.int/web/gaia/dpac/}
consortium). Funding for the DPAC has been provided by national institutions, in particular the institutions participating in the \textit{Gaia} Multilateral Agreement

The research shown here acknowledges use of the Hypatia Catalog Database, an online compilation of stellar abundance data as described in \citep{hinkel14}, which was supported by NASA's Nexus for Exoplanet System Science (NExSS) research coordination network and the Vanderbilt Initiative in Data-Intensive Astrophysics (VIDA).

J.T.C would like to thank SW and BW-C, and is supported by the Australian Government Research Training Program (RTP) Scholarship.
J.D.S and S.M acknowledges the support of the Australian Research Council through Discovery Project grant DP180101791.
S.B. acknowledges funds from the Australian Research Council (grants DP150100250 and DP160103747). Parts of this research were supported by the Australian Research Council (ARC) Centre of Excellence for All Sky Astrophysics in 3 Dimensions (ASTRO 3D), through project number CE170100013.
Y.S.T. is grateful to be supported by the NASA Hubble Fellowship grant HST-HF2-51425.001 awarded by the Space Telescope Science Institute.

\section*{Data Availability}
The data underlying this article are available in the article and in its online supplementary material.

%%%%%%%%%%%%%%%%%%%%%%%%%%%%%%%%%%%%%%%%%%%%%%%%%%



\bibliographystyle{mnras}
\bibliography{references} % if your bibtex file is called example.bib

\begin{thebibliography}{}
\makeatletter
\relax
\def\mn@urlcharsother{\let\do\@makeother \do\$\do\&\do\#\do\^\do\_\do\%\do\~}
\def\mn@doi{\begingroup\mn@urlcharsother \@ifnextchar [ {\mn@doi@}
  {\mn@doi@[]}}
\def\mn@doi@[#1]#2{\def\@tempa{#1}\ifx\@tempa\@empty \href
  {http://dx.doi.org/#2} {doi:#2}\else \href {http://dx.doi.org/#2} {#1}\fi
  \endgroup}
\def\mn@eprint#1#2{\mn@eprint@#1:#2::\@nil}
\def\mn@eprint@arXiv#1{\href {http://arxiv.org/abs/#1} {{\tt arXiv:#1}}}
\def\mn@eprint@dblp#1{\href {http://dblp.uni-trier.de/rec/bibtex/#1.xml}
  {dblp:#1}}
\def\mn@eprint@#1:#2:#3:#4\@nil{\def\@tempa {#1}\def\@tempb {#2}\def\@tempc
  {#3}\ifx \@tempc \@empty \let \@tempc \@tempb \let \@tempb \@tempa \fi \ifx
  \@tempb \@empty \def\@tempb {arXiv}\fi \@ifundefined
  {mn@eprint@\@tempb}{\@tempb:\@tempc}{\expandafter \expandafter \csname
  mn@eprint@\@tempb\endcsname \expandafter{\@tempc}}}

\bibitem[\protect\citeauthoryear{{Addison} et~al.,}{{Addison}
  et~al.}{2019}]{minervaAus}
{Addison} B.,  et~al., 2019, \mn@doi [\pasp] {10.1088/1538-3873/ab03aa}, \href
  {https://ui.adsabs.harvard.edu/abs/2019PASP..131k5003A} {131, 115003}

\bibitem[\protect\citeauthoryear{{Addison} et~al.,}{{Addison}
  et~al.}{2020}]{TOI257}
{Addison} B.~C.,  et~al., 2020, arXiv e-prints, \href
  {https://ui.adsabs.harvard.edu/abs/2020arXiv200107345A} {p. arXiv:2001.07345}

\bibitem[\protect\citeauthoryear{{Adibekyan} et~al.,}{{Adibekyan}
  et~al.}{2012}]{Adibekyan12}
{Adibekyan} V.~Z.,  et~al., 2012, \mn@doi [\aap] {10.1051/0004-6361/201219564},
  \href {https://ui.adsabs.harvard.edu/abs/2012A&A...543A..89A} {543, A89}

\bibitem[\protect\citeauthoryear{{Adibekyan} et~al.,}{{Adibekyan}
  et~al.}{2013}]{Adibekyan13}
{Adibekyan} V.~Z.,  et~al., 2013, \mn@doi [\aap] {10.1051/0004-6361/201321520},
  \href {https://ui.adsabs.harvard.edu/abs/2013A&A...554A..44A} {554, A44}

\bibitem[\protect\citeauthoryear{{Adibekyan}, {Figueira}  \&
  {Santos}}{{Adibekyan} et~al.}{2016}]{Adibekyan2016}
{Adibekyan} V.,  {Figueira} P.,   {Santos} N.~C.,  2016, \mn@doi [Origins of
  Life and Evolution of the Biosphere] {10.1007/s11084-016-9486-1}, \href
  {https://ui.adsabs.harvard.edu/abs/2016OLEB...46..351A} {46, 351}

\bibitem[\protect\citeauthoryear{{Adibekyan}, {Gon{\c{c}}alves da Silva},
  {Sousa}, {Santos}, {Delgado Mena}  \& {Hakobyan}}{{Adibekyan}
  et~al.}{2017}]{MgSiNeptunes}
{Adibekyan} V.,  {Gon{\c{c}}alves da Silva} H.~M.,  {Sousa} S.~G.,  {Santos}
  N.~C.,  {Delgado Mena} E.,   {Hakobyan} A.~A.,  2017, \mn@doi [Ap]
  {10.1007/s10511-017-9486-5}, \href
  {https://ui.adsabs.harvard.edu/abs/2017Ap.....60..325A} {60, 325}

\bibitem[\protect\citeauthoryear{{Alibert}}{{Alibert}}{2014}]{Alibert14}
{Alibert} Y.,  2014, \mn@doi [\aap] {10.1051/0004-6361/201322293}, \href
  {https://ui.adsabs.harvard.edu/abs/2014A&A...561A..41A} {561, A41}

\bibitem[\protect\citeauthoryear{{Amarsi} \& {Asplund}}{{Amarsi} \&
  {Asplund}}{2017}]{SiNLTE}
{Amarsi} A.~M.,  {Asplund} M.,  2017, \mn@doi [\mnras] {10.1093/mnras/stw2445},
  \href {https://ui.adsabs.harvard.edu/abs/2017MNRAS.464..264A} {464, 264}

\bibitem[\protect\citeauthoryear{{Amarsi}, {Asplund}, {Collet}  \&
  {Leenaarts}}{{Amarsi} et~al.}{2016a}]{ONLTE}
{Amarsi} A.~M.,  {Asplund} M.,  {Collet} R.,   {Leenaarts} J.,  2016a, \mn@doi
  [\mnras] {10.1093/mnras/stv2608}, \href
  {https://ui.adsabs.harvard.edu/abs/2016MNRAS.455.3735A} {455, 3735}

\bibitem[\protect\citeauthoryear{{Amarsi}, {Lind}, {Asplund}, {Barklem}  \&
  {Collet}}{{Amarsi} et~al.}{2016b}]{FeNLTE}
{Amarsi} A.~M.,  {Lind} K.,  {Asplund} M.,  {Barklem} P.~S.,   {Collet} R.,
  2016b, \mn@doi [\mnras] {10.1093/mnras/stw2077}, \href
  {https://ui.adsabs.harvard.edu/abs/2016MNRAS.463.1518A} {463, 1518}

\bibitem[\protect\citeauthoryear{{Anders} \& {Grevesse}}{{Anders} \&
  {Grevesse}}{1989}]{AndersGrevesse89}
{Anders} E.,  {Grevesse} N.,  1989, \mn@doi [\gca]
  {10.1016/0016-7037(89)90286-X}, \href
  {https://ui.adsabs.harvard.edu/abs/1989GeCoA..53..197A} {53, 197}

\bibitem[\protect\citeauthoryear{{Asplund}, {Grevesse}  \& {Sauval}}{{Asplund}
  et~al.}{2005}]{Asplund05}
{Asplund} M.,  {Grevesse} N.,   {Sauval} A.~J.,  2005, {The Solar Chemical
  Composition}.
p.~25

\bibitem[\protect\citeauthoryear{Asplund, Grevesse, Sauval  \& Scott}{Asplund
  et~al.}{2009}]{Asplund09}
Asplund M.,  Grevesse N.,  Sauval A.~J.,   Scott P.,  2009, \mn@doi [Annual
  Review of Astronomy and Astrophysics]
  {10.1146/annurev.astro.46.060407.145222}, 47, 481

\bibitem[\protect\citeauthoryear{{Astropy Collaboration} et~al.,}{{Astropy
  Collaboration} et~al.}{2013}]{astropy}
{Astropy Collaboration} et~al., 2013, \mn@doi [\aap]
  {10.1051/0004-6361/201322068}, \href
  {https://ui.adsabs.harvard.edu/abs/2013A&A...558A..33A} {558, A33}

\bibitem[\protect\citeauthoryear{{Barclay} et~al.,}{{Barclay}
  et~al.}{2013}]{Kepler69lol}
{Barclay} T.,  et~al., 2013, \mn@doi [\apj] {10.1088/0004-637X/768/2/101},
  \href {https://ui.adsabs.harvard.edu/abs/2013ApJ...768..101B} {768, 101}

\bibitem[\protect\citeauthoryear{{Barclay}, {Pepper}  \& {Quintana}}{{Barclay}
  et~al.}{2018}]{Barclay18}
{Barclay} T.,  {Pepper} J.,   {Quintana} E.~V.,  2018, \mn@doi [\apjs]
  {10.3847/1538-4365/aae3e9}, \href
  {https://ui.adsabs.harvard.edu/abs/2018ApJS..239....2B} {239, 2}

\bibitem[\protect\citeauthoryear{Batalha et~al.,}{Batalha
  et~al.}{2013}]{Batalha13}
Batalha N.~M.,  et~al., 2013, The Astrophysical Journal Supplement Series, 204,
  24

\bibitem[\protect\citeauthoryear{{Benz}, {Anic}, {Horner}  \& {Whitby}}{{Benz}
  et~al.}{2007}]{mercuryorigin}
{Benz} W.,  {Anic} A.,  {Horner} J.,   {Whitby} J.~A.,  2007, \mn@doi [\ssr]
  {10.1007/s11214-007-9284-1}, \href
  {https://ui.adsabs.harvard.edu/abs/2007SSRv..132..189B} {132, 189}

\bibitem[\protect\citeauthoryear{{Bitsch} \& {Battistini}}{{Bitsch} \&
  {Battistini}}{2020}]{Bitsch20}
{Bitsch} B.,  {Battistini} C.,  2020, \mn@doi [\aap]
  {10.1051/0004-6361/201936463}, \href
  {https://ui.adsabs.harvard.edu/abs/2020A&A...633A..10B} {633, A10}

\bibitem[\protect\citeauthoryear{{Bond}, {Lauretta}  \& {O'Brien}}{{Bond}
  et~al.}{2010a}]{Bond2010b}
{Bond} J.~C.,  {Lauretta} D.~S.,   {O'Brien} D.~P.,  2010a, \mn@doi [\icarus]
  {10.1016/j.icarus.2009.07.037}, \href
  {https://ui.adsabs.harvard.edu/abs/2010Icar..205..321B} {205, 321}

\bibitem[\protect\citeauthoryear{{Bond}, {O'Brien}  \& {Lauretta}}{{Bond}
  et~al.}{2010b}]{Bond10}
{Bond} J.~C.,  {O'Brien} D.~P.,   {Lauretta} D.~S.,  2010b, \mn@doi [\apj]
  {10.1088/0004-637X/715/2/1050}, \href
  {https://ui.adsabs.harvard.edu/abs/2010ApJ...715.1050B} {715, 1050}

\bibitem[\protect\citeauthoryear{{Bonomo} et~al.,}{{Bonomo}
  et~al.}{2019}]{bonomo19}
{Bonomo} A.~S.,  et~al., 2019, \mn@doi [Nature Astronomy]
  {10.1038/s41550-018-0684-9}, \href
  {https://ui.adsabs.harvard.edu/abs/2019NatAs...3..416B} {3, 416}

\bibitem[\protect\citeauthoryear{{Borucki} et~al.,}{{Borucki}
  et~al.}{2010}]{Kepler}
{Borucki} W.~J.,  et~al., 2010, \mn@doi [Science] {10.1126/science.1185402},
  \href {https://ui.adsabs.harvard.edu/abs/2010Sci...327..977B} {327, 977}

\bibitem[\protect\citeauthoryear{{Brewer} \& {Fischer}}{{Brewer} \&
  {Fischer}}{2016}]{Brewer16}
{Brewer} J.~M.,  {Fischer} D.~A.,  2016, \mn@doi [\apj]
  {10.3847/0004-637X/831/1/20}, \href
  {https://ui.adsabs.harvard.edu/abs/2016ApJ...831...20B} {831, 20}

\bibitem[\protect\citeauthoryear{Brewer, Fischer  \& Madhusudhan}{Brewer
  et~al.}{2017}]{COhotjupBrew}
Brewer J.~M.,  Fischer D.~A.,   Madhusudhan N.,  2017, \mn@doi [The
  Astronomical Journal] {10.3847/1538-3881/153/2/83}, 153, 83

\bibitem[\protect\citeauthoryear{{Brewer}, {Wang}, {Fischer}  \&
  {Foreman-Mackey}}{{Brewer} et~al.}{2018}]{Brewer2018}
{Brewer} J.~M.,  {Wang} S.,  {Fischer} D.~A.,   {Foreman-Mackey} D.,  2018,
  \mn@doi [\apjl] {10.3847/2041-8213/aae710}, \href
  {https://ui.adsabs.harvard.edu/abs/2018ApJ...867L...3B} {867, L3}

\bibitem[\protect\citeauthoryear{{Brugger}, {Mousis}, {Deleuil}  \&
  {Deschamps}}{{Brugger} et~al.}{2017}]{Brugger2017}
{Brugger} B.,  {Mousis} O.,  {Deleuil} M.,   {Deschamps} F.,  2017, \mn@doi
  [\apj] {10.3847/1538-4357/aa965a}, \href
  {https://ui.adsabs.harvard.edu/abs/2017ApJ...850...93B} {850, 93}

\bibitem[\protect\citeauthoryear{{Brugger}, {Mousis}, {Deleuil}  \&
  {Ronnet}}{{Brugger} et~al.}{2018}]{MercuryFesSi}
{Brugger} B.,  {Mousis} O.,  {Deleuil} M.,   {Ronnet} T.,  2018, in European
  Planetary Science Congress. pp EPSC2018--404

\bibitem[\protect\citeauthoryear{{Brugman}, {Phillips}  \& {Till}}{{Brugman}
  et~al.}{2020}]{Brugman2020}
{Brugman} K.~K.,  {Phillips} M.~G.,   {Till} C.~B.,  2020, in Exoplanets in Our
  Backyard: Solar System and Exoplanet Synergies on Planetary Formation,
  Evolution, and Habitability. p.~3016

\bibitem[\protect\citeauthoryear{{Buchhave} \& {Latham}}{{Buchhave} \&
  {Latham}}{2015}]{Buchhave2015}
{Buchhave} L.~A.,  {Latham} D.~W.,  2015, \mn@doi [\apj]
  {10.1088/0004-637X/808/2/187}, \href
  {https://ui.adsabs.harvard.edu/abs/2015ApJ...808..187B} {808, 187}

\bibitem[\protect\citeauthoryear{{Buchhave} et~al.,}{{Buchhave}
  et~al.}{2014}]{Buchhave2014}
{Buchhave} L.~A.,  et~al., 2014, \mn@doi [\nat] {10.1038/nature13254}, \href
  {https://ui.adsabs.harvard.edu/abs/2014Natur.509..593B} {509, 593}

\bibitem[\protect\citeauthoryear{{Buder} et~al.,}{{Buder}
  et~al.}{2018}]{GALAHDR2}
{Buder} S.,  et~al., 2018, \mn@doi [\mnras] {10.1093/mnras/sty1281}, \href
  {https://ui.adsabs.harvard.edu/\#abs/2018MNRAS.478.4513B} {478, 4513}

\bibitem[\protect\citeauthoryear{{Buhler}, {Knutson}, {Batygin}, {Fulton},
  {Fortney}, {Burrows}  \& {Wong}}{{Buhler} et~al.}{2016}]{Buhler16}
{Buhler} P.~B.,  {Knutson} H.~A.,  {Batygin} K.,  {Fulton} B.~J.,  {Fortney}
  J.~J.,  {Burrows} A.,   {Wong} I.,  2016, \mn@doi [\apj]
  {10.3847/0004-637X/821/1/26}, \href
  {https://ui.adsabs.harvard.edu/abs/2016ApJ...821...26B} {821, 26}

\bibitem[\protect\citeauthoryear{{Butters} et~al.,}{{Butters}
  et~al.}{2010}]{Butters2010}
{Butters} O.~W.,  et~al., 2010, \mn@doi [\aap] {10.1051/0004-6361/201015655},
  \href {https://ui.adsabs.harvard.edu/abs/2010A&A...520L..10B} {520, L10}

\bibitem[\protect\citeauthoryear{{Carter-Bond}, {O'Brien}, {Delgado Mena},
  {Israelian}, {Santos}  \& {Gonz{\'a}lez Hern{\'a}ndez}}{{Carter-Bond}
  et~al.}{2012}]{Bond2012}
{Carter-Bond} J.~C.,  {O'Brien} D.~P.,  {Delgado Mena} E.,  {Israelian} G.,
  {Santos} N.~C.,   {Gonz{\'a}lez Hern{\'a}ndez} J.~I.,  2012, \mn@doi [\apjl]
  {10.1088/2041-8205/747/1/L2}, \href
  {https://ui.adsabs.harvard.edu/abs/2012ApJ...747L...2C} {747, L2}

\bibitem[\protect\citeauthoryear{{Chen} \& {Kipping}}{{Chen} \&
  {Kipping}}{2017}]{ChenKipping}
{Chen} J.,  {Kipping} D.,  2017, \mn@doi [\apj] {10.3847/1538-4357/834/1/17},
  \href {https://ui.adsabs.harvard.edu/abs/2017ApJ...834...17C} {834, 17}

\bibitem[\protect\citeauthoryear{{Choi}, {Dotter}, {Conroy}, {Cantiello},
  {Paxton}  \& {Johnson}}{{Choi} et~al.}{2016}]{MIST}
{Choi} J.,  {Dotter} A.,  {Conroy} C.,  {Cantiello} M.,  {Paxton} B.,
  {Johnson} B.~D.,  2016, \mn@doi [\apj] {10.3847/0004-637X/823/2/102}, \href
  {https://ui.adsabs.harvard.edu/abs/2016ApJ...823..102C} {823, 102}

\bibitem[\protect\citeauthoryear{{Christiansen} et~al.,}{{Christiansen}
  et~al.}{2008}]{UNSWV320}
{Christiansen} J.~L.,  et~al., 2008, \mn@doi [\mnras]
  {10.1111/j.1365-2966.2008.13013.x}, \href
  {https://ui.adsabs.harvard.edu/abs/2008MNRAS.385.1749C} {385, 1749}

\bibitem[\protect\citeauthoryear{{Coughlin} et~al.,}{{Coughlin}
  et~al.}{2016}]{TR6}
{Coughlin} J.~L.,  et~al., 2016, \mn@doi [\apjs] {10.3847/0067-0049/224/1/12},
  \href {https://ui.adsabs.harvard.edu/abs/2016ApJS..224...12C} {224, 12}

\bibitem[\protect\citeauthoryear{{Cui} et~al.,}{{Cui} et~al.}{2012}]{LAMOST}
{Cui} X.-Q.,  et~al., 2012, \mn@doi [Research in Astronomy and Astrophysics]
  {10.1088/1674-4527/12/9/003}, \href
  {https://ui.adsabs.harvard.edu/abs/2012RAA....12.1197C} {12, 1197}

\bibitem[\protect\citeauthoryear{{Dalba} et~al.,}{{Dalba}
  et~al.}{2020}]{TESSKeck}
{Dalba} P.~A.,  et~al., 2020, \mn@doi [\aj] {10.3847/1538-3881/ab84e3}, \href
  {https://ui.adsabs.harvard.edu/abs/2020AJ....159..241D} {159, 241}

\bibitem[\protect\citeauthoryear{{Davis} et~al.,}{{Davis}
  et~al.}{2019}]{TESSLCO}
{Davis} A.~B.,  et~al., 2019, arXiv e-prints, \href
  {https://ui.adsabs.harvard.edu/abs/2019arXiv191210186D} {p. arXiv:1912.10186}

\bibitem[\protect\citeauthoryear{{De Silva} et~al.,}{{De Silva}
  et~al.}{2015}]{DeSilva15}
{De Silva} G.~M.,  et~al., 2015, \mn@doi [\mnras] {10.1093/mnras/stv327}, \href
  {https://ui.adsabs.harvard.edu/abs/2015MNRAS.449.2604D} {449, 2604}

\bibitem[\protect\citeauthoryear{{Delgado Mena}, {Israelian}, {Gonz{\'a}lez
  Hern{\'a}ndez}, {Bond}, {Santos}, {Udry}  \& {Mayor}}{{Delgado Mena}
  et~al.}{2010}]{DelgadoMena10}
{Delgado Mena} E.,  {Israelian} G.,  {Gonz{\'a}lez Hern{\'a}ndez} J.~I.,
  {Bond} J.~C.,  {Santos} N.~C.,  {Udry} S.,   {Mayor} M.,  2010, \mn@doi
  [\apj] {10.1088/0004-637X/725/2/2349}, \href
  {https://ui.adsabs.harvard.edu/abs/2010ApJ...725.2349D} {725, 2349}

\bibitem[\protect\citeauthoryear{{Dorn}, {Khan}, {Heng}, {Connolly}, {Alibert},
  {Benz}  \& {Tackley}}{{Dorn} et~al.}{2015}]{Dorn2015}
{Dorn} C.,  {Khan} A.,  {Heng} K.,  {Connolly} J. A.~D.,  {Alibert} Y.,  {Benz}
  W.,   {Tackley} P.,  2015, \mn@doi [\aap] {10.1051/0004-6361/201424915},
  \href {https://ui.adsabs.harvard.edu/abs/2015A&A...577A..83D} {577, A83}

\bibitem[\protect\citeauthoryear{{Dorn}, {Venturini}, {Khan}, {Heng},
  {Alibert}, {Helled}, {Rivoldini}  \& {Benz}}{{Dorn} et~al.}{2017a}]{Dorn2017}
{Dorn} C.,  {Venturini} J.,  {Khan} A.,  {Heng} K.,  {Alibert} Y.,  {Helled}
  R.,  {Rivoldini} A.,   {Benz} W.,  2017a, \mn@doi [\aap]
  {10.1051/0004-6361/201628708}, \href
  {https://ui.adsabs.harvard.edu/abs/2017A&A...597A..37D} {597, A37}

\bibitem[\protect\citeauthoryear{{Dorn}, {Hinkel}  \& {Venturini}}{{Dorn}
  et~al.}{2017b}]{Dorn2017b}
{Dorn} C.,  {Hinkel} N.~R.,   {Venturini} J.,  2017b, \mn@doi [\aap]
  {10.1051/0004-6361/201628749}, \href
  {https://ui.adsabs.harvard.edu/abs/2017A&A...597A..38D} {597, A38}

\bibitem[\protect\citeauthoryear{{Dorn}, {Harrison}, {Bonsor}  \&
  {Hands}}{{Dorn} et~al.}{2019}]{DornCaAlPaper}
{Dorn} C.,  {Harrison} J.~H.~D.,  {Bonsor} A.,   {Hands} T.~O.,  2019, \mn@doi
  [\mnras] {10.1093/mnras/sty3435}, \href
  {https://ui.adsabs.harvard.edu/\#abs/2019MNRAS.484..712D} {484, 712}

\bibitem[\protect\citeauthoryear{Duffy, Madhusudhan  \& Lee}{Duffy
  et~al.}{2015}]{Duffy15}
Duffy T.,  Madhusudhan N.,   Lee K.,  2015, in Schubert G.,  ed., , Treatise on
  Geophysics (Second Edition), second edition edn, Elsevier, Oxford, pp 149 --
  178, \mn@doi{https://doi.org/10.1016/B978-0-444-53802-4.00053-1}, \url
  {http://www.sciencedirect.com/science/article/pii/B9780444538024000531}

\bibitem[\protect\citeauthoryear{Eisner et~al.,}{Eisner
  et~al.}{2020}]{planethuntersTESS}
Eisner N.,  et~al., 2020, Monthly Notices of the Royal Astronomical Society,
  494, 750

\bibitem[\protect\citeauthoryear{{Elkins-Tanton} \& {Seager}}{{Elkins-Tanton}
  \& {Seager}}{2008}]{ElkinsTanton08}
{Elkins-Tanton} L.~T.,  {Seager} S.,  2008, \mn@doi [\apj] {10.1086/591433},
  \href {https://ui.adsabs.harvard.edu/abs/2008ApJ...685.1237E} {685, 1237}

\bibitem[\protect\citeauthoryear{{Endl} et~al.,}{{Endl} et~al.}{2016}]{RV5}
{Endl} M.,  et~al., 2016, \mn@doi [\apj] {10.3847/0004-637X/818/1/34}, \href
  {https://ui.adsabs.harvard.edu/abs/2016ApJ...818...34E} {818, 34}

\bibitem[\protect\citeauthoryear{{Feroz} \& {Hobson}}{{Feroz} \&
  {Hobson}}{2008}]{multinest08}
{Feroz} F.,  {Hobson} M.~P.,  2008, \mn@doi [\mnras]
  {10.1111/j.1365-2966.2007.12353.x}, \href
  {https://ui.adsabs.harvard.edu/abs/2008MNRAS.384..449F} {384, 449}

\bibitem[\protect\citeauthoryear{{Feroz}, {Hobson}  \& {Bridges}}{{Feroz}
  et~al.}{2009}]{multinest09}
{Feroz} F.,  {Hobson} M.~P.,   {Bridges} M.,  2009, \mn@doi [\mnras]
  {10.1111/j.1365-2966.2009.14548.x}, \href
  {https://ui.adsabs.harvard.edu/abs/2009MNRAS.398.1601F} {398, 1601}

\bibitem[\protect\citeauthoryear{{Feroz}, {Hobson}, {Cameron}  \&
  {Pettitt}}{{Feroz} et~al.}{2019}]{multinest19}
{Feroz} F.,  {Hobson} M.~P.,  {Cameron} E.,   {Pettitt} A.~N.,  2019, \mn@doi
  [The Open Journal of Astrophysics] {10.21105/astro.1306.2144}, \href
  {https://ui.adsabs.harvard.edu/abs/2019OJAp....2E..10F} {2, 10}

\bibitem[\protect\citeauthoryear{{Fischer} \& {Valenti}}{{Fischer} \&
  {Valenti}}{2005}]{Fischer2005}
{Fischer} D.~A.,  {Valenti} J.,  2005, \mn@doi [\apj] {10.1086/428383}, \href
  {https://ui.adsabs.harvard.edu/abs/2005ApJ...622.1102F} {622, 1102}

\bibitem[\protect\citeauthoryear{{Fischer} et~al.,}{{Fischer}
  et~al.}{2008}]{RV2}
{Fischer} D.~A.,  et~al., 2008, \mn@doi [\apj] {10.1086/525512}, \href
  {https://ui.adsabs.harvard.edu/abs/2008ApJ...675..790F} {675, 790}

\bibitem[\protect\citeauthoryear{{Fortney}}{{Fortney}}{2012}]{Fortney12}
{Fortney} J.~J.,  2012, \mn@doi [\apjl] {10.1088/2041-8205/747/2/L27}, \href
  {https://ui.adsabs.harvard.edu/abs/2012ApJ...747L..27F} {747, L27}

\bibitem[\protect\citeauthoryear{{Fortney} \& {Nettelmann}}{{Fortney} \&
  {Nettelmann}}{2010}]{Fortney10}
{Fortney} J.~J.,  {Nettelmann} N.,  2010, \mn@doi [\ssr]
  {10.1007/s11214-009-9582-x}, \href
  {https://ui.adsabs.harvard.edu/abs/2010SSRv..152..423F} {152, 423}

\bibitem[\protect\citeauthoryear{{Fuhrmann}}{{Fuhrmann}}{1998}]{Fuhrmann98}
{Fuhrmann} K.,  1998, \aap, \href
  {https://ui.adsabs.harvard.edu/abs/1998A&A...338..161F} {338, 161}

\bibitem[\protect\citeauthoryear{{Gaia Collaboration} et~al.,}{{Gaia
  Collaboration} et~al.}{2018}]{gaiadeearetwo}
{Gaia Collaboration} et~al., 2018, \mn@doi [\aap]
  {10.1051/0004-6361/201833051}, \href
  {https://ui.adsabs.harvard.edu/abs/2018A&A...616A...1G} {616, A1}

\bibitem[\protect\citeauthoryear{{Gao} et~al.,}{{Gao} et~al.}{2018}]{GALAHNLTE}
{Gao} X.,  et~al., 2018, \mn@doi [\mnras] {10.1093/mnras/sty2414}, \href
  {https://ui.adsabs.harvard.edu/abs/2018MNRAS.481.2666G} {481, 2666}

\bibitem[\protect\citeauthoryear{{Gilbert} et~al.,}{{Gilbert}
  et~al.}{2020}]{TESS700d}
{Gilbert} E.~A.,  et~al., 2020, arXiv e-prints, \href
  {https://ui.adsabs.harvard.edu/abs/2020arXiv200100952G} {p. arXiv:2001.00952}

\bibitem[\protect\citeauthoryear{Gillon et~al.,}{Gillon
  et~al.}{2017}]{Gillon17}
Gillon M.,  et~al., 2017, Nature, 542, 456

\bibitem[\protect\citeauthoryear{{Ginsburg} et~al.,}{{Ginsburg}
  et~al.}{2019}]{astroquery}
{Ginsburg} A.,  et~al., 2019, \mn@doi [\aj] {10.3847/1538-3881/aafc33}, \href
  {https://ui.adsabs.harvard.edu/abs/2019AJ....157...98G} {157, 98}

\bibitem[\protect\citeauthoryear{{Gonzalez}}{{Gonzalez}}{1997}]{hotjupfeh97}
{Gonzalez} G.,  1997, \mn@doi [\mnras] {10.1093/mnras/285.2.403}, \href
  {https://ui.adsabs.harvard.edu/abs/1997MNRAS.285..403G} {285, 403}

\bibitem[\protect\citeauthoryear{{Grevesse} \& {Sauval}}{{Grevesse} \&
  {Sauval}}{1998}]{GrevesseSauval98}
{Grevesse} N.,  {Sauval} A.~J.,  1998, \mn@doi [\ssr]
  {10.1023/A:1005161325181}, \href
  {https://ui.adsabs.harvard.edu/abs/1998SSRv...85..161G} {85, 161}

\bibitem[\protect\citeauthoryear{{Grevesse}, {Asplund}  \& {Sauval}}{{Grevesse}
  et~al.}{2007}]{Grevesse07}
{Grevesse} N.,  {Asplund} M.,   {Sauval} A.~J.,  2007, \mn@doi [\ssr]
  {10.1007/s11214-007-9173-7}, \href
  {https://ui.adsabs.harvard.edu/abs/2007SSRv..130..105G} {130, 105}

\bibitem[\protect\citeauthoryear{{Hardegree-Ullman}, {Zink}, {Christiansen},
  {Dressing}, {Ciardi}  \& {Schlieder}}{{Hardegree-Ullman}
  et~al.}{2020}]{KHU20}
{Hardegree-Ullman} K.~K.,  {Zink} J.~K.,  {Christiansen} J.~L.,  {Dressing}
  C.~D.,  {Ciardi} D.~R.,   {Schlieder} J.~E.,  2020, \mn@doi [\apjs]
  {10.3847/1538-4365/ab7230}, \href
  {https://ui.adsabs.harvard.edu/abs/2020ApJS..247...28H} {247, 28}

\bibitem[\protect\citeauthoryear{{Hellier} et~al.,}{{Hellier}
  et~al.}{2012}]{TR3}
{Hellier} C.,  et~al., 2012, \mn@doi [\mnras]
  {10.1111/j.1365-2966.2012.21780.x}, \href
  {https://ui.adsabs.harvard.edu/abs/2012MNRAS.426..739H} {426, 739}

\bibitem[\protect\citeauthoryear{{Hinkel} \& {Burger}}{{Hinkel} \&
  {Burger}}{2017a}]{HypatiaCat}
{Hinkel} N.~R.,  {Burger} D.,  2017a, arXiv e-prints, \href
  {https://ui.adsabs.harvard.edu/abs/2017arXiv171204944H} {p. arXiv:1712.04944}

\bibitem[\protect\citeauthoryear{{Hinkel} \& {Burger}}{{Hinkel} \&
  {Burger}}{2017b}]{Hinkel17}
{Hinkel} N.~R.,  {Burger} D.,  2017b, arXiv e-prints, \href
  {https://ui.adsabs.harvard.edu/abs/2017arXiv171204944H} {p. arXiv:1712.04944}

\bibitem[\protect\citeauthoryear{{Hinkel} \& {Unterborn}}{{Hinkel} \&
  {Unterborn}}{2018}]{Hinkel2018}
{Hinkel} N.~R.,  {Unterborn} C.~T.,  2018, \mn@doi [\apj]
  {10.3847/1538-4357/aaa5b4}, \href
  {https://ui.adsabs.harvard.edu/abs/2018ApJ...853...83H} {853, 83}

\bibitem[\protect\citeauthoryear{{Hinkel}, {Timmes}, {Young}, {Pagano}  \&
  {Turnbull}}{{Hinkel} et~al.}{2014}]{hinkel14}
{Hinkel} N.~R.,  {Timmes} F.~X.,  {Young} P.~A.,  {Pagano} M.~D.,   {Turnbull}
  M.~C.,  2014, \mn@doi [\aj] {10.1088/0004-6256/148/3/54}, \href
  {https://ui.adsabs.harvard.edu/abs/2014AJ....148...54H} {148, 54}

\bibitem[\protect\citeauthoryear{{Hinkel} et~al.,}{{Hinkel}
  et~al.}{2016}]{Hinkel16}
{Hinkel} N.~R.,  et~al., 2016, \mn@doi [\apjs] {10.3847/0067-0049/226/1/4},
  \href {https://ui.adsabs.harvard.edu/abs/2016ApJS..226....4H} {226, 4}

\bibitem[\protect\citeauthoryear{{Hinkel}, {Unterborn}, {Kane}, {Somers}  \&
  {Galvez}}{{Hinkel} et~al.}{2019}]{Hinkel2019}
{Hinkel} N.~R.,  {Unterborn} C.,  {Kane} S.~R.,  {Somers} G.,   {Galvez} R.,
  2019, \mn@doi [\apj] {10.3847/1538-4357/ab27c0}, \href
  {https://ui.adsabs.harvard.edu/abs/2019ApJ...880...49H} {880, 49}

\bibitem[\protect\citeauthoryear{{Hirose}, {Labrosse}  \& {Hernlund}}{{Hirose}
  et~al.}{2013}]{Hirose2013}
{Hirose} K.,  {Labrosse} S.,   {Hernlund} J.,  2013, \mn@doi [Annual Review of
  Earth and Planetary Sciences] {10.1146/annurev-earth-050212-124007}, \href
  {https://ui.adsabs.harvard.edu/abs/2013AREPS..41..657H} {41, 657}

\bibitem[\protect\citeauthoryear{{Horner} \& {Jones}}{{Horner} \&
  {Jones}}{2010}]{WhichExoEarth}
{Horner} J.,  {Jones} B.~W.,  2010, \mn@doi [International Journal of
  Astrobiology] {10.1017/S1473550410000261}, \href
  {https://ui.adsabs.harvard.edu/abs/2010IJAsB...9..273H} {9, 273}

\bibitem[\protect\citeauthoryear{{Horner} et~al.,}{{Horner}
  et~al.}{2020}]{SSreview}
{Horner} J.,  et~al., 2020, accepted to appear in PASP, available on arXiv
  e-prints, \href {https://ui.adsabs.harvard.edu/abs/2020arXiv200413209H} {p.
  arXiv:2004.13209}

\bibitem[\protect\citeauthoryear{{Huang} et~al.,}{{Huang}
  et~al.}{2018}]{Huang2018}
{Huang} C.~X.,  et~al., 2018, \mn@doi [\apjl] {10.3847/2041-8213/aaef91}, \href
  {https://ui.adsabs.harvard.edu/abs/2018ApJ...868L..39H} {868, L39}

\bibitem[\protect\citeauthoryear{{Huber} et~al.,}{{Huber}
  et~al.}{2019}]{bigboisTOI197b}
{Huber} D.,  et~al., 2019, \mn@doi [\aj] {10.3847/1538-3881/ab1488}, \href
  {https://ui.adsabs.harvard.edu/abs/2019AJ....157..245H} {157, 245}

\bibitem[\protect\citeauthoryear{{Hunter}}{{Hunter}}{2007}]{matplotlib}
{Hunter} J.~D.,  2007, \mn@doi [Computing in Science and Engineering]
  {10.1109/MCSE.2007.55}, \href
  {https://ui.adsabs.harvard.edu/abs/2007CSE.....9...90H} {9, 90}

\bibitem[\protect\citeauthoryear{{Johns}, {Marti}, {Huff}, {McCann},
  {Wittenmyer}, {Horner}  \& {Wright}}{{Johns} et~al.}{2018}]{GaiaRevisedRadii}
{Johns} D.,  {Marti} C.,  {Huff} M.,  {McCann} J.,  {Wittenmyer} R.~A.,
  {Horner} J.,   {Wright} D.~J.,  2018, \mn@doi [\apjs]
  {10.3847/1538-4365/aae5fb}, \href
  {https://ui.adsabs.harvard.edu/abs/2018ApJS..239...14J} {239, 14}

\bibitem[\protect\citeauthoryear{{Johnson} et~al.,}{{Johnson}
  et~al.}{2011}]{bigbois11}
{Johnson} J.~A.,  et~al., 2011, \mn@doi [\apjs] {10.1088/0067-0049/197/2/26},
  \href {https://ui.adsabs.harvard.edu/abs/2011ApJS..197...26J} {197, 26}

\bibitem[\protect\citeauthoryear{{Jones} et~al.,}{{Jones}
  et~al.}{2016}]{bigbois16}
{Jones} M.~I.,  et~al., 2016, \mn@doi [\aap] {10.1051/0004-6361/201628067},
  \href {https://ui.adsabs.harvard.edu/abs/2016A&A...590A..38J} {590, A38}

\bibitem[\protect\citeauthoryear{{Jord{\'a}n} et~al.,}{{Jord{\'a}n}
  et~al.}{2020}]{TOI677}
{Jord{\'a}n} A.,  et~al., 2020, \mn@doi [\aj] {10.3847/1538-3881/ab6f67}, \href
  {https://ui.adsabs.harvard.edu/abs/2020AJ....159..145J} {159, 145}

\bibitem[\protect\citeauthoryear{{Kite} \& {Ford}}{{Kite} \&
  {Ford}}{2018}]{waterworld18}
{Kite} E.~S.,  {Ford} E.~B.,  2018, \mn@doi [\apj] {10.3847/1538-4357/aad6e0},
  \href {https://ui.adsabs.harvard.edu/abs/2018ApJ...864...75K} {864, 75}

\bibitem[\protect\citeauthoryear{Kopparapu et~al.,}{Kopparapu
  et~al.}{2013}]{HZcalc}
Kopparapu R.~K.,  et~al., 2013, The Astrophysical Journal, 765, 131

\bibitem[\protect\citeauthoryear{{Kos} et~al.,}{{Kos}
  et~al.}{2017}]{GALAHpipeline}
{Kos} J.,  et~al., 2017, \mn@doi [\mnras] {10.1093/mnras/stw2064}, \href
  {https://ui.adsabs.harvard.edu/abs/2017MNRAS.464.1259K} {464, 1259}

\bibitem[\protect\citeauthoryear{{Kuchner} \& {Seager}}{{Kuchner} \&
  {Seager}}{2005}]{Kuchner05}
{Kuchner} M.~J.,  {Seager} S.,  2005, arXiv e-prints, \href
  {https://ui.adsabs.harvard.edu/abs/2005astro.ph..4214K} {pp
  astro--ph/0504214}

\bibitem[\protect\citeauthoryear{{Latham}, {Mazeh}, {Stefanik}, {Mayor}  \&
  {Burki}}{{Latham} et~al.}{1989}]{Latham89}
{Latham} D.~W.,  {Mazeh} T.,  {Stefanik} R.~P.,  {Mayor} M.,   {Burki} G.,
  1989, \mn@doi [\nat] {10.1038/339038a0}, \href
  {https://ui.adsabs.harvard.edu/abs/1989Natur.339...38L} {339, 38}

\bibitem[\protect\citeauthoryear{{Lawton} \& {Wright}}{{Lawton} \&
  {Wright}}{1989}]{gammaCeph}
{Lawton} A.~T.,  {Wright} P.,  1989, Journal of the British Interplanetary
  Society, \href {https://ui.adsabs.harvard.edu/abs/1989JBIS...42..335L} {42,
  335}

\bibitem[\protect\citeauthoryear{{Lewis} et~al.,}{{Lewis}
  et~al.}{2002}]{AAT2dF}
{Lewis} I.~J.,  et~al., 2002, \mn@doi [\mnras]
  {10.1046/j.1365-8711.2002.05333.x}, \href
  {https://ui.adsabs.harvard.edu/abs/2002MNRAS.333..279L} {333, 279}

\bibitem[\protect\citeauthoryear{{Ligi} et~al.,}{{Ligi}
  et~al.}{2019}]{Ligi2019}
{Ligi} R.,  et~al., 2019, \mn@doi [\aap] {10.1051/0004-6361/201936259}, \href
  {https://ui.adsabs.harvard.edu/abs/2019A&A...631A..92L} {631, A92}

\bibitem[\protect\citeauthoryear{{Lingam} \& {Loeb}}{{Lingam} \&
  {Loeb}}{2019}]{waterworld19}
{Lingam} M.,  {Loeb} A.,  2019, \mn@doi [International Journal of Astrobiology]
  {10.1017/S1473550418000083}, \href
  {https://ui.adsabs.harvard.edu/abs/2019IJAsB..18..112L} {18, 112}

\bibitem[\protect\citeauthoryear{{Lodders}}{{Lodders}}{2003}]{Lodders2003}
{Lodders} K.,  2003, \mn@doi [\apj] {10.1086/375492}, \href
  {https://ui.adsabs.harvard.edu/abs/2003ApJ...591.1220L} {591, 1220}

\bibitem[\protect\citeauthoryear{{Lodders}, {Palme}  \& {Gail}}{{Lodders}
  et~al.}{2009}]{Lodders09}
{Lodders} K.,  {Palme} H.,   {Gail} H.~P.,  2009, \mn@doi [Landolt
  B\&ouml;rnstein] {10.1007/978-3-540-88055-4_34}, \href
  {https://ui.adsabs.harvard.edu/abs/2009LanB...4B..712L} {4B, 712}

\bibitem[\protect\citeauthoryear{{Lovis} \& {Fischer}}{{Lovis} \&
  {Fischer}}{2010}]{RVmethod}
{Lovis} C.,  {Fischer} D.,  2010, {Radial Velocity Techniques for Exoplanets}.
pp 27--53

\bibitem[\protect\citeauthoryear{{Lovis} et~al.,}{{Lovis} et~al.}{2011}]{RV3}
{Lovis} C.,  et~al., 2011, \mn@doi [\aap] {10.1051/0004-6361/201015577}, \href
  {https://ui.adsabs.harvard.edu/abs/2011A&A...528A.112L} {528, A112}

\bibitem[\protect\citeauthoryear{Madhusudhan, Lee  \& Mousis}{Madhusudhan
  et~al.}{2012}]{Madhusudhan12}
Madhusudhan N.,  Lee K. K.~M.,   Mousis O.,  2012, The Astrophysical Journal
  Letters, 759, L40

\bibitem[\protect\citeauthoryear{Martell et~al.,}{Martell
  et~al.}{2016}]{GALAHDR1}
Martell S.~L.,  et~al., 2016, \mn@doi [Monthly Notices of the Royal
  Astronomical Society] {10.1093/mnras/stw2835}, 465, 3203

\bibitem[\protect\citeauthoryear{{Martell} et~al.,}{{Martell}
  et~al.}{2020}]{MartellLi}
{Martell} S.,  et~al., 2020, arXiv e-prints, \href
  {https://ui.adsabs.harvard.edu/abs/2020arXiv200602106M} {p. arXiv:2006.02106}

\bibitem[\protect\citeauthoryear{Mayor \& Queloz}{Mayor \&
  Queloz}{1995}]{Mayor1995}
Mayor M.,  Queloz D.,  1995, Nature, 378, 355

\bibitem[\protect\citeauthoryear{{McDonough}}{{McDonough}}{2003}]{McDonough2003}
{McDonough} W.~F.,  2003, \mn@doi [Treatise on Geochemistry]
  {10.1016/B0-08-043751-6/02015-6}, \href
  {https://ui.adsabs.harvard.edu/abs/2003TrGeo...2..547M} {2, 568}

\bibitem[\protect\citeauthoryear{{McDonough} \& {Sun}}{{McDonough} \&
  {Sun}}{1995}]{McDonough95}
{McDonough} W.~F.,  {Sun} S.~s.,  1995, \mn@doi [Chemical Geology]
  {10.1016/0009-2541(94)00140-4}, \href
  {https://ui.adsabs.harvard.edu/abs/1995ChGeo.120..223M} {120, 223}

\bibitem[\protect\citeauthoryear{McKerns, Strand, Sullivan, Fang  \&
  Aivazis}{McKerns et~al.}{2012}]{multiprocessing}
McKerns M.~M.,  Strand L.,  Sullivan T.,  Fang A.,   Aivazis M.~A.,  2012,
  arXiv preprint arXiv:1202.1056

\bibitem[\protect\citeauthoryear{McKinney et~al.}{McKinney
  et~al.}{2010}]{pandas}
McKinney W.,  et~al., 2010, in Proceedings of the 9th Python in Science
  Conference. pp 51--56

\bibitem[\protect\citeauthoryear{Miozzi, Morard, Antonangeli, Clark, Mezouar,
  Dorn, Rozel  \& Fiquet}{Miozzi et~al.}{2018}]{carbonworldEoS}
Miozzi F.,  Morard G.,  Antonangeli D.,  Clark A.~N.,  Mezouar M.,  Dorn C.,
  Rozel A.,   Fiquet G.,  2018, \mn@doi [Journal of Geophysical Research:
  Planets] {10.1029/2018JE005582}, 123, 2295

\bibitem[\protect\citeauthoryear{{Mocquet}, {Grasset}  \& {Sotin}}{{Mocquet}
  et~al.}{2014}]{Mocquet2014}
{Mocquet} A.,  {Grasset} O.,   {Sotin} C.,  2014, \mn@doi [Philosophical
  Transactions of the Royal Society of London Series A]
  {10.1098/rsta.2013.0164}, \href
  {https://ui.adsabs.harvard.edu/abs/2014RSPTA.37230164M} {372, 20130164}

\bibitem[\protect\citeauthoryear{{Mordasini} et~al.,}{{Mordasini}
  et~al.}{2011}]{HD103197}
{Mordasini} C.,  et~al., 2011, \mn@doi [\aap] {10.1051/0004-6361/200913521},
  \href {https://ui.adsabs.harvard.edu/abs/2011A&A...526A.111M} {526, A111}

\bibitem[\protect\citeauthoryear{{Moriarty}, {Madhusudhan}  \&
  {Fischer}}{{Moriarty} et~al.}{2014}]{Moriarty14}
{Moriarty} J.,  {Madhusudhan} N.,   {Fischer} D.,  2014, \mn@doi [\apj]
  {10.1088/0004-637X/787/1/81}, \href
  {https://ui.adsabs.harvard.edu/abs/2014ApJ...787...81M} {787, 81}

\bibitem[\protect\citeauthoryear{{Mortier}, {Santos}, {Sousa}, {Adibekyan},
  {Delgado Mena}, {Tsantaki}, {Israelian}  \& {Mayor}}{{Mortier}
  et~al.}{2013}]{Mortier13}
{Mortier} A.,  {Santos} N.~C.,  {Sousa} S.~G.,  {Adibekyan} V.~Z.,  {Delgado
  Mena} E.,  {Tsantaki} M.,  {Israelian} G.,   {Mayor} M.,  2013, \mn@doi
  [\aap] {10.1051/0004-6361/201321641}, \href
  {https://ui.adsabs.harvard.edu/abs/2013A&A...557A..70M} {557, A70}

\bibitem[\protect\citeauthoryear{{Morton}}{{Morton}}{2015}]{isochrones}
{Morton} T.~D.,  2015, {isochrones: Stellar model grid package} (\mn@eprint
  {ascl} {1503.010})

\bibitem[\protect\citeauthoryear{{Muirhead} et~al.,}{{Muirhead}
  et~al.}{2012}]{TR2}
{Muirhead} P.~S.,  et~al., 2012, \mn@doi [\apj] {10.1088/0004-637X/747/2/144},
  \href {https://ui.adsabs.harvard.edu/abs/2012ApJ...747..144M} {747, 144}

\bibitem[\protect\citeauthoryear{{Naef, D.} et~al.,}{{Naef, D.}
  et~al.}{2001}]{Naef01}
{Naef, D.} et~al., 2001, \mn@doi [Astronomy \& Astrophysics]
  {10.1051/0004-6361:20010853}, 375, L27

\bibitem[\protect\citeauthoryear{{Ness}, {Hogg}, {Rix}, {Ho}  \&
  {Zasowski}}{{Ness} et~al.}{2015}]{thecannon}
{Ness} M.,  {Hogg} D.~W.,  {Rix} H.~W.,  {Ho} A. Y.~Q.,   {Zasowski} G.,  2015,
  \mn@doi [\apj] {10.1088/0004-637X/808/1/16}, \href
  {https://ui.adsabs.harvard.edu/abs/2015ApJ...808...16N} {808, 16}

\bibitem[\protect\citeauthoryear{{Nielsen} et~al.,}{{Nielsen}
  et~al.}{2019a}]{WASP182b}
{Nielsen} L.~D.,  et~al., 2019a, \mn@doi [\mnras] {10.1093/mnras/stz2351},
  \href {https://ui.adsabs.harvard.edu/abs/2019MNRAS.489.2478N} {489, 2478}

\bibitem[\protect\citeauthoryear{{Nielsen} et~al.,}{{Nielsen}
  et~al.}{2019b}]{TOI120}
{Nielsen} L.~D.,  et~al., 2019b, \mn@doi [\aap] {10.1051/0004-6361/201834577},
  \href {https://ui.adsabs.harvard.edu/abs/2019A&A...623A.100N} {623, A100}

\bibitem[\protect\citeauthoryear{{Nisr}, {Meng}, {MacDowell}, {Yan},
  {Prakapenka}  \& {Shim}}{{Nisr} et~al.}{2017}]{Nisr2017}
{Nisr} C.,  {Meng} Y.,  {MacDowell} A.~A.,  {Yan} J.,  {Prakapenka} V.,
  {Shim} S.~H.,  2017, \mn@doi [Journal of Geophysical Research (Planets)]
  {10.1002/2016JE005158}, \href
  {https://ui.adsabs.harvard.edu/abs/2017JGRE..122..124N} {122, 124}

\bibitem[\protect\citeauthoryear{{Nittler}, {Chabot}, {Grove}  \&
  {Peplowski}}{{Nittler} et~al.}{2017}]{MercuryComposition}
{Nittler} L.~R.,  {Chabot} N.~L.,  {Grove} T.~L.,   {Peplowski} P.~N.,  2017,
  arXiv e-prints, \href {https://ui.adsabs.harvard.edu/abs/2017arXiv171202187N}
  {p. arXiv:1712.02187}

\bibitem[\protect\citeauthoryear{{Noack}, {Snellen}  \& {Rauer}}{{Noack}
  et~al.}{2017}]{waterworld17}
{Noack} L.,  {Snellen} I.,   {Rauer} H.,  2017, \mn@doi [\ssr]
  {10.1007/s11214-017-0413-1}, \href
  {https://ui.adsabs.harvard.edu/abs/2017SSRv..212..877N} {212, 877}

\bibitem[\protect\citeauthoryear{{Noyes} et~al.,}{{Noyes} et~al.}{2008}]{TR1}
{Noyes} R.~W.,  et~al., 2008, \mn@doi [\apjl] {10.1086/527358}, \href
  {https://ui.adsabs.harvard.edu/abs/2008ApJ...673L..79N} {673, L79}

\bibitem[\protect\citeauthoryear{Oliphant}{Oliphant}{2006}]{numpy1}
Oliphant T.~E.,  2006, A guide to NumPy

\bibitem[\protect\citeauthoryear{{Osborn} \& {Bayliss}}{{Osborn} \&
  {Bayliss}}{2019}]{osbornhotjup}
{Osborn} A.,  {Bayliss} D.,  2019, arXiv e-prints, \href
  {https://ui.adsabs.harvard.edu/abs/2019arXiv191105830O} {p. arXiv:1911.05830}

\bibitem[\protect\citeauthoryear{{Osorio} \& {Barklem}}{{Osorio} \&
  {Barklem}}{2016}]{MgNLTEb}
{Osorio} Y.,  {Barklem} P.~S.,  2016, \mn@doi [\aap]
  {10.1051/0004-6361/201526958}, \href
  {https://ui.adsabs.harvard.edu/abs/2016A&A...586A.120O} {586, A120}

\bibitem[\protect\citeauthoryear{{Osorio}, {Barklem}, {Lind}, {Belyaev},
  {Spielfiedel}, {Guitou}  \& {Feautrier}}{{Osorio} et~al.}{2015}]{MgNLTEa}
{Osorio} Y.,  {Barklem} P.~S.,  {Lind} K.,  {Belyaev} A.~K.,  {Spielfiedel} A.,
   {Guitou} M.,   {Feautrier} N.,  2015, \mn@doi [\aap]
  {10.1051/0004-6361/201525846}, \href
  {https://ui.adsabs.harvard.edu/abs/2015A&A...579A..53O} {579, A53}

\bibitem[\protect\citeauthoryear{{Petigura} \& {Marcy}}{{Petigura} \&
  {Marcy}}{2011}]{Petigura11}
{Petigura} E.~A.,  {Marcy} G.~W.,  2011, \mn@doi [\apj]
  {10.1088/0004-637X/735/1/41}, \href
  {https://ui.adsabs.harvard.edu/abs/2011ApJ...735...41P} {735, 41}

\bibitem[\protect\citeauthoryear{{Petigura} et~al.,}{{Petigura}
  et~al.}{2018}]{CKSIV}
{Petigura} E.~A.,  et~al., 2018, \mn@doi [\aj] {10.3847/1538-3881/aaa54c},
  \href {https://ui.adsabs.harvard.edu/abs/2018AJ....155...89P} {155, 89}

\bibitem[\protect\citeauthoryear{Putirka \& Rarick}{Putirka \&
  Rarick}{2019}]{exoplanetRockyComp}
Putirka K.~D.,  Rarick J.~C.,  2019, \mn@doi [American Mineralogist]
  {10.2138/am-2019-6787}, 104, 817

\bibitem[\protect\citeauthoryear{{Recio-Blanco} et~al.,}{{Recio-Blanco}
  et~al.}{2014}]{RecioBlanco14}
{Recio-Blanco} A.,  et~al., 2014, \mn@doi [\aap] {10.1051/0004-6361/201322944},
  \href {https://ui.adsabs.harvard.edu/abs/2014A&A...567A...5R} {567, A5}

\bibitem[\protect\citeauthoryear{{Reddy}, {Lambert}  \& {Allende
  Prieto}}{{Reddy} et~al.}{2006}]{Reddy06}
{Reddy} B.~E.,  {Lambert} D.~L.,   {Allende Prieto} C.,  2006, \mn@doi [\mnras]
  {10.1111/j.1365-2966.2006.10148.x}, \href
  {https://ui.adsabs.harvard.edu/abs/2006MNRAS.367.1329R} {367, 1329}

\bibitem[\protect\citeauthoryear{Ricker et~al.,}{Ricker et~al.}{2014}]{TESS}
Ricker G.~R.,  et~al., 2014, \mn@doi [Journal of Astronomical Telescopes,
  Instruments, and Systems] {10.1117/1.JATIS.1.1.014003}, 1, 1

\bibitem[\protect\citeauthoryear{Rogers \& Seager}{Rogers \&
  Seager}{2010}]{Rogers10}
Rogers L.~A.,  Seager S.,  2010, The Astrophysical Journal, 712, 974

\bibitem[\protect\citeauthoryear{{Rowe} et~al.,}{{Rowe} et~al.}{2014}]{TR5}
{Rowe} J.~F.,  et~al., 2014, \mn@doi [\apj] {10.1088/0004-637X/784/1/45}, \href
  {https://ui.adsabs.harvard.edu/abs/2014ApJ...784...45R} {784, 45}

\bibitem[\protect\citeauthoryear{{Salaris}, {Chieffi}  \&
  {Straniero}}{{Salaris} et~al.}{1993}]{FeHtoMH}
{Salaris} M.,  {Chieffi} A.,   {Straniero} O.,  1993, \mn@doi [\apj]
  {10.1086/173105}, \href
  {https://ui.adsabs.harvard.edu/abs/1993ApJ...414..580S} {414, 580}

\bibitem[\protect\citeauthoryear{{Santerne} et~al.,}{{Santerne}
  et~al.}{2014}]{Kepler420}
{Santerne} A.,  et~al., 2014, \mn@doi [\aap] {10.1051/0004-6361/201424158},
  \href {https://ui.adsabs.harvard.edu/abs/2014A&A...571A..37S} {571, A37}

\bibitem[\protect\citeauthoryear{{Santos}, {Israelian}  \& {Mayor}}{{Santos}
  et~al.}{2001}]{hotjupfeh01}
{Santos} N.~C.,  {Israelian} G.,   {Mayor} M.,  2001, \mn@doi [\aap]
  {10.1051/0004-6361:20010648}, \href
  {https://ui.adsabs.harvard.edu/abs/2001A&A...373.1019S} {373, 1019}

\bibitem[\protect\citeauthoryear{{Schubert}, {Anderson}, {Travis}  \&
  {Palguta}}{{Schubert} et~al.}{2007}]{Enceladus07}
{Schubert} G.,  {Anderson} J.~D.,  {Travis} B.~J.,   {Palguta} J.,  2007,
  \mn@doi [\icarus] {10.1016/j.icarus.2006.12.012}, \href
  {https://ui.adsabs.harvard.edu/abs/2007Icar..188..345S} {188, 345}

\bibitem[\protect\citeauthoryear{{Seager}, {Kuchner}, {Hier-Majumder}  \&
  {Militzer}}{{Seager} et~al.}{2007}]{Seager2007}
{Seager} S.,  {Kuchner} M.,  {Hier-Majumder} C.~A.,   {Militzer} B.,  2007,
  \mn@doi [\apj] {10.1086/521346}, \href
  {https://ui.adsabs.harvard.edu/abs/2007ApJ...669.1279S} {669, 1279}

\bibitem[\protect\citeauthoryear{Shallue \& Vanderburg}{Shallue \&
  Vanderburg}{2018}]{Shallue18}
Shallue C.~J.,  Vanderburg A.,  2018, The Astronomical Journal, 155, 94

\bibitem[\protect\citeauthoryear{{Sharma} et~al.,}{{Sharma}
  et~al.}{2018}]{TESSHERMES}
{Sharma} S.,  et~al., 2018, \mn@doi [\mnras] {10.1093/mnras/stx2582}, \href
  {https://ui.adsabs.harvard.edu/abs/2018MNRAS.473.2004S} {473, 2004}

\bibitem[\protect\citeauthoryear{{Sharma} et~al.,}{{Sharma}
  et~al.}{2019}]{K2HERMESsanjib}
{Sharma} S.,  et~al., 2019, \mn@doi [\mnras] {10.1093/mnras/stz2861}, \href
  {https://ui.adsabs.harvard.edu/abs/2019MNRAS.490.5335S} {490, 5335}

\bibitem[\protect\citeauthoryear{{Sheinis} et~al.,}{{Sheinis}
  et~al.}{2015}]{HERMES15}
{Sheinis} A.,  et~al., 2015, \mn@doi [Journal of Astronomical Telescopes,
  Instruments, and Systems] {10.1117/1.JATIS.1.3.035002}, \href
  {https://ui.adsabs.harvard.edu/abs/2015JATIS...1c5002S} {1, 035002}

\bibitem[\protect\citeauthoryear{{Skrutskie} et~al.,}{{Skrutskie}
  et~al.}{2006}]{TWOMASS}
{Skrutskie} M.~F.,  et~al., 2006, \mn@doi [\aj] {10.1086/498708}, \href
  {https://ui.adsabs.harvard.edu/abs/2006AJ....131.1163S} {131, 1163}

\bibitem[\protect\citeauthoryear{Smith et~al.,}{Smith et~al.}{2012}]{WASP61}
Smith A. M.~S.,  et~al., 2012, \mn@doi [Monthly Notices of the Royal
  Astronomical Society] {10.1111/j.1365-2966.2012.21780.x}, 426, 739

\bibitem[\protect\citeauthoryear{{Sousa} et~al.,}{{Sousa}
  et~al.}{2019}]{Sousa2019}
{Sousa} S.~G.,  et~al., 2019, \mn@doi [\mnras] {10.1093/mnras/stz664}, \href
  {https://ui.adsabs.harvard.edu/abs/2019MNRAS.485.3981S} {485, 3981}

\bibitem[\protect\citeauthoryear{{Stassun}, {Collins}  \& {Gaudi}}{{Stassun}
  et~al.}{2017}]{WASP61bparams}
{Stassun} K.~G.,  {Collins} K.~A.,   {Gaudi} B.~S.,  2017, \mn@doi [\aj]
  {10.3847/1538-3881/aa5df3}, \href
  {https://ui.adsabs.harvard.edu/abs/2017AJ....153..136S} {153, 136}

\bibitem[\protect\citeauthoryear{Stassun et~al.,}{Stassun
  et~al.}{2018}]{TIC_CTL}
Stassun K.~G.,  et~al., 2018, \mn@doi [The Astronomical Journal]
  {10.3847/1538-3881/aad050}, 156, 102

\bibitem[\protect\citeauthoryear{{Stassun} et~al.,}{{Stassun}
  et~al.}{2019}]{revisedTIC}
{Stassun} K.~G.,  et~al., 2019, \mn@doi [\aj] {10.3847/1538-3881/ab3467}, \href
  {https://ui.adsabs.harvard.edu/abs/2019AJ....158..138S} {158, 138}

\bibitem[\protect\citeauthoryear{{Steinmetz} et~al.,}{{Steinmetz}
  et~al.}{2006}]{RAVE}
{Steinmetz} M.,  et~al., 2006, \mn@doi [\aj] {10.1086/506564}, \href
  {https://ui.adsabs.harvard.edu/abs/2006AJ....132.1645S} {132, 1645}

\bibitem[\protect\citeauthoryear{{Suissa}, {Chen}  \& {Kipping}}{{Suissa}
  et~al.}{2018}]{legendoftherentwasway}
{Suissa} G.,  {Chen} J.,   {Kipping} D.,  2018, \mn@doi [\mnras]
  {10.1093/mnras/sty381}, \href
  {https://ui.adsabs.harvard.edu/abs/2018MNRAS.476.2613S} {476, 2613}

\bibitem[\protect\citeauthoryear{Sullivan et~al.,}{Sullivan
  et~al.}{2015}]{Sullivan15}
Sullivan P.~W.,  et~al., 2015, The Astrophysical Journal, 809, 77

\bibitem[\protect\citeauthoryear{{Tasker} et~al.,}{{Tasker}
  et~al.}{2017}]{Tasker2017}
{Tasker} E.,  et~al., 2017, \mn@doi [Nature Astronomy]
  {10.1038/s41550-017-0042}, \href
  {https://ui.adsabs.harvard.edu/abs/2017NatAs...1E..42T} {1, 0042}

\bibitem[\protect\citeauthoryear{{Taylor}}{{Taylor}}{2005}]{TOPCAT}
{Taylor} M.~B.,  2005, {TOPCAT \&amp; STIL: Starlink Table/VOTable Processing
  Software}.
p.~29

\bibitem[\protect\citeauthoryear{{Teske}, {Cunha}, {Schuler}, {Griffith}  \&
  {Smith}}{{Teske} et~al.}{2013}]{Teske13}
{Teske} J.~K.,  {Cunha} K.,  {Schuler} S.~C.,  {Griffith} C.~A.,   {Smith}
  V.~V.,  2013, \mn@doi [\apj] {10.1088/0004-637X/778/2/132}, \href
  {https://ui.adsabs.harvard.edu/abs/2013ApJ...778..132T} {778, 132}

\bibitem[\protect\citeauthoryear{{Teske}, {Cunha}, {Smith}, {Schuler}  \&
  {Griffith}}{{Teske} et~al.}{2014}]{Teske14}
{Teske} J.~K.,  {Cunha} K.,  {Smith} V.~V.,  {Schuler} S.~C.,   {Griffith}
  C.~A.,  2014, \mn@doi [\apj] {10.1088/0004-637X/788/1/39}, \href
  {https://ui.adsabs.harvard.edu/abs/2014ApJ...788...39T} {788, 39}

\bibitem[\protect\citeauthoryear{{Teske}, {Thorngren}, {Fortney}, {Hinkel}  \&
  {Brewer}}{{Teske} et~al.}{2019}]{Teske_metalrichhosts}
{Teske} J.~K.,  {Thorngren} D.,  {Fortney} J.~J.,  {Hinkel} N.,   {Brewer}
  J.~M.,  2019, \mn@doi [\aj] {10.3847/1538-3881/ab4f79}, \href
  {https://ui.adsabs.harvard.edu/abs/2019AJ....158..239T} {158, 239}

\bibitem[\protect\citeauthoryear{{Thiabaud}, {Marboeuf}, {Alibert}, {Cabral},
  {Leya}  \& {Mezger}}{{Thiabaud} et~al.}{2014a}]{Thiabaud2014}
{Thiabaud} A.,  {Marboeuf} U.,  {Alibert} Y.,  {Cabral} N.,  {Leya} I.,
  {Mezger} K.,  2014a, \mn@doi [\aap] {10.1051/0004-6361/201322208}, \href
  {https://ui.adsabs.harvard.edu/abs/2014A&A...562A..27T} {562, A27}

\bibitem[\protect\citeauthoryear{{Thiabaud}, {Marboeuf}, {Alibert}, {Cabral},
  {Leya}  \& {Mezger}}{{Thiabaud} et~al.}{2014b}]{Thiabaud14}
{Thiabaud} A.,  {Marboeuf} U.,  {Alibert} Y.,  {Cabral} N.,  {Leya} I.,
  {Mezger} K.,  2014b, \mn@doi [\aap] {10.1051/0004-6361/201322208}, \href
  {https://ui.adsabs.harvard.edu/abs/2014A&A...562A..27T} {562, A27}

\bibitem[\protect\citeauthoryear{{Thiabaud}, {Marboeuf}, {Alibert}, {Leya}  \&
  {Mezger}}{{Thiabaud} et~al.}{2015}]{Thiabaud2015}
{Thiabaud} A.,  {Marboeuf} U.,  {Alibert} Y.,  {Leya} I.,   {Mezger} K.,  2015,
  \mn@doi [\aap] {10.1051/0004-6361/201525963}, \href
  {https://ui.adsabs.harvard.edu/abs/2015A&A...580A..30T} {580, A30}

\bibitem[\protect\citeauthoryear{{Torres} et~al.,}{{Torres}
  et~al.}{2015}]{Torres15_KOIval}
{Torres} G.,  et~al., 2015, \mn@doi [\apj] {10.1088/0004-637X/800/2/99}, \href
  {https://ui.adsabs.harvard.edu/abs/2015ApJ...800...99T} {800, 99}

\bibitem[\protect\citeauthoryear{{Unterborn} \& {Panero}}{{Unterborn} \&
  {Panero}}{2017}]{Unterborn17}
{Unterborn} C.~T.,  {Panero} W.~R.,  2017, \mn@doi [\apj]
  {10.3847/1538-4357/aa7f79}, \href
  {https://ui.adsabs.harvard.edu/abs/2017ApJ...845...61U} {845, 61}

\bibitem[\protect\citeauthoryear{Unterborn \& Panero}{Unterborn \&
  Panero}{2019}]{UnterbornPanero2019}
Unterborn C.~T.,  Panero W.~R.,  2019, \mn@doi [Journal of Geophysical
  Research: Planets] {10.1029/2018JE005844}, 124, 1704

\bibitem[\protect\citeauthoryear{{Unterborn}, {Kabbes}, {Pigott}, {Reaman}  \&
  {Panero}}{{Unterborn} et~al.}{2014}]{Unterborn14}
{Unterborn} C.~T.,  {Kabbes} J.~E.,  {Pigott} J.~S.,  {Reaman} D.~M.,
  {Panero} W.~R.,  2014, \mn@doi [\apj] {10.1088/0004-637X/793/2/124}, \href
  {https://ui.adsabs.harvard.edu/abs/2014ApJ...793..124U} {793, 124}

\bibitem[\protect\citeauthoryear{{Unterborn}, {Dismukes}  \&
  {Panero}}{{Unterborn} et~al.}{2016}]{Unterborn2016}
{Unterborn} C.~T.,  {Dismukes} E.~E.,   {Panero} W.~R.,  2016, \mn@doi [\apj]
  {10.3847/0004-637X/819/1/32}, \href
  {https://ui.adsabs.harvard.edu/abs/2016ApJ...819...32U} {819, 32}

\bibitem[\protect\citeauthoryear{{Unterborn}, {Desch}, {Hinkel}  \&
  {Lorenzo}}{{Unterborn} et~al.}{2018a}]{Unterborn2018}
{Unterborn} C.~T.,  {Desch} S.~J.,  {Hinkel} N.~R.,   {Lorenzo} A.,  2018a,
  \mn@doi [Nature Astronomy] {10.1038/s41550-018-0411-6}, \href
  {https://ui.adsabs.harvard.edu/abs/2018NatAs...2..297U} {2, 297}

\bibitem[\protect\citeauthoryear{{Unterborn}, {Desch}  \& {Panero}}{{Unterborn}
  et~al.}{2018b}]{exoplex}
{Unterborn} C.~T.,  {Desch} S.~J.,   {Panero} W.~R.,  2018b, in AGU Fall
  Meeting Abstracts. pp P42A--02

\bibitem[\protect\citeauthoryear{{Valencia}, {O'Connell}  \&
  {Sasselov}}{{Valencia} et~al.}{2006}]{Valencia2006}
{Valencia} D.,  {O'Connell} R.~J.,   {Sasselov} D.,  2006, \mn@doi [\icarus]
  {10.1016/j.icarus.2005.11.021}, \href
  {https://ui.adsabs.harvard.edu/abs/2006Icar..181..545V} {181, 545}

\bibitem[\protect\citeauthoryear{{Valenti} \& {Piskunov}}{{Valenti} \&
  {Piskunov}}{1996}]{SMEold}
{Valenti} J.~A.,  {Piskunov} N.,  1996, \aaps, \href
  {https://ui.adsabs.harvard.edu/abs/1996A&AS..118..595V} {118, 595}

\bibitem[\protect\citeauthoryear{{Vanderspek} et~al.,}{{Vanderspek}
  et~al.}{2019}]{Vanderspek2019}
{Vanderspek} R.,  et~al., 2019, \mn@doi [\apjl] {10.3847/2041-8213/aafb7a},
  \href {https://ui.adsabs.harvard.edu/abs/2019ApJ...871L..24V} {871, L24}

\bibitem[\protect\citeauthoryear{{Virtanen} et~al.,}{{Virtanen}
  et~al.}{2020}]{scipy}
{Virtanen} P.,  et~al., 2020, \mn@doi [Nature Methods]
  {https://doi.org/10.1038/s41592-019-0686-2}, \href {https://rdcu.be/b08Wh}
  {17, 261}

\bibitem[\protect\citeauthoryear{{Vogt} et~al.,}{{Vogt} et~al.}{2010}]{RV1}
{Vogt} S.~S.,  et~al., 2010, \mn@doi [\apj] {10.1088/0004-637X/708/2/1366},
  \href {https://ui.adsabs.harvard.edu/abs/2010ApJ...708.1366V} {708, 1366}

\bibitem[\protect\citeauthoryear{Wang, Zhang, Zhang  \& Yi}{Wang
  et~al.}{2013}]{openblas13}
Wang Q.,  Zhang X.,  Zhang Y.,   Yi Q.,  2013, in SC'13: Proceedings of the
  International Conference on High Performance Computing, Networking, Storage
  and Analysis. pp 1--12

\bibitem[\protect\citeauthoryear{{Wang} et~al.,}{{Wang}
  et~al.}{2019a}]{SonghuTESS}
{Wang} S.,  et~al., 2019a, \mn@doi [\aj] {10.3847/1538-3881/aaf1b7}, \href
  {https://ui.adsabs.harvard.edu/abs/2019AJ....157...51W} {157, 51}

\bibitem[\protect\citeauthoryear{{Wang}, {Liu}, {Ireland}, {Brasser}, {Yong}
  \& {Lineweaver}}{{Wang} et~al.}{2019b}]{EarthWang}
{Wang} H.~S.,  {Liu} F.,  {Ireland} T.~R.,  {Brasser} R.,  {Yong} D.,
  {Lineweaver} C.~H.,  2019b, \mn@doi [\mnras] {10.1093/mnras/sty2749}, \href
  {https://ui.adsabs.harvard.edu/abs/2019MNRAS.482.2222W} {482, 2222}

\bibitem[\protect\citeauthoryear{{Wanke} \& {Dreibus}}{{Wanke} \&
  {Dreibus}}{1994}]{Wanke1994}
{Wanke} H.,  {Dreibus} G.,  1994, \mn@doi [Philosophical Transactions of the
  Royal Society of London Series A] {10.1098/rsta.1994.0132}, \href
  {https://ui.adsabs.harvard.edu/abs/1994RSPTA.349..285W} {349, 285}

\bibitem[\protect\citeauthoryear{{Wilson} \& {Militzer}}{{Wilson} \&
  {Militzer}}{2014}]{Wilson14}
{Wilson} H.~F.,  {Militzer} B.,  2014, \mn@doi [\apj]
  {10.1088/0004-637X/793/1/34}, \href
  {https://ui.adsabs.harvard.edu/abs/2014ApJ...793...34W} {793, 34}

\bibitem[\protect\citeauthoryear{{Wittenmyer} et~al.,}{{Wittenmyer}
  et~al.}{2014}]{RV4}
{Wittenmyer} R.~A.,  et~al., 2014, \mn@doi [\apj]
  {10.1088/0004-637X/783/2/103}, \href
  {https://ui.adsabs.harvard.edu/abs/2014ApJ...783..103W} {783, 103}

\bibitem[\protect\citeauthoryear{Wittenmyer et~al.,}{Wittenmyer
  et~al.}{2017}]{Wittenmyer17}
Wittenmyer R.~A.,  et~al., 2017, The Astronomical Journal, 154, 274

\bibitem[\protect\citeauthoryear{{Wittenmyer} et~al.,}{{Wittenmyer}
  et~al.}{2018}]{K2HERMESours}
{Wittenmyer} R.~A.,  et~al., 2018, \mn@doi [\aj] {10.3847/1538-3881/aaa3e4},
  \href {https://ui.adsabs.harvard.edu/abs/2018AJ....155...84W} {155, 84}

\bibitem[\protect\citeauthoryear{{Wittenmyer} et~al.,}{{Wittenmyer}
  et~al.}{2020}]{bigboisPPPS}
{Wittenmyer} R.~A.,  et~al., 2020, \mn@doi [\mnras] {10.1093/mnras/stz3378},
  \href {https://ui.adsabs.harvard.edu/abs/2020MNRAS.491.5248W} {491, 5248}

\bibitem[\protect\citeauthoryear{{Wolszczan} \& {Frail}}{{Wolszczan} \&
  {Frail}}{1992}]{Lich}
{Wolszczan} A.,  {Frail} D.~A.,  1992, \mn@doi [\nat] {10.1038/355145a0}, \href
  {https://ui.adsabs.harvard.edu/abs/1992Natur.355..145W} {355, 145}

\bibitem[\protect\citeauthoryear{Xianyi, Qian  \& Yunquan}{Xianyi
  et~al.}{2012}]{openblas12}
Xianyi Z.,  Qian W.,   Yunquan Z.,  2012, in 2012 IEEE 18th international
  conference on parallel and distributed systems. pp 684--691

\bibitem[\protect\citeauthoryear{{Zeng}, {Sasselov}  \& {Jacobsen}}{{Zeng}
  et~al.}{2016}]{zengPREM}
{Zeng} L.,  {Sasselov} D.~D.,   {Jacobsen} S.~B.,  2016, \mn@doi [\apj]
  {10.3847/0004-637X/819/2/127}, \href
  {https://ui.adsabs.harvard.edu/abs/2016ApJ...819..127Z} {819, 127}

\bibitem[\protect\citeauthoryear{{Zolotov}, {Tobie}, {Postberg}, {Magee},
  {Waite}  \& {Esposito}}{{Zolotov} et~al.}{2011}]{Enceladus11}
{Zolotov} M.~Y.,  {Tobie} G.,  {Postberg} F.,  {Magee} B.,  {Waite} J.~H.,
  {Esposito} L.,  2011, in EPSC-DPS Joint Meeting 2011. p.~1330

\bibitem[\protect\citeauthoryear{{van der Walt}, {Colbert}  \&
  {Varoquaux}}{{van der Walt} et~al.}{2011}]{numpy2}
{van der Walt} S.,  {Colbert} S.~C.,   {Varoquaux} G.,  2011, \mn@doi
  [Computing in Science and Engineering] {10.1109/MCSE.2011.37}, \href
  {https://ui.adsabs.harvard.edu/abs/2011CSE....13b..22V} {13, 22}

\makeatother
\end{thebibliography}

% \begin{thebibliography}{99}
% \bibitem[\protect\citeauthoryear{Author}{2012}]{Author2012}
% Author A.~N., 2013, Journal of Improbable Astronomy, 1, 1
% \bibitem[\protect\citeauthoryear{Others}{2013}]{Others2013}
% Others S., 2012, Journal of Interesting Stuff, 17, 198
% \end{thebibliography}

%%%%%%%%%%%%%%%%%%%%%%%%%%%%%%%%%%%%%%%%%%%%%%%%%%

%%%%%%%%%%%%%%%%% APPENDICES %%%%%%%%%%%%%%%%%%%%%

\appendix

\begin{table*}
\begin{center}
\caption{Column names, units, data types and descriptions for the GALAH-TESS physical parameters table}
\begin{tabular}{*{4}{l}} 

\hline
Name & Units & Data type & Description\\
\hline
\hline
\textsc{tic\_id} & & int64 & \textit{TESS} Input Catalog (TIC) Identifier\\
\textsc{twomass} & & int64 & Two Micron All-Sky Survey (2MASS) Identifier\\
\textsc{gaiadr2} & & long64 & GAIA DR2 Identifier\\
\textsc{priority} & & float64 & TIC v8 priority \\
\textsc{ra} & deg & float64 & J2000 right ascension from 2MASS\\
\textsc{dec} & deg & float64 & J2000 declination from 2MASS\\
\textsc{teff} & K & int64 & GALAH DR2 Effective temperature\\
\textsc{e\_teff} & K & int64 & Uncertainty in \textsc{teff}\\
\textsc{logg} & dex & float64 & GALAH DR2 Surface gravity\\
\textsc{e\_logg} & dex & float64 & Uncertainty in \textsc{logg}\\
\textsc{m\_h} & dex & float64 & GALAH DR2 overall metallicity\\
\textsc{e\_mh} & dex & float64 & Uncertainty in \textsc{m\_h}\\
\textsc{alpha\_fe} & dex & float64 & [$\alpha$/Fe] abundance\\
\textsc{e\_alpha\_fe} & dex & float64 & Uncertainty in \textsc{alpha\_fe}\\
\textsc{Vmag} & mag & float64 & V magnitude from TIC\\
\textsc{e\_Vmag} & mag & float64 & Uncertainty in \textsc{Vmag}\\
\textsc{Tmag} & mag & float64 & \textit{TESS} magnitude from TIC\\
\textsc{e\_Tmag} & mag & float64 & Uncertainty in \textsc{Tmag}\\
\textsc{Hmag} & mag & float64 & 2MASS H magnitude from TIC\\
\textsc{e\_Hmag} & mag & float64 & Uncertainty in \textsc{Hmag}\\
\textsc{Jmag} & mag & float64 & 2MASS J magnitude from TIC\\
\textsc{e\_Jmag} & mag & float64 & Uncertainty in \textsc{Jmag}\\
\textsc{Kmag} & mag & float64 & 2MASS K magnitude from TIC\\
\textsc{e\_Kmag} & mag & float64 & Uncertainty in \textsc{Kmag}\\
\textsc{Gmag} & mag & float64 & GAIA G magnitude from TIC\\
\textsc{e\_Gmag} & mag & float64 & Uncertainty in \textsc{Gmag}\\
\textsc{GRPmag} & mag & float64 & GAIA G\textunderscore{RP} magnitude from TIC\\
\textsc{e\_GRPmag} & mag & float64 & Uncertainty in \textsc{GRPmag}\\
\textsc{GBPmag} & mag & float64 & GAIA G\textunderscore{BP} magnitude from TIC\\
\textsc{e\_GBPmag} & mag & float64 & Uncertainty in \textsc{GBPmag}\\
\textsc{plx} & mas & float64 & Parallax from TIC\\
\textsc{e\_plx} & mas & float64 & Uncertainty in \textsc{plx}\\
\textsc{dist} & pc & float64 & Distance from TIC\\
\textsc{e\_dist} & pc & float64 & Uncertainty in \textsc{dist}\\
\textsc{radius} & $R_\odot$ & float64 & isochrone Stellar radius\\
\textsc{e\_radius} & $R_\odot$ & float64 & Uncertainty in \textsc{radius}\\
\textsc{mass} & $M_\odot$ & float64 & isochrone Stellar mass\\
\textsc{e\_mass} & $M_\odot$ & float64 & Uncertainty in \textsc{mass}\\
\textsc{rho} & gcm$^{-3}$ & float64 & isochrone Stellar density\\
\textsc{e\_rho} & gcm$^{-3}$ & float64 & Uncertainty in \textsc{rho}\\
\textsc{lum} & $L_\odot$ & float64 & Stellar luminosity\\
\textsc{e\_lum} & $L_\odot$ & float64 & Uncertainty in \textsc{lum}\\
\textsc{age} & Gyr & float64 & isochrone Stellar age\\
\textsc{e\_age} & Gyr & float64 & Uncertainty in \textsc{age}\\
\textsc{eep} & & int64 & MIST isochrone equivalent evolutionary phase\\
\textsc{e\_eep} & & int64 & Uncertainty in \textsc{eep}\\
\textsc{rv} & kms$^{-1}$ & float64 & GALAH DR2 Radial velocity from internal cross-correlation against data\\
\textsc{e\_rv} & kms$^{-1}$ & float64 & Uncertainty in \textsc{rv}\\
\textsc{vsini} & kms$^{-1}$ & float64 & GALAH DR2 Line of sight rotational velocity\\
\textsc{e\_vsini} & kms$^{-1}$ & float64 & Uncertainty in \textsc{vsini}\\
\textsc{vmic} & kms$^{-1}$ & float64 & GALAH DR2 Microturbulence velocity\\
\textsc{e\_vmic} & kms$^{-1}$ & float64 & Uncertainty in \textsc{vmic}\\
\textsc{hzRecVen} & AU & float64 & Recent Venus Habitable Zone\\
\textsc{hzRunGrn} & AU & float64 & Runaway Greenhouse Habitable Zone\\
\textsc{hzMoiGrn} & AU & float64 & Moist Greenhouse Habitable Zone\\
\textsc{hzMaxGrn} & AU & float64 & Maximum Greenhouse Habitable Zone\\
\textsc{hzEarMar} & AU & float64 & Early Mars Habitable Zone\\
\textsc{x\_h} & dex & float64 & [X/H] abundance for element X\\
\textsc{e\_x\_h} & dex & float64 & [X/H] Uncertainty in \textsc{x\_h}\\
\textsc{c\_o} & & float64 & (C/O) abundance ratio\\
\textsc{e\_c\_o} & & float64 & Uncertainty in \textsc{c\_o}\\
\textsc{mg\_si} & & float64 & (Mg/Si) abundance ratio\\
\textsc{e\_mg\_si} & & float64 & Uncertainty in \textsc{mg\_si}\\
\textsc{fe\_mg} & & float64 & (Fe/Mg) abundance ratio\\
\textsc{e\_fe\_mg} & & float64 & Uncertainty in \textsc{fe\_mg}\\
\textsc{fe\_si} & & float64 & (Fe/Si) abundance ratio\\
\textsc{e\_fe\_si} & & float64 & Uncertainty in \textsc{fe\_si}\\

\hline

\end{tabular}
\end{center}
\end{table*}

%%%%%%%%%%%%%%%%%%%%%%%%%%%%%%%%%%%%%%%%%%%%%%%%%%

% Don't change these lines
\bsp % typesetting comment
\label{lastpage}
\end{document}